\documentclass[twocolumn,secnumarabic,amssymb, aps, prl]{revtex4-1}
\usepackage{graphicx}
\usepackage{amsmath}
\usepackage{amsfonts}
\usepackage{multirow}
\usepackage[colorlinks, linkcolor=blue, anchorcolor=blue,citecolor=blue]{hyperref}

\begin{document}
	\title{Non-Fermi-Liquid/Marginal-Fermi-Liquid Signatures Induced by Van Hove Singularity}%
	\author{Yi-Hui Xing$^{1,2}$}
	\author{Wu-Ming Liu$^{2}$}
	\email{wliu@iphy.ac.cn}
	\author{Xiao-Tian Zhang$^{1,3}$}
	\email{xtzhang1@bjtu.edu.cn}
	\affiliation{$^{1}$School of Physical Sciences and Engineering, Beijing Jiaotong University, Beijing 100044, China;}
	\affiliation{$^{2}$Beijing National Laboratory for Condensed Matter Physics, Institute of Physics, Chinese Academy of Sciences, Beijing, 100190, China;}
	\affiliation{$^{3}$Kavli Institute for Theoretical Sciences, University of Chinese Academy of Sciences, Beijing, 100190, China.}

	\begin{abstract}
	Strange metal behavior is widely observed in cuprates, ruthenates, and twisted bilayer graphene.
We study quantum critical transport at a two-dimensional Lifshitz transition, 
where the Fermi surface hosts a van Hove singularity and changes from convex to concave.
Self-consistently solving the fermion-boson coupled system, 
we demonstrate a linear-in-$1/\omega$ optical conductivity driven by the interplay between impurity and critical scattering.
The resistivity exhibits a persistent linear-in-$T$ in the quantum critical regime down to $T\rightarrow 0$.
Spatially fluctuating Yukawa interaction further extends the linear-in-$T$ regime, 
providing a unified mechanism for strange metallic transport at the Lifshitz transition.
	\end{abstract}
	
	\maketitle
	
	The strange metal, initially discovered in high-temperature superconductors such as cuprates\cite{Greene2020}, 
iron pnictides\cite{Shibauchi2014} and heavy fermion metals\cite{Paschen2020},
has more recently been observed in materials such as twisted bilayer graphene\cite{Cao2020,Jaoui2022} and dichalcogenides\cite{Ghiotto2021}.
The strange metal exhibits an ubiquitous linear-in-$T$ resistivity in a broad temperature regime
that extends from very high temperatures down to near zero temperature.
In addition, the specific heat shows a logarithmic enhancement $\sim T\ln T$ compared to the Fermi liquid metals.
The prevailing consensus is that understanding the strange metallicity
is the key to unravelling the mystery of high-$T_{\rm c}$ superconductivity itself.
One scenario for the high-$T_{\rm c}$ cuprates is that
the enigmatic pseudogap phase terminates at a critical doping rate $p_{\rm c}$
where a putative quantum critical point (QCP) is masked by the superconducting dome\cite{Michon2019,Grissonnanche2021,Zhu2022}.
The anomalous strange metal then dwells in the finite-temperature quantum critical regime mediated by a Pomeranchuk-type QCP.
	
	Intriguingly, in several cuprate materials\cite{Benhabib2015} 
$p_{\rm c}$ is very close to, if not exactly at, the Lifshitz transition point $p^\ast$\cite{Volovik2017} 
where the Fermi surface undergoes a topological transition 
and the van Hove singularity (VHS) crosses the Fermi surface.
The Fermi surface topology imposes a strict constraint on the pseudogap region, 
requiring $p_{\rm c} \lesssim p^\ast$\cite{DoironLeyraud2017}. 
Consistent with this constraint, numerical calculations find that the
pseudogap endpoint satisfies $p_{\rm c}\leq p^\ast$, with the two
coinciding in the strong-coupling regime \cite{PhysRevX.8.021048}.
This perspective is also shared by recent experiments on twisted bilayer graphene\cite{Cao2020}:
The strange metal transport is restricted to half-filling points, 
at which the Hall coefficient changes sign\cite{Andrea2019} indicating a Lifshitz transition\cite{Kim2016,PhysRevX.8.041041}.
Theoretically, the linear-in-$T$ resistivity is predicted near a Lifshitz transition, 
arising from the scattering of electrons off an overdamped critical boson 
in the presence of both impurity and umklapp scattering\cite{PALee2021,PhysRevB.108.134502,PhysRevB.110.094520}.
However, the linear-in-$T$ resistivity gives way to the Fermi liquid behavior in the low-$T$ limit
due to the finite distance towards the Lifshitz transition point.
Therefore, it is important to clarify the low-energy NFL behaviors and the associated transport properties 
exactly at the Lifshitz transition in a controlled theoretical framework.
More progressively, whether the interplay between the topological VHS and the QCP
can lead to the strange metal behaviors in the mediated quantum critical regime.

\begin{figure}
	\centering
	\includegraphics[width=\linewidth]{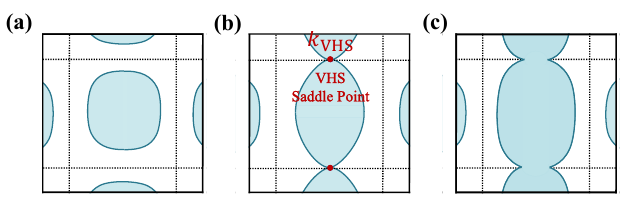}
	\caption{\textbf{Evolution of the Fermi surface geometry across the 2D Lifshitz transition.}
	(a) Convex Fermi surface; (b) Lifshitz transition point; (c) Concave Fermi surface.
	The solid blue lines represent the Fermi surface enclosing the occupied regions with a shaded blue color.
	The VHS saddle points are plotted by red dots at $k_{v}$. 
	}
	\label{fig1}
\end{figure}

In this paper, we study quantum transport in two-dimensional systems 
at the Lifshitz transition and near a Pomeranchuk instability, 
where the Fermi surface undergoes a convex-to-concave reconstruction. 
The VHS and critical fluctuations are intertwined through Yukawa interactions.
Within a self-consistent low-energy framework in the presence of impurity scattering, 
we uncover a distinct non-Fermi liquid (NFL) state at the VHS saddle point, 
featuring a specific heat $C \sim T\ln T$ that is insensitive to details of many-body correlations.
Transport arises from both impurity and normal electron-electron (\emph{ee}) scattering in the extra channel.
The optical conductivity exhibits a linear-in-$\omega$ dependence dominated by the VHS, 
while contributions from the convex Fermi surface vanish.
At finite temperatures, the $\emph{dc}$ resistivity is linear-in-$T$ down to $T=0$,
indicating robust strange metallicity that is not short-circuited by other Fermi surface regions.
Further incorporating spatially random Yukawa interaction reproduces a marginal Fermi liquid near the VHS, 
establishing a unified framework for strange-metal transport extending beyond the quantum critical regime.\\

\section{Non-Fermi Liquids at VHS}
As shown in Fig.~\ref{fig1}, the Lifshitz transition represents a particular type of 
convex-to-concave transition for a 2D Fermi surface.
At the Lifshitz transition [Fig.~\ref{fig1}(b)], the Fermi surface consists of two parts: the generic convex Fermi surface 
and the VHS saddle points at ${\bm k}= {\bm k}_{v}$.
We investigate the fate of a Fermi liquid at VHS saddle points
subjected to coupling with critical bosons via Yukawa interactions.
The critical boson carries a zero momentum, simultaneously coupling to all patches on the Fermi surface.
We focus on the single-patch low-energy effective theory at the VHS saddle point, which reads
\begin{equation}
	\begin{aligned}
		& \mathcal{L}= \mathcal{L}_0 + \mathcal{L}_w  + \mathcal{L}_J, \\
		&\mathcal{L}_0 =  \sum_{m}\psi^\dagger_m(\partial_\tau-\partial_x^2+a\partial_y^2)\psi_m+\sum_l \phi_l^*(\Delta -\partial_\tau^2 - \bm{\nabla}^2 )\phi_l,  \\
		&\mathcal{L}_w = \frac{1}{\sqrt{N}}\sum_{mn}w_{mn}(\bm{x})\psi_m^\dagger({\bm x},\tau)\psi_n({\bm x},\tau),\\
		&  \mathcal{L}_{J}= \frac{1}{\sqrt{NN'}}\sum_{mnl}J_{mnl}\psi^\dagger_m({\bm x},\tau)\psi_n({\bm x},\tau)\phi_l({\bm x},\tau).
	\end{aligned}
	\label{eff}
\end{equation}
Here $\phi$ and $\psi$ are the boson and fermion field operators
with flavor indices $m,n=1,2...,N$ and $l=1,2...,N^\prime$, respectively. 
The fermionic dispersion takes the saddle-point form
$
\epsilon_{\bm k}\simeq t a_0^2 k_x^2 - a t a_0^2 k_y^2 \quad (a>0),
$
where $t$ denotes the hopping parameter and $a_0$ the lattice constant. We set
$t=a_0=1$ throughout the main text and discuss the role of physical units in
Appendix~B.
The spatially uniform coupling $J_{mnl}$ has been promoted to a random all-to-all Yukawa interaction\cite{Esterlis2019,Esterlis2021} in flavor spaces inspired by SYK\cite{RevModPhys.94.035004} in order to facilitate a controlled large-$N$ expansion.
This Yukawa term represents the leading projection of the microscopic $d$-wave Pomeranchuk form factor
$\cos k_x-\cos k_y$ onto the VHS patches at $(0,\pm\pi)$ \cite{Auvray2019,Benhabib2015}.
Its leading constant part has been absorbed into $J_{mnl}$, while the $O(k^2)$ corrections are irrelevant for the leading scaling.
$w_{mn}(\bm{x})$ denotes the fermionic potential disorder that gives rise to the impurity scattering
with a rate $\Gamma\equiv |w|^2\Lambda_\theta/2\pi$, where $\Lambda_\theta$ being a dimensionless UV cutoff associated with the logarithmically divergent density of states.

We begin by evaluating the self-energy diagrams 
in Fig.~\ref{fig3}(a,b) in the large-$N$ limit.
The one-loop boson bubble diagram in Fig.~\ref{fig3}(a) is calculated 
at zero temperature and in a static limit defined by $|\Omega_m|\ll {\rm max}\big(|\epsilon_{\bm q}|,\Gamma \big)$.
The boson self-energy obtained from the VHS deviates from the standard Landau damping by a factor $f(x)$, 
which is given by
\begin{equation}
	\Pi_{J}(i\Omega_m,\bm{q})\simeq  - \tilde{J}^2 \frac{|\Omega_m|}{|\epsilon_{\bm q}|} f\Big(\frac{\Gamma}{|\epsilon_{\bm q}|}\Big)
	\label{bo-se-J}
\end{equation}
where $\tilde{J}^2=|J|^2N/N'$ and the factor has the asymptotics $f(0) \sim 1,\ f(x\rightarrow \infty)\sim x^{-1}$.
In the strong impurity scattering regime $\Gamma\gg |\epsilon_{\bm q}|$, 
we obtain an Ohmic damping $\Pi_{J}(i\Omega_m,\bm{q})\sim |\Omega_m|/\Gamma$, indicative of a dynamical exponent $z = 2$. 
In contrast, for modest scattering $|\Omega_m| \ll \Gamma\ll |\epsilon_{\bm q}|$, the boson self-energy reduces to the
Landau-damping form $\sim |\Omega_m|/|\epsilon_{\bm q}|$.
This form, characterized by $z = 4$ associated with the saddle-point dispersion at VHS,
persists all the way to the extreme weak-scattering limit $\Gamma\ll {\rm min}(|\Omega_m|,|\epsilon_{\bm q}|)$.

\begin{figure}
	\centering
	\includegraphics[width=\linewidth]{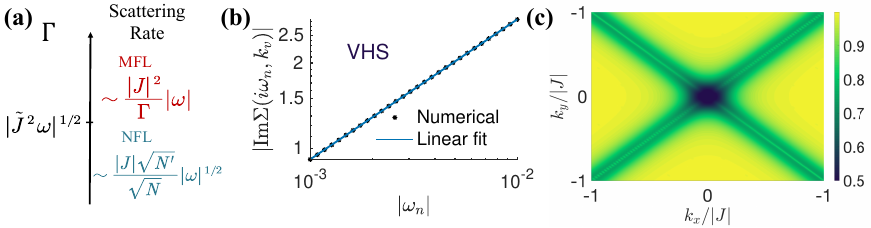}
	\caption{\textbf{Quasiparticle scattering rate in the vicinity of the VHS.} 
	(a) The NFL in the clean limit $\Gamma \ll |\tilde{J}^2\omega|^{1/2}$ (blue) crossovers to the MFL in the dirty limit $\Gamma \gg |\tilde{J}^2\omega|^{1/2}$ (red).
	(b) The plot of $\ln |{\rm Im}\,\Sigma(i\omega,{\bm k})|$ versus $\ln |\omega|$ at the VHS, where the fitted slope is labeled as $0.4945$.
	(c) Numerically extracted frequency dependence of the imaginary part of the fermion self-energy near the VHS. The color bar indicates the slope of the $\ln |{\rm Im}\,\Sigma(i\omega,{\bm k})| - \ln |\omega|$ map. Asterisks denote the numerical data, and solid lines represent the linear fits.	
	}
	\label{fig2}
\end{figure}

Our main result is the emergence of multiple NFL regimes both at the VHS and its vicinity in the clean limit $\Gamma^2 \ll \tilde{J}^2 |\omega|$.
By feeding the boson self-energy in Eq.~(\ref{bo-se-J}) to the diagram Fig.~\ref{fig3}(b),
the fermion self-energy is derived as
\begin{equation}
	\Sigma_{J}(i\omega_n,{\bm k}_v)\simeq -i |J|\sqrt{\frac{N'}{N}}{\rm sgn}(\omega_n)|\omega_n|^{1/2}.
	\label{fesecl}
\end{equation}
The NFL self-energy overwrites the original dynamic term at low frequencies $|\omega|<|J|^2N'/N$ as plotted in Fig.~\ref{fig2}(a).
The quasiparticle broadening ${\rm Im}\Sigma_R(\omega) \sim |\omega|^{\alpha}$ 
is characterized by an anomalous exponent $\alpha=1/2$ and an enhanced weight $\sim \sqrt{N^\prime/N}$.
The analytic result is further verified numerically in a self-consistent manner which is demonstrated in Fig.~\ref{fig2}(b).
This exponent differs from the $\alpha=2/3$ found for the NFL at the convex 2D Fermi surface,
where the dispersion is linear ${\epsilon}_{\bm k}\simeq v_F k_x +\kappa k_y^2$\cite{Metlitski2010,Esterlis2019,Esterlis2021}.
The smaller exponent indicates that the VHS has a greater quasiparticle scattering rate in the low-energy limit, 
making it a ``hotspot'' compared to ``coldspots.'' 
The clean limit $\Gamma^2 \ll \tilde{J}^2|\omega|$ literally indicates that 
the impurity scattering rate is much smaller that the \emph{ee} scattering: $\Gamma\ll |{\rm Im}\Sigma_{J}|$.
Therefore, we can safely drop the impurity potential in Eq.(\ref{eff}) and 
confirm that the one-loop self-energies are self-consistent for a clean system with $w=0$.
We provide a rigorous analytic proof in Supplemental Material Sec.~1.A.1 and illustrate the numerically obtained momentum-resolved distribution of the exponent $\alpha$ in Fig.~\ref{fig2}(c).

In contrast, the dirty limit $\Gamma^2\gg \tilde{J}^2|\omega|$ amounts to 
a stronger impurity scattering $|{\rm Im}\Sigma_{J}|\ll \Gamma$.
As depicted in Fig.~\ref{fig2}(a), the fermion self-energy crossovers to a MFL
\begin{equation}
	\Sigma_{J}(i\omega_n,{\bm k}_v)\simeq -i\frac{|J|^2}{\Gamma}\omega_n\ln\frac{\Gamma^2}{\tilde{J}^2 |\omega_n|}.
	\label{sef_J_dirt}
\end{equation}
which is qualitatively same as the MFL on the convex 2D Fermi surface\cite{Patel2023}. 
In the dirty limit, the renormalized Green function is approximated as $G^{-1}(i\omega_n,{\bm k})\simeq  i\Gamma{\rm sgn}(\omega_n)-\epsilon_{\bm k}$, 
from which the self-consistency is readily justified as detailed in Supplemental Material Sec.~S1.A.2.

We close by noting that, in the dirty limit at finite temperature,
$\Gamma^2 \gg \tilde{J}^2 T$, the electronic specific-heat coefficient
exhibits a logarithmic divergence.
The specific heat contains a regular background contribution as well as a correction generated by the electron-boson coupling. 
In the clean limit at finite temperature, $\Gamma^2\ll \tilde{J}^2 T$, the critical interaction correction, $C_{\rm int}\sim \sqrt{T}$, overwhelms the fermionic van Hove contribution, $C_{\rm VHS}\sim T\ln(\Lambda_{\rm U}^2/T)$. 
In the dirty limit at finite temperature, $\Gamma^2\gg \tilde{J}^2 T$, corresponding to sufficiently low temperatures, the interaction correction instead behaves as $T\ln(\Gamma/T)$, as detailed in Supplemental Material Sec.~S2. 
Thus, in this regime, the self-energy-induced many-body contribution has the same temperature dependence as the regular van Hove background and is thermodynamically indistinguishable from it~\cite{Shen2022}.\\

\section{Optical Conductivity}
\begin{figure}
	\centering
	\includegraphics[width=0.9\linewidth]{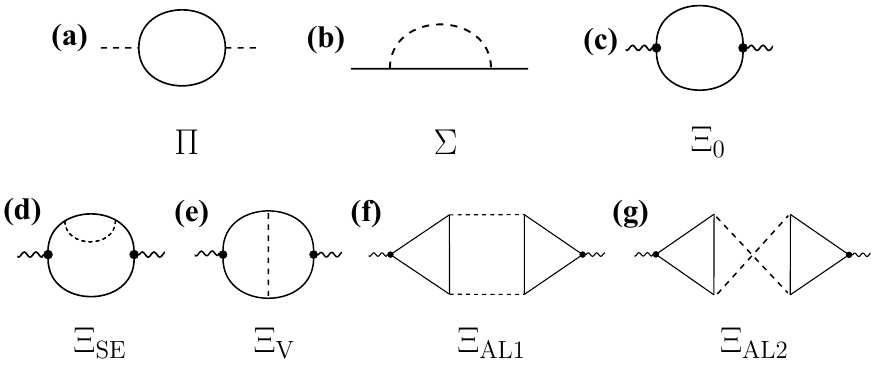}
	\caption{\textbf{Feynman diagrams of the self-energy corrections and current-current correlations.} 
	Solid, dashed and wavy lines are the fermion propagators, boson propagators and current operators, respectively. Solid dots represent the group velocity. (a)-(b) are boson polarization $\Pi$ and fermion self-energy $\Sigma$, respectively. The solid lines can be regarded as fully renormalized propagators. (c)-(g) are the current-current correlation diagrams. (c) one-loop polarization bubble $\Xi_0$; (d) self-energy diagram $\Xi_{\rm SE}$; (e) vertex diagram $\Xi_{\rm V}$, and (f)-(g) are the two Aslamazov-Larkin diagrams $\Xi_{\rm AL1}$,$\Xi_{\rm AL2}$.
	}
	\label{fig3}
\end{figure}

The frequency and/or temperature scaling of the optical conductivity represents a key experimental knob 
for the \emph{ee} interactions\cite{Maslov2016}, especially for identifying the NFL and MFL behaviors\cite{YBKim1994,Guo2022,Patel2023}.
It sensitively depends on both the Galilean invariance of the low-energy dispersion\cite{Chubukov2011} 
and the geometry of the 2D Fermi surface\cite{Songci2023}.
To characterize the peculiar NFL phases of matter at Lifshitz transition, 
we evaluate the optical conductivity arisen from the VHS saddle point via 
the Kubo formula $\mathrm{Re}[\sigma(\omega,T)]=\mathrm{Im}[\Xi(i\omega_n,0)_{i\omega_n\rightarrow \omega+i0^+}]/\omega$. 
In the large-$N$ limit, the current correlation $\Xi(i\omega_n,0)$ is perturbatively evaluated in $o(|J|^2)$,
which receives contributions from Feynman diagrams in Fig.~\ref{fig3}(c-g).
The normal \emph{ee} interaction (dashed lines) is mediated by the Landau-damped critical boson given in Eq.(\ref{bo-se-J}).
A leading order in ${o}(|J|^2)$, the summation over Fig.~\ref{fig3}(d-g) leads to a compact expression 
\begin{equation}
\begin{aligned}
\Xi(i\omega_n) = &\frac{N\tilde{J}^2|J|^2}{4(|\omega_n|+2\Gamma)^2} \int_q D_+(q) D_-(q)  \int_{k}^\ast \int_{k^\prime -q}^\ast  (\Delta v)^2. \\
\end{aligned}
\label{totxi}
\end{equation}
We provide the detailed derivation and explanation in appendix.~B 
and highlight here the critical dependence of the optical conductivity on changes in the group velocity
for a small-${\bm q}$ elastic scattering process $({\bm k},{\bm k}^\prime) \rightarrow ({\bm k}+{\bm q},{\bm k}^\prime-{\bm q})$
\begin{equation}
 \Delta v =v_{{\bm k}+{\bm q}}-v_{\bm k} +v_{{\bm k}'-{\bm q}}-v_{{\bm k}'}.
\end{equation}
For a generic 2D convex Fermi surface as depicted in Fig.~\ref{fig4}(a),
there is only one inequivalent intersection point indicating that the injecting ${\bm k}, {\bm k}^\prime$ are at most swapped.
This swap scattering channel leads to $\Delta v = 0$
and a vanishing optical conductivity at the leading order according to Eq.(\ref{totxi}).
At the Lifshitz transition in Fig.~\ref{fig4}(b), 
there emerges an extra intersecting point ${\bm k}$ situated near the VHS at the Brillouin zone boundary.
This extra channel is nothing but the pseudo umklapp scattering process\cite{Chubukov2011}. 
For the swap channel near the VHS, we demonstrate a similar cancellation among the leading-order diagrams;
Whereas, the extra channel amounts to a small-${\bm q}$ exchange with $(\Delta v)^2 \sim q^2 $
which then leads to a non-vanishing optical conductivity.

A simple power-counting based on the self-energy in Eq.(\ref{fesecl}) then leads to the optical conductivity in the clean limit $\Gamma^2 \ll \tilde{J}^2|\omega|$
that scales as $\big[{\rm Re}\sigma^J(\omega;T=0)\big]^{-1} \sim q^2 \times {\rm Im}\Sigma^{\rm R}_J(\omega,{\bm k}) \sim |\omega|^{2/z} |\omega|^{1/2} \sim |\omega|$.
Performing the full diagrammatical calculation in Eq.(\ref{totxi}) in a limit of weak impurity scattering $\Gamma\ll |\omega|$ yields
\begin{equation}
{\rm Re}\big[ \sigma^J(\Gamma^2/\tilde{J}^2 \ll  |\omega| ,\Gamma\ll |\omega|;T=0)\big] \simeq \frac{N|J|^2\Lambda_\theta}{16\pi^3\sqrt{a}}\frac{1}{|\omega|}.
\label{sig_T0_Gm0}
\end{equation}
This result can be smoothly continued to the clean limit $\Gamma\rightarrow 0$ since the impurity scattering rate 
is negligible compared to all relevant energy scales.
Accordingly, in the regime $\Gamma\ll|\omega|\ll T$, an independent calculation of the optical conductivity
yields
\begin{equation}
{\rm Re}\big[ \sigma^J(\Gamma \ll |\omega|\ll T )\big]
\simeq \frac{N|J|^2T}{\omega^2},
\label{sigma_T}
\end{equation}
up to logarithmic corrections. 
The zero- and finite-temperature optical conductivity can be summarized by 
the generalized Drude formula as
\begin{equation}
\operatorname{Re}\sigma^J(\omega,T)
\sim
\frac{\tau_{\rm tr}^{-1}(\omega,T)}{\omega^2}, \qquad  \tau_{\rm tr}^{-1}(\omega,T) \sim {\rm max}(|\omega|,T).
\label{sigma_scaling_form}
\end{equation}
Here the $\omega^{-2}$ denominator follows from the generic
finite-frequency structure of the current response in the collisionless regime, 
whereas the nature of the NFL metallic state is encoded in the transport scattering rate $\tau_{\rm tr}^{-1}$.
It exhibits peculiar scaling form
that should be contrasted with a conventional Fermi liquid, for which
$\tau^{-1}_{\mathrm{tr}}(\omega,T)\sim\omega^2+(2\pi T)^2$.

In summary, in a Galilean-invariant system with
a parabolic dispersion, the electric current is proportional to the total
momentum, so that, momentum-conserving
\emph{ee} scattering cannot relax the current and produce no
regular dissipative optical response.
The Galilean invariance is broken at the VHS, therefore the normal \emph{ee} scattering in the extra channel
can relax the component of the current orthogonal to momentum,
producing a nonvanishing dissipative optical response at finite frequency.
The component overlapping with the conserved momentum remains protected,
however, and gives rise to a zero-frequency Drude delta function and,
hence, vanishing dc resistivity in the clean limit. A finite dc
resistivity requires an explicit momentum-relaxation mechanism, such as
impurity scattering, Umklapp processes, or electron--phonon scattering.
\\

\section{Linear-in-$T$ Resistivity}
\begin{figure}
	\centering
	\includegraphics[width=0.9\linewidth]{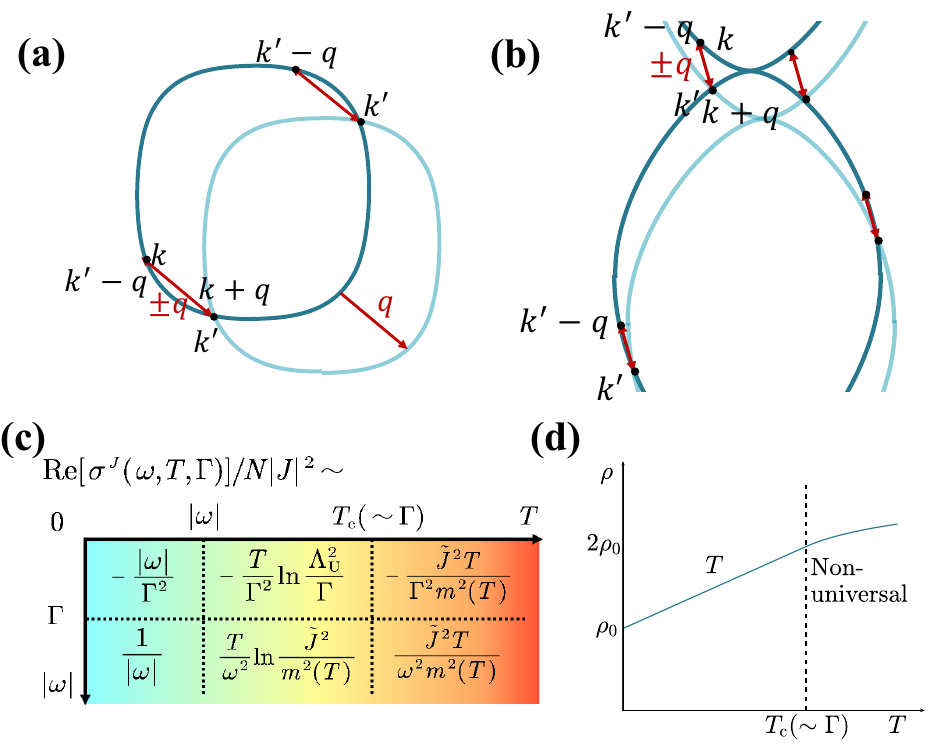}
	\caption{\textbf{Schematic figures for the scattering process and transport properties.}
		(a)-(b) Schematic figures for the small-${\bm q}$ scattering processes: $({\bm k},{\bm k}^\prime) \rightarrow ({\bm k}+{\bm q},{\bm k}^\prime-{\bm q})$. (c) illustrates the swap channel for a convex Fermi surface, and (b) illustrates the extra channel at the convex-to-concave transition. 
		(c) Real part of optical conductivity as a function of the temperature $T$ and frequency $|\omega|$ domains. The blue, yellow and red colors indicate the low-$T$ regime, finite-$T$ quantum critical regime and high-$T$ non-universal regime, respectively.
		(d) The resistivity $\rho(T)$ in extended temperature range. $\rho_0$ is the residual resistivity at $T=0$. $\rho(T)-\rho_0$ is linear-in-$T$ within the quantum critical regime $T\ll T_c$ and crossovers to a non-universal regime at higher temperatures.}
	\label{fig4}
\end{figure}

Recently, spatially random interactions $J'(\bm{x})$ have been proposed
as a general mechanism for momentum relaxation and, more importantly, for
generating the long-sought linear-in-$T$ resistivity on a generic convex
2D Fermi surface \cite{Patel2023}, although their microscopic
origin remains unclear.
In contrast, we demonstrate that, for a 2D quantum
critical system at a Lifshitz transition, the spatially uniform interaction
$J$, together with impurity scattering, is sufficient to generate strange
metallicity.
We focus on the normal \emph{ee} scattering in extra channel 
with a small-${\bm q}$ momentum transfer at the zone boundary.
In addition, the impurity scattering is introduced to relax the momentum
and to avoid the suppressed transport contribution from the VHS ``hotspot'' 
which is known as the short-circuit issue\cite{Hlubina1995}.
The linear-in-$T$ resistivity, illustrated in Fig.~\ref{fig4}(d), dwelling in the finite temperature quantum critical regime extends down to $T=0$.
The strange metallicity is primarily attributed to the peculiar Fermi surface geometry at Lifshitz transition 
and the associated non-Galilean-invariant electronic dispersion at the VHS saddle point.

The finite-$T$ quantum critical regime is defined by the prevalence of quantum critical scaling
and the condition that thermal fluctuations are negligible, as quantified by a boson thermal mass satisfying $m^2(T)\ll \tilde{J}^2$.
The dangerously irrelevant boson interactions gives rise to 
$m^2(T)\sim \tilde{J}^2 T/\Gamma$ (up to logarithmic corrections as detailed in Supplemental Material Sec.~S4.B.2),
which dictates a crossover temperature $T_c\sim \Gamma$ above which non-universal behaviors are expected.
Analytical expressions of optical conductivity in distinct regimes, specified by the impurity scattering strength and temperature,
are demonstrated in Fig.~\ref{fig4}(c).
We discover a broad finite-$T$ region $|\omega|\ll T\ll \Gamma (\sim T_c)$
where the \emph{dc} conductivity is given by 
\begin{equation}
{\rm Re}\big[\sigma^{J}\big(\omega=0,T\ll \Gamma \big)\big]\sim -\frac{N|J|^2T}{\Gamma^2}\ln\frac{\Lambda_{\rm U}^2}{\Gamma}.
\label{sigma_J_T_2}
\end{equation}
Note that the linear-in-$T$ behavior can be extrapolated to $T=0$ when the $dc$ limit $\omega=0$ is taken.
This result can also be obtained from
Eq.~\eqref{sigma_T} by replacing $\omega$ with $\omega+i\Gamma$ and
subsequently taking the dc limit.
In the opposite limit $ T\gg T_{c}$, thermal fluctuations disrupt the quantum critical scaling,
which manifests as the explicit presence of boson mass in the optical conductivity 
\begin{equation}
	{\rm Re}\big[\sigma^{J}\big(T\gg T_c \big)\big]\sim N|J|^2 \tilde{J}^2\frac{ {\rm sgn}(|\omega|-\Gamma)}{{\rm max}(\Gamma^2,\omega^2)} \frac{T}{m^2(T)}.
\end{equation}
The seemingly higher-order term $o(|J|^4)$ is reduced by the boson mass
which leads to a non-universal saturated \emph{dc} conductivity at the leading-order $\mathrm{Re}\big[\sigma^{J}(\omega=0,T\gg T_c)\big]\sim - N |J|^2/\Gamma$.

The linear-in-$T$ optical conductivity leads to the linear-in-$T$ 
resistivity in the quantum critical regime extending all the way to $T=0$, which is described by
\begin{equation}
\begin{aligned}
	\rho(T\ll T_c) 
	=&\frac{1}{{\rm Re}[\sigma^0(\omega=0,T\big)]+{\rm Re}[\sigma^J(\omega=0,T\ll T_c\big)]}\\
	=& \frac{1}{N\Lambda_{\rm U}^2/\Gamma-N|J|^2T/\Gamma^2} \simeq \frac{\Gamma}{N\Lambda_{\rm U}^2}+\frac{|J|^2}{N\Lambda_{\rm U}^4}T.	
\end{aligned}
	\label{rho_J}
\end{equation}
Here, the impurity contribution represented by Fig.~\ref{fig3}(a) provides the leading residual conductivity,
${\rm Re}[\sigma^0(\omega=0,T)]=N\Lambda_{\rm U}^2/\Gamma$,
whereas the linear-in-$T$ correction to the resistivity arises entirely from
\emph{ee} scattering in Eq.(\ref{sigma_J_T_2}). 
The last line of Eq.~\eqref{rho_J} follows from a perturbative
expansion valid when the temperature-dependent correction is much smaller than the residual
resistivity,
\begin{equation}
\rho(T)-\rho_0\ll\rho_0,
\qquad
\rho_0={\rm Re}[\sigma^0(\omega=0,T)]^{-1}.
\end{equation}
which is also known as the Matthiessen's rule.
This requirement yields
$
T\ll\Gamma\left(\frac{\Lambda_{\rm U}}{|J|}\right)^2.
$
In the quantum-critical regime, $T\ll T_c\sim\Gamma$, this condition is
automatically satisfied provided that $\Lambda_{\rm U}\gg|J|$.

We conclude by remarking on the competing effects of the VHS hot spots and the Fermi surface cold spots on the total resistivity.
The VHS saddle point acts as a ``hotspot'' due to its enhanced quasiparticle scattering rate 
$\sim |\omega|^{1/2}$ in the clean limit, and the rest of the Fermi-surface points exhibit the scattering rate $\sim |\omega|^{2/3}$.
This situation is closely analogous to the short-circuit problem encountered near an antiferromagnetic QCP \cite{Hlubina1995}. 
Numerical and analytical studies \cite{Rosch1999} have shown that, in the presence of relatively strong impurity scattering, $\Gamma\gg T$, transport is no longer dominated by the weakly scattered cold regions; instead, the NFL hot spots generated by AFM critical fluctuations can control the total resistivity over a broad temperature range. 
In the present problem, the quantum-critical regime $T\ll T_c$ obtained from Eq.~(\ref{sigma_J_T_2}), with $T_c\sim\Gamma$, corresponds precisely to the same strong-impurity condition $\Gamma\gg T$.
Impurity scattering therefore suppresses the parametric distinction between the scattering rates at the VHS and on the convex portions of the Fermi surface, preventing the conventional short-circuit mechanism.
It remains necessary, however, to show that the transport corrections from ordinary convex Fermi-surface regions, excluding the convex points connected
to the VHS through the extra scattering channel, are subleading to the VHS
contribution. For a normal \emph{ee} interaction of strength $J$,
the putative $T^{4/3}$ optical-conductivity correction
\cite{LeeNagaosa1992,YBKim1994} has a vanishing coefficient on a convex 2D
Fermi surface at leading order, $\mathcal{O}(|J|^2\Gamma^{-2})$. Nonvanishing
contributions from the convex regions may nevertheless arise through three
types of processes. First, higher-order diagrams contribute only at
$\mathcal{O}(|J|^3\Gamma^{-3})$ or beyond. Second, processes involving states
slightly away from the Fermi surface occur at
$\mathcal{O}(|J|^2\Gamma^{-2})$, but scale as $T^{8/3}$ and are further
suppressed by a factor of $k_F^{-1}$ \cite{Songci2023}. Third, large-${\bm Q}$
Umklapp processes, with $Q\sim k_F$, can be mediated by noncritical
electron--electron interactions, such as the screened Coulomb interaction.
Although these processes arise at the same leading diagrammatic order, their
contribution also scales as $T^{8/3}$ and is additionally suppressed by a
factor of $Q^{-4}$. Consequently, despite the larger phase space occupied by
the convex regions, all three corrections are parametrically subleading to
the critical VHS contribution and do not modify the leading linear-in-$T$
resistivity.
Therefore, the linear-in-$T$ resistivity arising from the VHS is the dominant contribution at Lifshitz transition
This argument is justified numerically by \emph{P. A. Lee} for a system close to, but not exactly at, 
the Lifshitz transition with a finite distance $\Delta_q\ne 0$\cite{PALee2021}.
The anti-nodal points proximate to the VHS give rise to a linear-in-$T$ resistivity 
that dominates over a finite temperature range in the presence of impurity scattering. 
The upper bound of this range is set by $\Gamma$, consistent with our analytic result in Fig.~\ref{fig4}(d);
whereas, at lower energies, $T \ll \Delta_q^3$, Fermi-liquid behavior is recovered.
A naive $\Delta_q \rightarrow 0$ limit, as noted in Ref.~\cite{PALee2021}, results in an unphysical divergence in the scattering rate. 
The Lifshitz transition point $\Delta_q = 0$ therefore requires a systematic analysis, as carried out in this study.

\section{Spatially random Yukawa interaction}

For the 2D system at Lifshitz transition where the Fermi surface undergoes a convex-to-concave transition,
it's necessary to determine the fate of NFL at the VHS saddle point
when subjected to the spatially random Yukawa interaction $J^\prime({\bm x})$.
We are motivated to examine whether the VHS in turn short-circuits the established strange metallicity arising from the MFL\cite{Patel2023}.
The interplay between the interactions $J$ and $J^\prime$ is discussed in appendix.~A for the clean system $w =0$.

We start with the low-energy effective theory $ \mathcal{L}=\mathcal{L}_0 + \mathcal{L}_w + \mathcal{L}_{J^\prime}$
with the spatially random interaction term $\mathcal{L}_{J^\prime} =\frac{1}{\sqrt{NN'}}\sum_{mnl}J_{mnl}'(\bm{x})\psi^\dagger_m(\bm{x},\tau)\psi_n(\bm{x},\tau)\phi_l(\bm{x},\tau)$.
$J'(\bm{x})$ obeys the Gaussian distribution in both flavor space and real space.
We derive the fermion self-energy self-consistently which  takes the form of MFL with $z = 2$
regardless of the strength of impurity scattering, {\sl i.e.} $ -{\rm Im}\Sigma_{J^\prime}(\omega,0) \sim |J^\prime|^2 |\omega|,\ \forall \Gamma \ge 0.$
This is \emph{not} different from the one from generic convex Fermi surface points,
since the spatial random interaction is insensitive to the form of fermionic dispersions.
The resistivity should receive equal contributions from both the VHS saddle point and convex Fermi surface points.
Whereas, the momentum-independent boson self-energy generated by the
spatially random Yukawa interaction plays a crucial role in thermodynamics,
which leads to $C(T) \sim T\ln(1/T)$.
Furthermore, we evaluate the optical conductivity at $T=0$ and the \emph{dc} resistivity at finite-$T$
with $w\ne 0$ and without potential disorder $w=0$ (see Supplemental Material Sec.~S3.B and Sec.~S4.C, respectively.)
The conductivity exhibits a linear-in-$\omega$ in the low-temperature limit $T\ll |\omega|$,
and a linear-in-$T$ behavior at finite temperatures $|\omega|\ll T$
even when the thermal fluctuations disrupt the quantum critical scaling,
which reads ${\rm Re}\big[ \sigma^{J^\prime}(\omega=0,T)\big] \sim  (N^2/N^\prime)\Lambda_{\rm U}^2 |J'|^4\Gamma^{-2}\frac{T^2}{m^2(T)}
\sim N\Lambda_{\rm U}^2 |J'|^2\Gamma^{-2}T$.
As a result, we demonstrate that the resistivity arising from the VHS 
exhibits a persistent linear-in-$T$ feature not only in the quantum critical regime,
but also, unexceptionally, extending to the non-universal regime
\begin{equation}
	\rho\big(T\ll \Gamma/|J'|^2\big) \sim  \frac{\Gamma}{N\Lambda_{\rm U}^2}+\frac{|J'|^2 T}{N\Lambda_{\rm U}^2}.
\end{equation}
Thus, spatial randomness substantially broadens the strange-metal regime.

\section{Discussion and Outlook}
We have demonstrated the strange metal transport at the Lifshitz transition near a Pomeranchuk-type of QCP
in the presence of spatially uniform and/or random Yukawa interactions in 2+1D.
The key physical processes are the normal \emph{ee} scattering in extra channel 
(equivalently the pseudo umklapp scattering) and impurity scattering, 
which additively determine the resistivity in the low-temperature regime. 
Beyond this temperature range, the Matthiessen rule breaks down, 
suggesting that distinct transport phenomena might emerge even before electron-phonon coupling comes into play.
While our analysis shows that quantum critical transport from the VHS hotspots is not short-circuited by cold regions, a full numerical treatment of the Boltzmann equation for a 2D Fermi surface evolving from convex to concave would provide valuable insight. Such a computation should incorporate both pseudo-umklapp and impurity scattering on equal footing. Nevertheless, the applicability of the Boltzmann description itself must be examined in the quantum critical regime. When normal \emph{ee} scattering is treated self-consistently within the Migdal-Eliashberg framework, the Prange-Kadanoff argument ensures the validity of the Boltzmann equation even in the absence of well-defined quasiparticles. However, recent studies suggest that the Migdal-Eliashberg theory becomes unstable in 2+1D systems, particularly within the Yukawa-SYK formulation~\cite{Haoyu2023,Haoyu2024}.

Experimentally, the effective mass inferred from heat capacity exhibits a divergent peak near the critical doping $p_{\rm c}$ or the Lifshitz transition $p^\ast$~\cite{Michon2019}, 
whereas the cyclotron mass in cuprates deviates markedly from the thermodynamic mass across the same doping range~\cite{Legros2022}. 
Kohns theorem dictates that in a translationally invariant, Galilean system, interactions do not renormalize the cyclotron mass. 
Hence, any viable theory for cuprates must break at least one of these symmetries. 
Recent works~\cite{Guo2024,Patel2023} attribute this to translational symmetry breaking from potential disorder or random Yukawa interaction $J^\prime({\bm x})$.
Alternatively, it is of great interest to explore the perspective that the Galilean invariance is broken at the Lifshitz transition.\\

\section{Acknowledgements}

Xiao-Tian Zhang acknowledges helpful discussions with Fu-Chun Zhang, Patrick. A. Lee and Andrey V. Chubukov. Xiao-Tian Zhang is supported by NSFC Grant No.~12404178 and by Prof. Fu-Chun Zhang funded by China's Ministry of Science and Technology (Grant No.~2022YFA1403900) and NSFC (Grant No.~11920101005). \\
Yi-Hui Xing acknowledges financial support from Prof. Shang Liu funded by the Chinese
Academy of Sciences (CAS) under Grant No. ~SBR-150
and a startup fund from the Institute of Physics, CAS.\\
Wu-Ming Liu is supported by the National Key R\&D Program of China under Grants No.~2021YFA1400900,~2021YFA0718300, and 2021YFA1402100, NSFC under Grants No.~12174461,~12234012,~12334012, and 52327808, and the Space Application System of China Manned Space Program.



\bibliography{paper}

\begin{thebibliography}{96}%
\makeatletter
\providecommand \@ifxundefined [1]{%
 \@ifx{#1\undefined}
}%
\providecommand \@ifnum [1]{%
 \ifnum #1\expandafter \@firstoftwo
 \else \expandafter \@secondoftwo
 \fi
}%
\providecommand \@ifx [1]{%
 \ifx #1\expandafter \@firstoftwo
 \else \expandafter \@secondoftwo
 \fi
}%
\providecommand \natexlab [1]{#1}%
\providecommand \enquote  [1]{``#1''}%
\providecommand \bibnamefont  [1]{#1}%
\providecommand \bibfnamefont [1]{#1}%
\providecommand \citenamefont [1]{#1}%
\providecommand \href@noop [0]{\@secondoftwo}%
\providecommand \href [0]{\begingroup \@sanitize@url \@href}%
\providecommand \@href[1]{\@@startlink{#1}\@@href}%
\providecommand \@@href[1]{\endgroup#1\@@endlink}%
\providecommand \@sanitize@url [0]{\catcode `\\12\catcode `\$12\catcode
  `\&12\catcode `\#12\catcode `\^12\catcode `\_12\catcode `\%12\relax}%
\providecommand \@@startlink[1]{}%
\providecommand \@@endlink[0]{}%
\providecommand \url  [0]{\begingroup\@sanitize@url \@url }%
\providecommand \@url [1]{\endgroup\@href {#1}{\urlprefix }}%
\providecommand \urlprefix  [0]{URL }%
\providecommand \Eprint [0]{\href }%
\providecommand \doibase [0]{http://dx.doi.org/}%
\providecommand \selectlanguage [0]{\@gobble}%
\providecommand \bibinfo  [0]{\@secondoftwo}%
\providecommand \bibfield  [0]{\@secondoftwo}%
\providecommand \translation [1]{[#1]}%
\providecommand \BibitemOpen [0]{}%
\providecommand \bibitemStop [0]{}%
\providecommand \bibitemNoStop [0]{.\EOS\space}%
\providecommand \EOS [0]{\spacefactor3000\relax}%
\providecommand \BibitemShut  [1]{\csname bibitem#1\endcsname}%
\let\auto@bib@innerbib\@empty
\bibitem [{\citenamefont {Reiss}\ \emph {et~al.}(2020)\citenamefont {Reiss},
  \citenamefont {Graf}, \citenamefont {Haghighirad}, \citenamefont {Knafo},
  \citenamefont {Drigo}, \citenamefont {Bristow}, \citenamefont {Schofield},\
  and\ \citenamefont {Coldea}}]{reiss2020quenched}%
  \BibitemOpen
  \bibfield  {author} {\bibinfo {author} {\bibfnamefont {P.}~\bibnamefont
  {Reiss}}, \bibinfo {author} {\bibfnamefont {D.}~\bibnamefont {Graf}},
  \bibinfo {author} {\bibfnamefont {A.~A.}\ \bibnamefont {Haghighirad}},
  \bibinfo {author} {\bibfnamefont {W.}~\bibnamefont {Knafo}}, \bibinfo
  {author} {\bibfnamefont {L.}~\bibnamefont {Drigo}}, \bibinfo {author}
  {\bibfnamefont {M.}~\bibnamefont {Bristow}}, \bibinfo {author} {\bibfnamefont
  {A.~J.}\ \bibnamefont {Schofield}}, \ and\ \bibinfo {author} {\bibfnamefont
  {A.~I.}\ \bibnamefont {Coldea}},\ }\href {\doibase 10.1038/s41567-019-0694-2}
  {\bibfield  {journal} {\bibinfo  {journal} {Nat. Phys.}\ }\textbf {\bibinfo
  {volume} {16}},\ \bibinfo {pages} {89} (\bibinfo {year} {2020})}\BibitemShut
  {NoStop}%
\bibitem [{\citenamefont {Lee}\ and\ \citenamefont
  {Phillips}(2012)}]{PhysRevB.86.245113}%
  \BibitemOpen
  \bibfield  {author} {\bibinfo {author} {\bibfnamefont {W.-C.}\ \bibnamefont
  {Lee}}\ and\ \bibinfo {author} {\bibfnamefont {P.~W.}\ \bibnamefont
  {Phillips}},\ }\href {\doibase 10.1103/PhysRevB.86.245113} {\bibfield
  {journal} {\bibinfo  {journal} {Phys. Rev. B}\ }\textbf {\bibinfo {volume}
  {86}},\ \bibinfo {pages} {245113} (\bibinfo {year} {2012})}\BibitemShut
  {NoStop}%
\bibitem [{\citenamefont {Chen}\ \emph
  {et~al.}(2019{\natexlab{a}})\citenamefont {Chen}, \citenamefont {Wang},
  \citenamefont {Li}, \citenamefont {Feng}, \citenamefont {Dai}, \citenamefont
  {Xu},\ and\ \citenamefont {Si}}]{chen2019heavy}%
  \BibitemOpen
  \bibfield  {author} {\bibinfo {author} {\bibfnamefont {J.}~\bibnamefont
  {Chen}}, \bibinfo {author} {\bibfnamefont {Z.}~\bibnamefont {Wang}}, \bibinfo
  {author} {\bibfnamefont {Y.}~\bibnamefont {Li}}, \bibinfo {author}
  {\bibfnamefont {C.}~\bibnamefont {Feng}}, \bibinfo {author} {\bibfnamefont
  {J.}~\bibnamefont {Dai}}, \bibinfo {author} {\bibfnamefont {Z.}~\bibnamefont
  {Xu}}, \ and\ \bibinfo {author} {\bibfnamefont {Q.}~\bibnamefont {Si}},\
  }\href {\doibase 10.1038/s41598-019-48662-8} {\bibfield  {journal} {\bibinfo
  {journal} {Scientific Reports}\ }\textbf {\bibinfo {volume} {9}},\ \bibinfo
  {pages} {12307} (\bibinfo {year} {2019}{\natexlab{a}})}\BibitemShut {NoStop}%
\bibitem [{\citenamefont {Seiro}\ \emph {et~al.}(2018)\citenamefont {Seiro},
  \citenamefont {Jiao}, \citenamefont {Kirchner}, \citenamefont {Hartmann},
  \citenamefont {Friedemann}, \citenamefont {Krellner}, \citenamefont {Geibel},
  \citenamefont {Si}, \citenamefont {Steglich},\ and\ \citenamefont
  {Wirth}}]{seiro2018evolution}%
  \BibitemOpen
  \bibfield  {author} {\bibinfo {author} {\bibfnamefont {S.}~\bibnamefont
  {Seiro}}, \bibinfo {author} {\bibfnamefont {L.}~\bibnamefont {Jiao}},
  \bibinfo {author} {\bibfnamefont {S.}~\bibnamefont {Kirchner}}, \bibinfo
  {author} {\bibfnamefont {S.}~\bibnamefont {Hartmann}}, \bibinfo {author}
  {\bibfnamefont {S.}~\bibnamefont {Friedemann}}, \bibinfo {author}
  {\bibfnamefont {C.}~\bibnamefont {Krellner}}, \bibinfo {author}
  {\bibfnamefont {C.}~\bibnamefont {Geibel}}, \bibinfo {author} {\bibfnamefont
  {Q.}~\bibnamefont {Si}}, \bibinfo {author} {\bibfnamefont {F.}~\bibnamefont
  {Steglich}}, \ and\ \bibinfo {author} {\bibfnamefont {S.}~\bibnamefont
  {Wirth}},\ }\href {\doibase 10.1038/s41467-018-05801-5} {\bibfield  {journal}
  {\bibinfo  {journal} {Nat. Commun.}\ }\textbf {\bibinfo {volume} {9}},\
  \bibinfo {pages} {3324} (\bibinfo {year} {2018})}\BibitemShut {NoStop}%
\bibitem [{\citenamefont {Xu}\ \emph {et~al.}(2020)\citenamefont {Xu},
  \citenamefont {Wu}, \citenamefont {Jian},\ and\ \citenamefont
  {Xu}}]{PhysRevB.101.205426}%
  \BibitemOpen
  \bibfield  {author} {\bibinfo {author} {\bibfnamefont {Y.}~\bibnamefont
  {Xu}}, \bibinfo {author} {\bibfnamefont {X.-C.}\ \bibnamefont {Wu}}, \bibinfo
  {author} {\bibfnamefont {C.-M.}\ \bibnamefont {Jian}}, \ and\ \bibinfo
  {author} {\bibfnamefont {C.}~\bibnamefont {Xu}},\ }\href {\doibase
  10.1103/PhysRevB.101.205426} {\bibfield  {journal} {\bibinfo  {journal}
  {Phys. Rev. B}\ }\textbf {\bibinfo {volume} {101}},\ \bibinfo {pages}
  {205426} (\bibinfo {year} {2020})}\BibitemShut {NoStop}%
\bibitem [{\citenamefont {Wang}\ \emph {et~al.}(2022)\citenamefont {Wang},
  \citenamefont {Yu}, \citenamefont {Kwan}, \citenamefont {Jia}, \citenamefont
  {Lei}, \citenamefont {Klemenz}, \citenamefont {Cevallos}, \citenamefont
  {Singha}, \citenamefont {Devakul}, \citenamefont {Watanabe} \emph
  {et~al.}}]{wang2022one}%
  \BibitemOpen
  \bibfield  {author} {\bibinfo {author} {\bibfnamefont {P.}~\bibnamefont
  {Wang}}, \bibinfo {author} {\bibfnamefont {G.}~\bibnamefont {Yu}}, \bibinfo
  {author} {\bibfnamefont {Y.~H.}\ \bibnamefont {Kwan}}, \bibinfo {author}
  {\bibfnamefont {Y.}~\bibnamefont {Jia}}, \bibinfo {author} {\bibfnamefont
  {S.}~\bibnamefont {Lei}}, \bibinfo {author} {\bibfnamefont {S.}~\bibnamefont
  {Klemenz}}, \bibinfo {author} {\bibfnamefont {F.~A.}\ \bibnamefont
  {Cevallos}}, \bibinfo {author} {\bibfnamefont {R.}~\bibnamefont {Singha}},
  \bibinfo {author} {\bibfnamefont {T.}~\bibnamefont {Devakul}}, \bibinfo
  {author} {\bibfnamefont {K.}~\bibnamefont {Watanabe}},  \emph {et~al.},\
  }\href {\doibase 10.1038/s41586-022-04514-6} {\bibfield  {journal} {\bibinfo
  {journal} {Nature}\ }\textbf {\bibinfo {volume} {605}},\ \bibinfo {pages}
  {57} (\bibinfo {year} {2022})}\BibitemShut {NoStop}%
\bibitem [{\citenamefont {Cao}\ \emph {et~al.}(2020)\citenamefont {Cao},
  \citenamefont {Chowdhury}, \citenamefont {Rodan-Legrain}, \citenamefont
  {Rubies-Bigorda}, \citenamefont {Watanabe}, \citenamefont {Taniguchi},
  \citenamefont {Senthil},\ and\ \citenamefont {Jarillo-Herrero}}]{Cao2020}%
  \BibitemOpen
  \bibfield  {author} {\bibinfo {author} {\bibfnamefont {Y.}~\bibnamefont
  {Cao}}, \bibinfo {author} {\bibfnamefont {D.}~\bibnamefont {Chowdhury}},
  \bibinfo {author} {\bibfnamefont {D.}~\bibnamefont {Rodan-Legrain}}, \bibinfo
  {author} {\bibfnamefont {O.}~\bibnamefont {Rubies-Bigorda}}, \bibinfo
  {author} {\bibfnamefont {K.}~\bibnamefont {Watanabe}}, \bibinfo {author}
  {\bibfnamefont {T.}~\bibnamefont {Taniguchi}}, \bibinfo {author}
  {\bibfnamefont {T.}~\bibnamefont {Senthil}}, \ and\ \bibinfo {author}
  {\bibfnamefont {P.}~\bibnamefont {Jarillo-Herrero}},\ }\href {\doibase
  10.1103/PhysRevLett.124.076801} {\bibfield  {journal} {\bibinfo  {journal}
  {Phys. Rev. Lett.}\ }\textbf {\bibinfo {volume} {124}},\ \bibinfo {pages}
  {076801} (\bibinfo {year} {2020})}\BibitemShut {NoStop}%
\bibitem [{\citenamefont {Varma}\ \emph {et~al.}(1989)\citenamefont {Varma},
  \citenamefont {Littlewood}, \citenamefont {Schmitt-Rink}, \citenamefont
  {Abrahams},\ and\ \citenamefont {Ruckenstein}}]{Varma1989}%
  \BibitemOpen
  \bibfield  {author} {\bibinfo {author} {\bibfnamefont {C.~M.}\ \bibnamefont
  {Varma}}, \bibinfo {author} {\bibfnamefont {P.~B.}\ \bibnamefont
  {Littlewood}}, \bibinfo {author} {\bibfnamefont {S.}~\bibnamefont
  {Schmitt-Rink}}, \bibinfo {author} {\bibfnamefont {E.}~\bibnamefont
  {Abrahams}}, \ and\ \bibinfo {author} {\bibfnamefont {A.~E.}\ \bibnamefont
  {Ruckenstein}},\ }\href {\doibase 10.1103/PhysRevLett.63.1996} {\bibfield
  {journal} {\bibinfo  {journal} {Phys. Rev. Lett.}\ }\textbf {\bibinfo
  {volume} {63}},\ \bibinfo {pages} {1996} (\bibinfo {year}
  {1989})}\BibitemShut {NoStop}%
\bibitem [{\citenamefont {Kakehashi}\ and\ \citenamefont
  {Fulde}(2005)}]{PhysRevLett.94.156401}%
  \BibitemOpen
  \bibfield  {author} {\bibinfo {author} {\bibfnamefont {Y.}~\bibnamefont
  {Kakehashi}}\ and\ \bibinfo {author} {\bibfnamefont {P.}~\bibnamefont
  {Fulde}},\ }\href {\doibase 10.1103/PhysRevLett.94.156401} {\bibfield
  {journal} {\bibinfo  {journal} {Phys. Rev. Lett.}\ }\textbf {\bibinfo
  {volume} {94}},\ \bibinfo {pages} {156401} (\bibinfo {year}
  {2005})}\BibitemShut {NoStop}%
\bibitem [{\citenamefont {Senthil}(2008)}]{CFM}%
  \BibitemOpen
  \bibfield  {author} {\bibinfo {author} {\bibfnamefont {T.}~\bibnamefont
  {Senthil}},\ }\href {\doibase 10.1103/PhysRevB.78.035103} {\bibfield
  {journal} {\bibinfo  {journal} {Phys. Rev. B}\ }\textbf {\bibinfo {volume}
  {78}},\ \bibinfo {pages} {035103} (\bibinfo {year} {2008})}\BibitemShut
  {NoStop}%
\bibitem [{\citenamefont {Patel}\ \emph {et~al.}(2023)\citenamefont {Patel},
  \citenamefont {Guo}, \citenamefont {Esterlis},\ and\ \citenamefont
  {Sachdev}}]{Patel2023}%
  \BibitemOpen
  \bibfield  {author} {\bibinfo {author} {\bibfnamefont {A.~A.}\ \bibnamefont
  {Patel}}, \bibinfo {author} {\bibfnamefont {H.}~\bibnamefont {Guo}}, \bibinfo
  {author} {\bibfnamefont {I.}~\bibnamefont {Esterlis}}, \ and\ \bibinfo
  {author} {\bibfnamefont {S.}~\bibnamefont {Sachdev}},\ }\href {\doibase
  10.1126/science.abq6011} {\bibfield  {journal} {\bibinfo  {journal}
  {Science}\ }\textbf {\bibinfo {volume} {381}},\ \bibinfo {pages} {790}
  (\bibinfo {year} {2023})}\BibitemShut {NoStop}%
\bibitem [{\citenamefont {Esterlis}\ \emph {et~al.}(2021)\citenamefont
  {Esterlis}, \citenamefont {Guo}, \citenamefont {Patel},\ and\ \citenamefont
  {Sachdev}}]{Esterlis2021}%
  \BibitemOpen
  \bibfield  {author} {\bibinfo {author} {\bibfnamefont {I.}~\bibnamefont
  {Esterlis}}, \bibinfo {author} {\bibfnamefont {H.}~\bibnamefont {Guo}},
  \bibinfo {author} {\bibfnamefont {A.~A.}\ \bibnamefont {Patel}}, \ and\
  \bibinfo {author} {\bibfnamefont {S.}~\bibnamefont {Sachdev}},\ }\href
  {\doibase 10.1103/PhysRevB.103.235129} {\bibfield  {journal} {\bibinfo
  {journal} {Phys. Rev. B}\ }\textbf {\bibinfo {volume} {103}},\ \bibinfo
  {pages} {235129} (\bibinfo {year} {2021})}\BibitemShut {NoStop}%
\bibitem [{\citenamefont {Guo}\ \emph {et~al.}(2022)\citenamefont {Guo},
  \citenamefont {Patel}, \citenamefont {Esterlis},\ and\ \citenamefont
  {Sachdev}}]{Guo2022}%
  \BibitemOpen
  \bibfield  {author} {\bibinfo {author} {\bibfnamefont {H.}~\bibnamefont
  {Guo}}, \bibinfo {author} {\bibfnamefont {A.~A.}\ \bibnamefont {Patel}},
  \bibinfo {author} {\bibfnamefont {I.}~\bibnamefont {Esterlis}}, \ and\
  \bibinfo {author} {\bibfnamefont {S.}~\bibnamefont {Sachdev}},\ }\href
  {\doibase 10.1103/PhysRevB.106.115151} {\bibfield  {journal} {\bibinfo
  {journal} {Phys. Rev. B}\ }\textbf {\bibinfo {volume} {106}},\ \bibinfo
  {pages} {115151} (\bibinfo {year} {2022})}\BibitemShut {NoStop}%
\bibitem [{\citenamefont {Lee}(2009)}]{PhysRevB.80.165102}%
  \BibitemOpen
  \bibfield  {author} {\bibinfo {author} {\bibfnamefont {S.-S.}\ \bibnamefont
  {Lee}},\ }\href {\doibase 10.1103/PhysRevB.80.165102} {\bibfield  {journal}
  {\bibinfo  {journal} {Phys. Rev. B}\ }\textbf {\bibinfo {volume} {80}},\
  \bibinfo {pages} {165102} (\bibinfo {year} {2009})}\BibitemShut {NoStop}%
\bibitem [{\citenamefont {Holder}\ and\ \citenamefont
  {Metzner}(2014)}]{PhysRevB.90.161106}%
  \BibitemOpen
  \bibfield  {author} {\bibinfo {author} {\bibfnamefont {T.}~\bibnamefont
  {Holder}}\ and\ \bibinfo {author} {\bibfnamefont {W.}~\bibnamefont
  {Metzner}},\ }\href {\doibase 10.1103/PhysRevB.90.161106} {\bibfield
  {journal} {\bibinfo  {journal} {Phys. Rev. B}\ }\textbf {\bibinfo {volume}
  {90}},\ \bibinfo {pages} {161106} (\bibinfo {year} {2014})}\BibitemShut
  {NoStop}%
\bibitem [{\citenamefont {Rao}\ and\ \citenamefont
  {Piazza}(2023)}]{PhysRevLett.130.083603}%
  \BibitemOpen
  \bibfield  {author} {\bibinfo {author} {\bibfnamefont {P.}~\bibnamefont
  {Rao}}\ and\ \bibinfo {author} {\bibfnamefont {F.}~\bibnamefont {Piazza}},\
  }\href {\doibase 10.1103/PhysRevLett.130.083603} {\bibfield  {journal}
  {\bibinfo  {journal} {Phys. Rev. Lett.}\ }\textbf {\bibinfo {volume} {130}},\
  \bibinfo {pages} {083603} (\bibinfo {year} {2023})}\BibitemShut {NoStop}%
\bibitem [{\citenamefont {Polshyn}\ \emph {et~al.}(2019)\citenamefont
  {Polshyn}, \citenamefont {Yankowitz}, \citenamefont {Chen}, \citenamefont
  {Zhang}, \citenamefont {Watanabe}, \citenamefont {Taniguchi}, \citenamefont
  {Dean},\ and\ \citenamefont {Young}}]{Andrea2019}%
  \BibitemOpen
  \bibfield  {author} {\bibinfo {author} {\bibfnamefont {H.}~\bibnamefont
  {Polshyn}}, \bibinfo {author} {\bibfnamefont {M.}~\bibnamefont {Yankowitz}},
  \bibinfo {author} {\bibfnamefont {S.}~\bibnamefont {Chen}}, \bibinfo {author}
  {\bibfnamefont {Y.}~\bibnamefont {Zhang}}, \bibinfo {author} {\bibfnamefont
  {K.}~\bibnamefont {Watanabe}}, \bibinfo {author} {\bibfnamefont
  {T.}~\bibnamefont {Taniguchi}}, \bibinfo {author} {\bibfnamefont {C.~R.}\
  \bibnamefont {Dean}}, \ and\ \bibinfo {author} {\bibfnamefont {A.~F.}\
  \bibnamefont {Young}},\ }\href {\doibase 10.1038/s41567-019-0596-3}
  {\bibfield  {journal} {\bibinfo  {journal} {Nat. Phys.}\ }\textbf {\bibinfo
  {volume} {15}},\ \bibinfo {pages} {1011} (\bibinfo {year}
  {2019})}\BibitemShut {NoStop}%
\bibitem [{\citenamefont {Mousatov}\ \emph {et~al.}(2020)\citenamefont
  {Mousatov}, \citenamefont {Berg},\ and\ \citenamefont
  {Hartnoll}}]{Hartnoll2020}%
  \BibitemOpen
  \bibfield  {author} {\bibinfo {author} {\bibfnamefont {C.~H.}\ \bibnamefont
  {Mousatov}}, \bibinfo {author} {\bibfnamefont {E.}~\bibnamefont {Berg}}, \
  and\ \bibinfo {author} {\bibfnamefont {S.~A.}\ \bibnamefont {Hartnoll}},\
  }\href {\doibase 10.1073/pnas.1915224117} {\bibfield  {journal} {\bibinfo
  {journal} {Proc. Natl. Acad. Sci.}\ }\textbf {\bibinfo {volume} {117}},\
  \bibinfo {pages} {2852} (\bibinfo {year} {2020})}\BibitemShut {NoStop}%
\bibitem [{\citenamefont {Xu}\ \emph {et~al.}(2021{\natexlab{a}})\citenamefont
  {Xu}, \citenamefont {Herman}, \citenamefont {Granata}, \citenamefont
  {Destraz}, \citenamefont {Das}, \citenamefont {Vonka}, \citenamefont
  {Gerber}, \citenamefont {Spring}, \citenamefont {Gibert}, \citenamefont
  {Schilling} \emph {et~al.}}]{Xu2021}%
  \BibitemOpen
  \bibfield  {author} {\bibinfo {author} {\bibfnamefont {Y.}~\bibnamefont
  {Xu}}, \bibinfo {author} {\bibfnamefont {F.}~\bibnamefont {Herman}}, \bibinfo
  {author} {\bibfnamefont {V.}~\bibnamefont {Granata}}, \bibinfo {author}
  {\bibfnamefont {D.}~\bibnamefont {Destraz}}, \bibinfo {author} {\bibfnamefont
  {L.}~\bibnamefont {Das}}, \bibinfo {author} {\bibfnamefont {J.}~\bibnamefont
  {Vonka}}, \bibinfo {author} {\bibfnamefont {S.}~\bibnamefont {Gerber}},
  \bibinfo {author} {\bibfnamefont {J.}~\bibnamefont {Spring}}, \bibinfo
  {author} {\bibfnamefont {M.}~\bibnamefont {Gibert}}, \bibinfo {author}
  {\bibfnamefont {A.}~\bibnamefont {Schilling}},  \emph {et~al.},\ }\href
  {\doibase 10.1038/s42005-020-00504-0} {\bibfield  {journal} {\bibinfo
  {journal} {Communications Physics}\ }\textbf {\bibinfo {volume} {4}},\
  \bibinfo {pages} {1} (\bibinfo {year} {2021}{\natexlab{a}})}\BibitemShut
  {NoStop}%
\bibitem [{\citenamefont {Wang}\ \emph {et~al.}(2020)\citenamefont {Wang},
  \citenamefont {Shih}, \citenamefont {Ghiotto}, \citenamefont {Xian},
  \citenamefont {Rhodes}, \citenamefont {Tan}, \citenamefont {Claassen},
  \citenamefont {Kennes}, \citenamefont {Bai}, \citenamefont {Kim} \emph
  {et~al.}}]{wang2020correlated}%
  \BibitemOpen
  \bibfield  {author} {\bibinfo {author} {\bibfnamefont {L.}~\bibnamefont
  {Wang}}, \bibinfo {author} {\bibfnamefont {E.-M.}\ \bibnamefont {Shih}},
  \bibinfo {author} {\bibfnamefont {A.}~\bibnamefont {Ghiotto}}, \bibinfo
  {author} {\bibfnamefont {L.}~\bibnamefont {Xian}}, \bibinfo {author}
  {\bibfnamefont {D.~A.}\ \bibnamefont {Rhodes}}, \bibinfo {author}
  {\bibfnamefont {C.}~\bibnamefont {Tan}}, \bibinfo {author} {\bibfnamefont
  {M.}~\bibnamefont {Claassen}}, \bibinfo {author} {\bibfnamefont {D.~M.}\
  \bibnamefont {Kennes}}, \bibinfo {author} {\bibfnamefont {Y.}~\bibnamefont
  {Bai}}, \bibinfo {author} {\bibfnamefont {B.}~\bibnamefont {Kim}},  \emph
  {et~al.},\ }\href {\doibase 10.1038/s41563-020-0708-6} {\bibfield  {journal}
  {\bibinfo  {journal} {Nature materials}\ }\textbf {\bibinfo {volume} {19}},\
  \bibinfo {pages} {861} (\bibinfo {year} {2020})}\BibitemShut {NoStop}%
\bibitem [{\citenamefont {Yudin}\ \emph {et~al.}(2014)\citenamefont {Yudin},
  \citenamefont {Hirschmeier}, \citenamefont {Hafermann}, \citenamefont
  {Eriksson}, \citenamefont {Lichtenstein},\ and\ \citenamefont
  {Katsnelson}}]{PhysRevLett.112.070403}%
  \BibitemOpen
  \bibfield  {author} {\bibinfo {author} {\bibfnamefont {D.}~\bibnamefont
  {Yudin}}, \bibinfo {author} {\bibfnamefont {D.}~\bibnamefont {Hirschmeier}},
  \bibinfo {author} {\bibfnamefont {H.}~\bibnamefont {Hafermann}}, \bibinfo
  {author} {\bibfnamefont {O.}~\bibnamefont {Eriksson}}, \bibinfo {author}
  {\bibfnamefont {A.~I.}\ \bibnamefont {Lichtenstein}}, \ and\ \bibinfo
  {author} {\bibfnamefont {M.~I.}\ \bibnamefont {Katsnelson}},\ }\href
  {\doibase 10.1103/PhysRevLett.112.070403} {\bibfield  {journal} {\bibinfo
  {journal} {Phys. Rev. Lett.}\ }\textbf {\bibinfo {volume} {112}},\ \bibinfo
  {pages} {070403} (\bibinfo {year} {2014})}\BibitemShut {NoStop}%
\bibitem [{\citenamefont {Karp}\ \emph {et~al.}(2020)\citenamefont {Karp},
  \citenamefont {Bramberger}, \citenamefont {Grundner}, \citenamefont
  {Schollw\"ock}, \citenamefont {Millis},\ and\ \citenamefont
  {Zingl}}]{PhysRevLett.125.166401}%
  \BibitemOpen
  \bibfield  {author} {\bibinfo {author} {\bibfnamefont {J.}~\bibnamefont
  {Karp}}, \bibinfo {author} {\bibfnamefont {M.}~\bibnamefont {Bramberger}},
  \bibinfo {author} {\bibfnamefont {M.}~\bibnamefont {Grundner}}, \bibinfo
  {author} {\bibfnamefont {U.}~\bibnamefont {Schollw\"ock}}, \bibinfo {author}
  {\bibfnamefont {A.~J.}\ \bibnamefont {Millis}}, \ and\ \bibinfo {author}
  {\bibfnamefont {M.}~\bibnamefont {Zingl}},\ }\href {\doibase
  10.1103/PhysRevLett.125.166401} {\bibfield  {journal} {\bibinfo  {journal}
  {Phys. Rev. Lett.}\ }\textbf {\bibinfo {volume} {125}},\ \bibinfo {pages}
  {166401} (\bibinfo {year} {2020})}\BibitemShut {NoStop}%
\bibitem [{\citenamefont {Kang}\ \emph {et~al.}(2022)\citenamefont {Kang},
  \citenamefont {Fang}, \citenamefont {Kim}, \citenamefont {Ortiz},
  \citenamefont {Ryu}, \citenamefont {Kim}, \citenamefont {Yoo}, \citenamefont
  {Sangiovanni}, \citenamefont {Di~Sante}, \citenamefont {Park} \emph
  {et~al.}}]{kang2022twofold}%
  \BibitemOpen
  \bibfield  {author} {\bibinfo {author} {\bibfnamefont {M.}~\bibnamefont
  {Kang}}, \bibinfo {author} {\bibfnamefont {S.}~\bibnamefont {Fang}}, \bibinfo
  {author} {\bibfnamefont {J.-K.}\ \bibnamefont {Kim}}, \bibinfo {author}
  {\bibfnamefont {B.~R.}\ \bibnamefont {Ortiz}}, \bibinfo {author}
  {\bibfnamefont {S.~H.}\ \bibnamefont {Ryu}}, \bibinfo {author} {\bibfnamefont
  {J.}~\bibnamefont {Kim}}, \bibinfo {author} {\bibfnamefont {J.}~\bibnamefont
  {Yoo}}, \bibinfo {author} {\bibfnamefont {G.}~\bibnamefont {Sangiovanni}},
  \bibinfo {author} {\bibfnamefont {D.}~\bibnamefont {Di~Sante}}, \bibinfo
  {author} {\bibfnamefont {B.-G.}\ \bibnamefont {Park}},  \emph {et~al.},\
  }\href {\doibase 10.1038/s41567-021-01451-5} {\bibfield  {journal} {\bibinfo
  {journal} {Nat. Phys.}\ }\textbf {\bibinfo {volume} {18}},\ \bibinfo {pages}
  {301} (\bibinfo {year} {2022})}\BibitemShut {NoStop}%
\bibitem [{\citenamefont {Wu}\ \emph {et~al.}(2021)\citenamefont {Wu},
  \citenamefont {Zhang}, \citenamefont {Watanabe}, \citenamefont {Taniguchi},\
  and\ \citenamefont {Andrei}}]{wu2021chern}%
  \BibitemOpen
  \bibfield  {author} {\bibinfo {author} {\bibfnamefont {S.}~\bibnamefont
  {Wu}}, \bibinfo {author} {\bibfnamefont {Z.}~\bibnamefont {Zhang}}, \bibinfo
  {author} {\bibfnamefont {K.}~\bibnamefont {Watanabe}}, \bibinfo {author}
  {\bibfnamefont {T.}~\bibnamefont {Taniguchi}}, \ and\ \bibinfo {author}
  {\bibfnamefont {E.~Y.}\ \bibnamefont {Andrei}},\ }\href {\doibase
  10.1038/s41563-020-00911-2} {\bibfield  {journal} {\bibinfo  {journal}
  {Nature materials}\ }\textbf {\bibinfo {volume} {20}},\ \bibinfo {pages}
  {488} (\bibinfo {year} {2021})}\BibitemShut {NoStop}%
\bibitem [{\citenamefont {Luo}\ \emph {et~al.}(2023)\citenamefont {Luo},
  \citenamefont {Han}, \citenamefont {Liu}, \citenamefont {Chen}, \citenamefont
  {Huang}, \citenamefont {Huai}, \citenamefont {Li}, \citenamefont {Wang},
  \citenamefont {Shen}, \citenamefont {Ding} \emph {et~al.}}]{luo2023unique}%
  \BibitemOpen
  \bibfield  {author} {\bibinfo {author} {\bibfnamefont {Y.}~\bibnamefont
  {Luo}}, \bibinfo {author} {\bibfnamefont {Y.}~\bibnamefont {Han}}, \bibinfo
  {author} {\bibfnamefont {J.}~\bibnamefont {Liu}}, \bibinfo {author}
  {\bibfnamefont {H.}~\bibnamefont {Chen}}, \bibinfo {author} {\bibfnamefont
  {Z.}~\bibnamefont {Huang}}, \bibinfo {author} {\bibfnamefont
  {L.}~\bibnamefont {Huai}}, \bibinfo {author} {\bibfnamefont {H.}~\bibnamefont
  {Li}}, \bibinfo {author} {\bibfnamefont {B.}~\bibnamefont {Wang}}, \bibinfo
  {author} {\bibfnamefont {J.}~\bibnamefont {Shen}}, \bibinfo {author}
  {\bibfnamefont {S.}~\bibnamefont {Ding}},  \emph {et~al.},\ }\href {\doibase
  10.1038/s41467-023-39500-7} {\bibfield  {journal} {\bibinfo  {journal} {Nat.
  Commun.}\ }\textbf {\bibinfo {volume} {14}},\ \bibinfo {pages} {3819}
  (\bibinfo {year} {2023})}\BibitemShut {NoStop}%
\bibitem [{\citenamefont {Sano}\ \emph {et~al.}(2016)\citenamefont {Sano},
  \citenamefont {Koretsune}, \citenamefont {Tadano}, \citenamefont {Akashi},\
  and\ \citenamefont {Arita}}]{PhysRevB.93.094525}%
  \BibitemOpen
  \bibfield  {author} {\bibinfo {author} {\bibfnamefont {W.}~\bibnamefont
  {Sano}}, \bibinfo {author} {\bibfnamefont {T.}~\bibnamefont {Koretsune}},
  \bibinfo {author} {\bibfnamefont {T.}~\bibnamefont {Tadano}}, \bibinfo
  {author} {\bibfnamefont {R.}~\bibnamefont {Akashi}}, \ and\ \bibinfo {author}
  {\bibfnamefont {R.}~\bibnamefont {Arita}},\ }\href {\doibase
  10.1103/PhysRevB.93.094525} {\bibfield  {journal} {\bibinfo  {journal} {Phys.
  Rev. B}\ }\textbf {\bibinfo {volume} {93}},\ \bibinfo {pages} {094525}
  (\bibinfo {year} {2016})}\BibitemShut {NoStop}%
\bibitem [{\citenamefont {Irkhin}\ \emph {et~al.}(2002)\citenamefont {Irkhin},
  \citenamefont {Katanin},\ and\ \citenamefont
  {Katsnelson}}]{PhysRevLett.89.076401}%
  \BibitemOpen
  \bibfield  {author} {\bibinfo {author} {\bibfnamefont {V.~Y.}\ \bibnamefont
  {Irkhin}}, \bibinfo {author} {\bibfnamefont {A.~A.}\ \bibnamefont {Katanin}},
  \ and\ \bibinfo {author} {\bibfnamefont {M.~I.}\ \bibnamefont {Katsnelson}},\
  }\href {\doibase 10.1103/PhysRevLett.89.076401} {\bibfield  {journal}
  {\bibinfo  {journal} {Phys. Rev. Lett.}\ }\textbf {\bibinfo {volume} {89}},\
  \bibinfo {pages} {076401} (\bibinfo {year} {2002})}\BibitemShut {NoStop}%
\bibitem [{\citenamefont {Wu}\ \emph {et~al.}(2023{\natexlab{a}})\citenamefont
  {Wu}, \citenamefont {Wu},\ and\ \citenamefont
  {Yao}}]{PhysRevLett.130.126001}%
  \BibitemOpen
  \bibfield  {author} {\bibinfo {author} {\bibfnamefont {Y.-M.}\ \bibnamefont
  {Wu}}, \bibinfo {author} {\bibfnamefont {Z.}~\bibnamefont {Wu}}, \ and\
  \bibinfo {author} {\bibfnamefont {H.}~\bibnamefont {Yao}},\ }\href {\doibase
  10.1103/PhysRevLett.130.126001} {\bibfield  {journal} {\bibinfo  {journal}
  {Phys. Rev. Lett.}\ }\textbf {\bibinfo {volume} {130}},\ \bibinfo {pages}
  {126001} (\bibinfo {year} {2023}{\natexlab{a}})}\BibitemShut {NoStop}%
\bibitem [{\citenamefont {Castro}\ \emph {et~al.}(2023)\citenamefont {Castro},
  \citenamefont {Shaffer}, \citenamefont {Wu},\ and\ \citenamefont
  {Santos}}]{PhysRevLett.131.026601}%
  \BibitemOpen
  \bibfield  {author} {\bibinfo {author} {\bibfnamefont {P.}~\bibnamefont
  {Castro}}, \bibinfo {author} {\bibfnamefont {D.}~\bibnamefont {Shaffer}},
  \bibinfo {author} {\bibfnamefont {Y.-M.}\ \bibnamefont {Wu}}, \ and\ \bibinfo
  {author} {\bibfnamefont {L.~H.}\ \bibnamefont {Santos}},\ }\href {\doibase
  10.1103/PhysRevLett.131.026601} {\bibfield  {journal} {\bibinfo  {journal}
  {Phys. Rev. Lett.}\ }\textbf {\bibinfo {volume} {131}},\ \bibinfo {pages}
  {026601} (\bibinfo {year} {2023})}\BibitemShut {NoStop}%
\bibitem [{\citenamefont {Sachdev}\ and\ \citenamefont {Ye}(1993)}]{SY1993}%
  \BibitemOpen
  \bibfield  {author} {\bibinfo {author} {\bibfnamefont {S.}~\bibnamefont
  {Sachdev}}\ and\ \bibinfo {author} {\bibfnamefont {J.}~\bibnamefont {Ye}},\
  }\href {\doibase 10.1103/PhysRevLett.70.3339} {\bibfield  {journal} {\bibinfo
   {journal} {Phys. Rev. Lett.}\ }\textbf {\bibinfo {volume} {70}},\ \bibinfo
  {pages} {3339} (\bibinfo {year} {1993})}\BibitemShut {NoStop}%
\bibitem [{\citenamefont {Wang}(2020)}]{PhysRevLett.124.017002}%
  \BibitemOpen
  \bibfield  {author} {\bibinfo {author} {\bibfnamefont {Y.}~\bibnamefont
  {Wang}},\ }\href {\doibase 10.1103/PhysRevLett.124.017002} {\bibfield
  {journal} {\bibinfo  {journal} {Phys. Rev. Lett.}\ }\textbf {\bibinfo
  {volume} {124}},\ \bibinfo {pages} {017002} (\bibinfo {year}
  {2020})}\BibitemShut {NoStop}%
\bibitem [{\citenamefont {Chowdhury}\ \emph {et~al.}(2018)\citenamefont
  {Chowdhury}, \citenamefont {Werman}, \citenamefont {Berg},\ and\
  \citenamefont {Senthil}}]{PhysRevX.8.031024}%
  \BibitemOpen
  \bibfield  {author} {\bibinfo {author} {\bibfnamefont {D.}~\bibnamefont
  {Chowdhury}}, \bibinfo {author} {\bibfnamefont {Y.}~\bibnamefont {Werman}},
  \bibinfo {author} {\bibfnamefont {E.}~\bibnamefont {Berg}}, \ and\ \bibinfo
  {author} {\bibfnamefont {T.}~\bibnamefont {Senthil}},\ }\href {\doibase
  10.1103/PhysRevX.8.031024} {\bibfield  {journal} {\bibinfo  {journal} {Phys.
  Rev. X}\ }\textbf {\bibinfo {volume} {8}},\ \bibinfo {pages} {031024}
  (\bibinfo {year} {2018})}\BibitemShut {NoStop}%
\bibitem [{\citenamefont {Wu}\ \emph {et~al.}(2023{\natexlab{b}})\citenamefont
  {Wu}, \citenamefont {Nosov}, \citenamefont {Patel},\ and\ \citenamefont
  {Raghu}}]{Raghu2023}%
  \BibitemOpen
  \bibfield  {author} {\bibinfo {author} {\bibfnamefont {Y.-M.}\ \bibnamefont
  {Wu}}, \bibinfo {author} {\bibfnamefont {P.~A.}\ \bibnamefont {Nosov}},
  \bibinfo {author} {\bibfnamefont {A.~A.}\ \bibnamefont {Patel}}, \ and\
  \bibinfo {author} {\bibfnamefont {S.}~\bibnamefont {Raghu}},\ }\href
  {\doibase 10.1103/PhysRevLett.130.026001} {\bibfield  {journal} {\bibinfo
  {journal} {Phys. Rev. Lett.}\ }\textbf {\bibinfo {volume} {130}},\ \bibinfo
  {pages} {026001} (\bibinfo {year} {2023}{\natexlab{b}})}\BibitemShut
  {NoStop}%
\bibitem [{\citenamefont {Sachdev}(2015)}]{PhysRevX.5.041025}%
  \BibitemOpen
  \bibfield  {author} {\bibinfo {author} {\bibfnamefont {S.}~\bibnamefont
  {Sachdev}},\ }\href {\doibase 10.1103/PhysRevX.5.041025} {\bibfield
  {journal} {\bibinfo  {journal} {Phys. Rev. X}\ }\textbf {\bibinfo {volume}
  {5}},\ \bibinfo {pages} {041025} (\bibinfo {year} {2015})}\BibitemShut
  {NoStop}%
\bibitem [{\citenamefont {Chen}\ \emph
  {et~al.}(2019{\natexlab{b}})\citenamefont {Chen}, \citenamefont {Hashimoto},
  \citenamefont {He}, \citenamefont {Song}, \citenamefont {Xu}, \citenamefont
  {He}, \citenamefont {Devereaux}, \citenamefont {Eisaki}, \citenamefont {Lu},
  \citenamefont {Zaanen} \emph {et~al.}}]{chen2019incoherent}%
  \BibitemOpen
  \bibfield  {author} {\bibinfo {author} {\bibfnamefont {S.-D.}\ \bibnamefont
  {Chen}}, \bibinfo {author} {\bibfnamefont {M.}~\bibnamefont {Hashimoto}},
  \bibinfo {author} {\bibfnamefont {Y.}~\bibnamefont {He}}, \bibinfo {author}
  {\bibfnamefont {D.}~\bibnamefont {Song}}, \bibinfo {author} {\bibfnamefont
  {K.-J.}\ \bibnamefont {Xu}}, \bibinfo {author} {\bibfnamefont {J.-F.}\
  \bibnamefont {He}}, \bibinfo {author} {\bibfnamefont {T.~P.}\ \bibnamefont
  {Devereaux}}, \bibinfo {author} {\bibfnamefont {H.}~\bibnamefont {Eisaki}},
  \bibinfo {author} {\bibfnamefont {D.-H.}\ \bibnamefont {Lu}}, \bibinfo
  {author} {\bibfnamefont {J.}~\bibnamefont {Zaanen}},  \emph {et~al.},\ }\href
  {\doibase 10.1126/science.aaw8850} {\bibfield  {journal} {\bibinfo  {journal}
  {Science}\ }\textbf {\bibinfo {volume} {366}},\ \bibinfo {pages} {1099}
  (\bibinfo {year} {2019}{\natexlab{b}})}\BibitemShut {NoStop}%
\bibitem [{\citenamefont {Chowdhury}\ \emph {et~al.}(2022)\citenamefont
  {Chowdhury}, \citenamefont {Georges}, \citenamefont {Parcollet},\ and\
  \citenamefont {Sachdev}}]{RevModPhys.94.035004}%
  \BibitemOpen
  \bibfield  {author} {\bibinfo {author} {\bibfnamefont {D.}~\bibnamefont
  {Chowdhury}}, \bibinfo {author} {\bibfnamefont {A.}~\bibnamefont {Georges}},
  \bibinfo {author} {\bibfnamefont {O.}~\bibnamefont {Parcollet}}, \ and\
  \bibinfo {author} {\bibfnamefont {S.}~\bibnamefont {Sachdev}},\ }\href
  {\doibase 10.1103/RevModPhys.94.035004} {\bibfield  {journal} {\bibinfo
  {journal} {Rev. Mod. Phys.}\ }\textbf {\bibinfo {volume} {94}},\ \bibinfo
  {pages} {035004} (\bibinfo {year} {2022})}\BibitemShut {NoStop}%
\bibitem [{phi()}]{phi4}%
  \BibitemOpen
  \href@noop {} {}\bibinfo {note} {There is an implicit assumption that
  high-order magnon-mangon interactions is much smaller than electron-spin
  exchange interactions due to the renormalization of the effective mass of
  higher-order magnon-magnon interactions is different from that at zero
  temperature. However, at the mean-field level, we do not require this
  assumption.}\BibitemShut {Stop}%
\bibitem [{\citenamefont {Zhou}\ and\ \citenamefont
  {Wang}(2022)}]{zhou2022chern}%
  \BibitemOpen
  \bibfield  {author} {\bibinfo {author} {\bibfnamefont {S.}~\bibnamefont
  {Zhou}}\ and\ \bibinfo {author} {\bibfnamefont {Z.}~\bibnamefont {Wang}},\
  }\href {\doibase 10.1038/s41467-022-34832-2} {\bibfield  {journal} {\bibinfo
  {journal} {Nat. Commun.}\ }\textbf {\bibinfo {volume} {13}},\ \bibinfo
  {pages} {7288} (\bibinfo {year} {2022})}\BibitemShut {NoStop}%
\bibitem [{\citenamefont {Hamidian}\ \emph {et~al.}(2016)\citenamefont
  {Hamidian}, \citenamefont {Edkins}, \citenamefont {Joo}, \citenamefont
  {Kostin}, \citenamefont {Eisaki}, \citenamefont {Uchida}, \citenamefont
  {Lawler}, \citenamefont {Kim}, \citenamefont {Mackenzie}, \citenamefont
  {Fujita} \emph {et~al.}}]{hamidian2016detection}%
  \BibitemOpen
  \bibfield  {author} {\bibinfo {author} {\bibfnamefont {M.}~\bibnamefont
  {Hamidian}}, \bibinfo {author} {\bibfnamefont {S.}~\bibnamefont {Edkins}},
  \bibinfo {author} {\bibfnamefont {S.~H.}\ \bibnamefont {Joo}}, \bibinfo
  {author} {\bibfnamefont {A.}~\bibnamefont {Kostin}}, \bibinfo {author}
  {\bibfnamefont {H.}~\bibnamefont {Eisaki}}, \bibinfo {author} {\bibfnamefont
  {S.}~\bibnamefont {Uchida}}, \bibinfo {author} {\bibfnamefont
  {M.}~\bibnamefont {Lawler}}, \bibinfo {author} {\bibfnamefont {E.-A.}\
  \bibnamefont {Kim}}, \bibinfo {author} {\bibfnamefont {A.}~\bibnamefont
  {Mackenzie}}, \bibinfo {author} {\bibfnamefont {K.}~\bibnamefont {Fujita}},
  \emph {et~al.},\ }\href {\doibase 10.1038/nature17411} {\bibfield  {journal}
  {\bibinfo  {journal} {Nature}\ }\textbf {\bibinfo {volume} {532}},\ \bibinfo
  {pages} {343} (\bibinfo {year} {2016})}\BibitemShut {NoStop}%
\bibitem [{\citenamefont {Wang}\ \emph {et~al.}(2015)\citenamefont {Wang},
  \citenamefont {Agterberg},\ and\ \citenamefont
  {Chubukov}}]{PhysRevLett.114.197001}%
  \BibitemOpen
  \bibfield  {author} {\bibinfo {author} {\bibfnamefont {Y.}~\bibnamefont
  {Wang}}, \bibinfo {author} {\bibfnamefont {D.~F.}\ \bibnamefont {Agterberg}},
  \ and\ \bibinfo {author} {\bibfnamefont {A.}~\bibnamefont {Chubukov}},\
  }\href {\doibase 10.1103/PhysRevLett.114.197001} {\bibfield  {journal}
  {\bibinfo  {journal} {Phys. Rev. Lett.}\ }\textbf {\bibinfo {volume} {114}},\
  \bibinfo {pages} {197001} (\bibinfo {year} {2015})}\BibitemShut {NoStop}%
\bibitem [{\citenamefont {Webster}\ and\ \citenamefont
  {Yan}(2018)}]{PhysRevB.98.144411}%
  \BibitemOpen
  \bibfield  {author} {\bibinfo {author} {\bibfnamefont {L.}~\bibnamefont
  {Webster}}\ and\ \bibinfo {author} {\bibfnamefont {J.-A.}\ \bibnamefont
  {Yan}},\ }\href {\doibase 10.1103/PhysRevB.98.144411} {\bibfield  {journal}
  {\bibinfo  {journal} {Phys. Rev. B}\ }\textbf {\bibinfo {volume} {98}},\
  \bibinfo {pages} {144411} (\bibinfo {year} {2018})}\BibitemShut {NoStop}%
\bibitem [{\citenamefont {M\ae{}land}\ and\ \citenamefont
  {Sudb\o{}}(2022)}]{PhysRevResearch.4.L032025}%
  \BibitemOpen
  \bibfield  {author} {\bibinfo {author} {\bibfnamefont {K.}~\bibnamefont
  {M\ae{}land}}\ and\ \bibinfo {author} {\bibfnamefont {A.}~\bibnamefont
  {Sudb\o{}}},\ }\href {\doibase 10.1103/PhysRevResearch.4.L032025} {\bibfield
  {journal} {\bibinfo  {journal} {Phys. Rev. Res.}\ }\textbf {\bibinfo {volume}
  {4}},\ \bibinfo {pages} {L032025} (\bibinfo {year} {2022})}\BibitemShut
  {NoStop}%
\bibitem [{ElE()}]{ElEnSp}%
  \BibitemOpen
  \href@noop {} {}\bibinfo {note} {In fact, there is an additional term
  $\frac{JS}{\sqrt{2S}}\sum_i(c^\dagger_{i\downarrow}c_{i\downarrow}-c^\dagger_{i\uparrow}c_{i\uparrow})$
  in $H_{em}$ that can lead to different energy spectra for electrons with
  different spins. However, this can be addressed by introducing a magnetic
  field $B$ to the normal metal layer (with new tuning parameter
  $\Delta=k_z^2+4K/\bar{J}a^2-2B/S\bar{J}a^2$) or changing the chemical
  potential. Furthermore, in the subsequent scaling analysis, it can be proven
  that the NFL/MFL behavior is not sensitive to the specific form of the energy
  spectrum (, it only depends on the power-law behavior of momentum after a
  Taylor expansion). So, even when considering scattering between patches near
  VHPS of Fermi surface with different spin, it does not affect our
  conclusions.}\BibitemShut {Stop}%
\bibitem [{EP()}]{EP}%
  \BibitemOpen
  \href@noop {} {}\bibinfo {note} {Unless the Fermi surface is nested, in which
  case the contribution of flat Fermi surface is also important
  \cite{PhysRevB.64.165107}.}\BibitemShut {Stop}%
\bibitem [{EnE()}]{EnEl}%
  \BibitemOpen
  \href@noop {} {}\bibinfo {note} {For a square lattice considering only
  nearest-neighbor hopping, the VHPs occur at half-filling, and the electron
  dispersions around VHPs can be written as $\pm t(k_x^2-k_y^2)$. Dispersion
  near VHP in Ref. \cite{PhysRevLett.130.126001} is
  $\tilde{t}(k_x'^2-3k_y'^2)/2$.}\BibitemShut {Stop}%
\bibitem [{\citenamefont {Isobe}\ \emph {et~al.}(2018)\citenamefont {Isobe},
  \citenamefont {Yuan},\ and\ \citenamefont {Fu}}]{PhysRevX.8.041041}%
  \BibitemOpen
  \bibfield  {author} {\bibinfo {author} {\bibfnamefont {H.}~\bibnamefont
  {Isobe}}, \bibinfo {author} {\bibfnamefont {N.~F.~Q.}\ \bibnamefont {Yuan}},
  \ and\ \bibinfo {author} {\bibfnamefont {L.}~\bibnamefont {Fu}},\ }\href
  {\doibase 10.1103/PhysRevX.8.041041} {\bibfield  {journal} {\bibinfo
  {journal} {Phys. Rev. X}\ }\textbf {\bibinfo {volume} {8}},\ \bibinfo {pages}
  {041041} (\bibinfo {year} {2018})}\BibitemShut {NoStop}%
\bibitem [{\citenamefont {Yao}\ and\ \citenamefont
  {Yang}(2015)}]{PhysRevB.92.035132}%
  \BibitemOpen
  \bibfield  {author} {\bibinfo {author} {\bibfnamefont {H.}~\bibnamefont
  {Yao}}\ and\ \bibinfo {author} {\bibfnamefont {F.}~\bibnamefont {Yang}},\
  }\href {\doibase 10.1103/PhysRevB.92.035132} {\bibfield  {journal} {\bibinfo
  {journal} {Phys. Rev. B}\ }\textbf {\bibinfo {volume} {92}},\ \bibinfo
  {pages} {035132} (\bibinfo {year} {2015})}\BibitemShut {NoStop}%
\bibitem [{SM()}]{SM}%
  \BibitemOpen
  \href@noop {} {}\bibinfo {note} {See Supplemental Material for the derivation
  of the self-energies of fermion and boson (S1), the specific heat (S2), the
  transport properties in the clean limit (S3), the transport properties in the
  dirty limit (S4) and the crossover exhibited by both types of Yukawa
  interactions (S5).}\BibitemShut {Stop}%
\bibitem [{\citenamefont {Affleck}\ and\ \citenamefont
  {Ludwig}(1993)}]{PhysRevB.48.7297}%
  \BibitemOpen
  \bibfield  {author} {\bibinfo {author} {\bibfnamefont {I.}~\bibnamefont
  {Affleck}}\ and\ \bibinfo {author} {\bibfnamefont {A.~W.~W.}\ \bibnamefont
  {Ludwig}},\ }\href {\doibase 10.1103/PhysRevB.48.7297} {\bibfield  {journal}
  {\bibinfo  {journal} {Phys. Rev. B}\ }\textbf {\bibinfo {volume} {48}},\
  \bibinfo {pages} {7297} (\bibinfo {year} {1993})}\BibitemShut {NoStop}%
\bibitem [{\citenamefont {Gerlach}\ \emph {et~al.}(2017)\citenamefont
  {Gerlach}, \citenamefont {Schattner}, \citenamefont {Berg},\ and\
  \citenamefont {Trebst}}]{PhysRevB.95.035124}%
  \BibitemOpen
  \bibfield  {author} {\bibinfo {author} {\bibfnamefont {M.~H.}\ \bibnamefont
  {Gerlach}}, \bibinfo {author} {\bibfnamefont {Y.}~\bibnamefont {Schattner}},
  \bibinfo {author} {\bibfnamefont {E.}~\bibnamefont {Berg}}, \ and\ \bibinfo
  {author} {\bibfnamefont {S.}~\bibnamefont {Trebst}},\ }\href {\doibase
  10.1103/PhysRevB.95.035124} {\bibfield  {journal} {\bibinfo  {journal} {Phys.
  Rev. B}\ }\textbf {\bibinfo {volume} {95}},\ \bibinfo {pages} {035124}
  (\bibinfo {year} {2017})}\BibitemShut {NoStop}%
\bibitem [{\citenamefont {Driskell}\ \emph {et~al.}(2021)\citenamefont
  {Driskell}, \citenamefont {Lederer}, \citenamefont {Bauer}, \citenamefont
  {Trebst},\ and\ \citenamefont {Kim}}]{PhysRevLett.127.046601}%
  \BibitemOpen
  \bibfield  {author} {\bibinfo {author} {\bibfnamefont {G.}~\bibnamefont
  {Driskell}}, \bibinfo {author} {\bibfnamefont {S.}~\bibnamefont {Lederer}},
  \bibinfo {author} {\bibfnamefont {C.}~\bibnamefont {Bauer}}, \bibinfo
  {author} {\bibfnamefont {S.}~\bibnamefont {Trebst}}, \ and\ \bibinfo {author}
  {\bibfnamefont {E.-A.}\ \bibnamefont {Kim}},\ }\href {\doibase
  10.1103/PhysRevLett.127.046601} {\bibfield  {journal} {\bibinfo  {journal}
  {Phys. Rev. Lett.}\ }\textbf {\bibinfo {volume} {127}},\ \bibinfo {pages}
  {046601} (\bibinfo {year} {2021})}\BibitemShut {NoStop}%
\bibitem [{\citenamefont {Hertz}(1976)}]{Hertz1976}%
  \BibitemOpen
  \bibfield  {author} {\bibinfo {author} {\bibfnamefont {J.~A.}\ \bibnamefont
  {Hertz}},\ }\href {\doibase 10.1103/PhysRevB.14.1165} {\bibfield  {journal}
  {\bibinfo  {journal} {Phys. Rev. B}\ }\textbf {\bibinfo {volume} {14}},\
  \bibinfo {pages} {1165} (\bibinfo {year} {1976})}\BibitemShut {NoStop}%
\bibitem [{\citenamefont {Lederer}\ \emph {et~al.}(2017)\citenamefont
  {Lederer}, \citenamefont {Schattner}, \citenamefont {Berg},\ and\
  \citenamefont {Kivelson}}]{doi:10.1073/pnas.1620651114}%
  \BibitemOpen
  \bibfield  {author} {\bibinfo {author} {\bibfnamefont {S.}~\bibnamefont
  {Lederer}}, \bibinfo {author} {\bibfnamefont {Y.}~\bibnamefont {Schattner}},
  \bibinfo {author} {\bibfnamefont {E.}~\bibnamefont {Berg}}, \ and\ \bibinfo
  {author} {\bibfnamefont {S.~A.}\ \bibnamefont {Kivelson}},\ }\href {\doibase
  10.1073/pnas.1620651114} {\bibfield  {journal} {\bibinfo  {journal} {Proc.
  Natl. Acad. Sci.}\ }\textbf {\bibinfo {volume} {114}},\ \bibinfo {pages}
  {4905} (\bibinfo {year} {2017})}\BibitemShut {NoStop}%
\bibitem [{\citenamefont {Lederer}\ \emph {et~al.}(2015)\citenamefont
  {Lederer}, \citenamefont {Schattner}, \citenamefont {Berg},\ and\
  \citenamefont {Kivelson}}]{PhysRevLett.114.097001}%
  \BibitemOpen
  \bibfield  {author} {\bibinfo {author} {\bibfnamefont {S.}~\bibnamefont
  {Lederer}}, \bibinfo {author} {\bibfnamefont {Y.}~\bibnamefont {Schattner}},
  \bibinfo {author} {\bibfnamefont {E.}~\bibnamefont {Berg}}, \ and\ \bibinfo
  {author} {\bibfnamefont {S.~A.}\ \bibnamefont {Kivelson}},\ }\href {\doibase
  10.1103/PhysRevLett.114.097001} {\bibfield  {journal} {\bibinfo  {journal}
  {Phys. Rev. Lett.}\ }\textbf {\bibinfo {volume} {114}},\ \bibinfo {pages}
  {097001} (\bibinfo {year} {2015})}\BibitemShut {NoStop}%
\bibitem [{\citenamefont {Radovan}\ \emph {et~al.}(2003)\citenamefont
  {Radovan}, \citenamefont {Fortune}, \citenamefont {Murphy}, \citenamefont
  {Hannahs}, \citenamefont {Palm}, \citenamefont {Tozer},\ and\ \citenamefont
  {Hall}}]{radovan2003magnetic}%
  \BibitemOpen
  \bibfield  {author} {\bibinfo {author} {\bibfnamefont {H.~A.}\ \bibnamefont
  {Radovan}}, \bibinfo {author} {\bibfnamefont {N.~A.}\ \bibnamefont
  {Fortune}}, \bibinfo {author} {\bibfnamefont {T.~P.}\ \bibnamefont {Murphy}},
  \bibinfo {author} {\bibfnamefont {S.~T.}\ \bibnamefont {Hannahs}}, \bibinfo
  {author} {\bibfnamefont {E.~C.}\ \bibnamefont {Palm}}, \bibinfo {author}
  {\bibfnamefont {S.~W.}\ \bibnamefont {Tozer}}, \ and\ \bibinfo {author}
  {\bibfnamefont {D.}~\bibnamefont {Hall}},\ }\href {\doibase
  10.1038/nature01842} {\bibfield  {journal} {\bibinfo  {journal} {Nature}\
  }\textbf {\bibinfo {volume} {425}},\ \bibinfo {pages} {51} (\bibinfo {year}
  {2003})}\BibitemShut {NoStop}%
\bibitem [{\citenamefont {Agterberg}\ \emph {et~al.}(2020)\citenamefont
  {Agterberg}, \citenamefont {Davis}, \citenamefont {Edkins}, \citenamefont
  {Fradkin}, \citenamefont {Van~Harlingen}, \citenamefont {Kivelson},
  \citenamefont {Lee}, \citenamefont {Radzihovsky}, \citenamefont {Tranquada},\
  and\ \citenamefont {Wang}}]{annurev-conmatphys-031119-050711}%
  \BibitemOpen
  \bibfield  {author} {\bibinfo {author} {\bibfnamefont {D.~F.}\ \bibnamefont
  {Agterberg}}, \bibinfo {author} {\bibfnamefont {J.~S.}\ \bibnamefont
  {Davis}}, \bibinfo {author} {\bibfnamefont {S.~D.}\ \bibnamefont {Edkins}},
  \bibinfo {author} {\bibfnamefont {E.}~\bibnamefont {Fradkin}}, \bibinfo
  {author} {\bibfnamefont {D.~J.}\ \bibnamefont {Van~Harlingen}}, \bibinfo
  {author} {\bibfnamefont {S.~A.}\ \bibnamefont {Kivelson}}, \bibinfo {author}
  {\bibfnamefont {P.~A.}\ \bibnamefont {Lee}}, \bibinfo {author} {\bibfnamefont
  {L.}~\bibnamefont {Radzihovsky}}, \bibinfo {author} {\bibfnamefont {J.~M.}\
  \bibnamefont {Tranquada}}, \ and\ \bibinfo {author} {\bibfnamefont
  {Y.}~\bibnamefont {Wang}},\ }\href {\doibase
  10.1146/annurev-conmatphys-031119-050711} {\bibfield  {journal} {\bibinfo
  {journal} {Annu. Rev. Condens. Matter Phys.}\ }\textbf {\bibinfo {volume}
  {11}},\ \bibinfo {pages} {231} (\bibinfo {year} {2020})}\BibitemShut
  {NoStop}%
\bibitem [{\citenamefont {Li}\ \emph {et~al.}(2010)\citenamefont {Li},
  \citenamefont {Luican}, \citenamefont {Lopes~dos Santos}, \citenamefont
  {Castro~Neto}, \citenamefont {Reina}, \citenamefont {Kong},\ and\
  \citenamefont {Andrei}}]{li2010observation}%
  \BibitemOpen
  \bibfield  {author} {\bibinfo {author} {\bibfnamefont {G.}~\bibnamefont
  {Li}}, \bibinfo {author} {\bibfnamefont {A.}~\bibnamefont {Luican}}, \bibinfo
  {author} {\bibfnamefont {J.~M.~B.}\ \bibnamefont {Lopes~dos Santos}},
  \bibinfo {author} {\bibfnamefont {A.~H.}\ \bibnamefont {Castro~Neto}},
  \bibinfo {author} {\bibfnamefont {A.}~\bibnamefont {Reina}}, \bibinfo
  {author} {\bibfnamefont {J.}~\bibnamefont {Kong}}, \ and\ \bibinfo {author}
  {\bibfnamefont {E.~Y.}\ \bibnamefont {Andrei}},\ }\href {\doibase
  10.1038/nphys1463} {\bibfield  {journal} {\bibinfo  {journal} {Nat. Phys.}\
  }\textbf {\bibinfo {volume} {6}},\ \bibinfo {pages} {109} (\bibinfo {year}
  {2010})}\BibitemShut {NoStop}%
\bibitem [{\citenamefont {Yan}\ \emph {et~al.}(2012)\citenamefont {Yan},
  \citenamefont {Liu}, \citenamefont {Dou}, \citenamefont {Meng}, \citenamefont
  {Feng}, \citenamefont {Chu}, \citenamefont {Zhang}, \citenamefont {Liu},
  \citenamefont {Nie},\ and\ \citenamefont {He}}]{PhysRevLett.109.126801}%
  \BibitemOpen
  \bibfield  {author} {\bibinfo {author} {\bibfnamefont {W.}~\bibnamefont
  {Yan}}, \bibinfo {author} {\bibfnamefont {M.}~\bibnamefont {Liu}}, \bibinfo
  {author} {\bibfnamefont {R.-F.}\ \bibnamefont {Dou}}, \bibinfo {author}
  {\bibfnamefont {L.}~\bibnamefont {Meng}}, \bibinfo {author} {\bibfnamefont
  {L.}~\bibnamefont {Feng}}, \bibinfo {author} {\bibfnamefont {Z.-D.}\
  \bibnamefont {Chu}}, \bibinfo {author} {\bibfnamefont {Y.}~\bibnamefont
  {Zhang}}, \bibinfo {author} {\bibfnamefont {Z.}~\bibnamefont {Liu}}, \bibinfo
  {author} {\bibfnamefont {J.-C.}\ \bibnamefont {Nie}}, \ and\ \bibinfo
  {author} {\bibfnamefont {L.}~\bibnamefont {He}},\ }\href {\doibase
  10.1103/PhysRevLett.109.126801} {\bibfield  {journal} {\bibinfo  {journal}
  {Phys. Rev. Lett.}\ }\textbf {\bibinfo {volume} {109}},\ \bibinfo {pages}
  {126801} (\bibinfo {year} {2012})}\BibitemShut {NoStop}%
\bibitem [{\citenamefont {Yin}\ \emph {et~al.}(2016)\citenamefont {Yin},
  \citenamefont {Wang}, \citenamefont {Peng}, \citenamefont {Tan},
  \citenamefont {Liao}, \citenamefont {Lin}, \citenamefont {Sun}, \citenamefont
  {Koh}, \citenamefont {Chen}, \citenamefont {Peng},\ and\ \citenamefont
  {Liu}}]{yin2016selectively}%
  \BibitemOpen
  \bibfield  {author} {\bibinfo {author} {\bibfnamefont {J.}~\bibnamefont
  {Yin}}, \bibinfo {author} {\bibfnamefont {H.}~\bibnamefont {Wang}}, \bibinfo
  {author} {\bibfnamefont {H.}~\bibnamefont {Peng}}, \bibinfo {author}
  {\bibfnamefont {Z.}~\bibnamefont {Tan}}, \bibinfo {author} {\bibfnamefont
  {L.}~\bibnamefont {Liao}}, \bibinfo {author} {\bibfnamefont {L.}~\bibnamefont
  {Lin}}, \bibinfo {author} {\bibfnamefont {X.}~\bibnamefont {Sun}}, \bibinfo
  {author} {\bibfnamefont {A.~L.}\ \bibnamefont {Koh}}, \bibinfo {author}
  {\bibfnamefont {Y.}~\bibnamefont {Chen}}, \bibinfo {author} {\bibfnamefont
  {H.}~\bibnamefont {Peng}}, \ and\ \bibinfo {author} {\bibfnamefont
  {Z.}~\bibnamefont {Liu}},\ }\href {\doibase 10.1038/ncomms10699} {\bibfield
  {journal} {\bibinfo  {journal} {Nat. Commun.}\ }\textbf {\bibinfo {volume}
  {7}},\ \bibinfo {pages} {10699} (\bibinfo {year} {2016})}\BibitemShut
  {NoStop}%
\bibitem [{\citenamefont {Xu}\ \emph {et~al.}(2021{\natexlab{b}})\citenamefont
  {Xu}, \citenamefont {Al~Ezzi}, \citenamefont {Balakrishnan}, \citenamefont
  {Garcia-Ruiz}, \citenamefont {Tsim}, \citenamefont {Mullan}, \citenamefont
  {Barrier}, \citenamefont {Xin}, \citenamefont {Piot} \emph
  {et~al.}}]{xu2021tunable}%
  \BibitemOpen
  \bibfield  {author} {\bibinfo {author} {\bibfnamefont {S.}~\bibnamefont
  {Xu}}, \bibinfo {author} {\bibfnamefont {M.~M.}\ \bibnamefont {Al~Ezzi}},
  \bibinfo {author} {\bibfnamefont {N.}~\bibnamefont {Balakrishnan}}, \bibinfo
  {author} {\bibfnamefont {A.}~\bibnamefont {Garcia-Ruiz}}, \bibinfo {author}
  {\bibfnamefont {B.}~\bibnamefont {Tsim}}, \bibinfo {author} {\bibfnamefont
  {C.}~\bibnamefont {Mullan}}, \bibinfo {author} {\bibfnamefont
  {J.}~\bibnamefont {Barrier}}, \bibinfo {author} {\bibfnamefont
  {N.}~\bibnamefont {Xin}}, \bibinfo {author} {\bibfnamefont {B.~A.}\
  \bibnamefont {Piot}},  \emph {et~al.},\ }\href {\doibase
  10.1038/s41567-021-01172-9} {\bibfield  {journal} {\bibinfo  {journal} {Nat.
  Phys.}\ }\textbf {\bibinfo {volume} {17}},\ \bibinfo {pages} {619} (\bibinfo
  {year} {2021}{\natexlab{b}})}\BibitemShut {NoStop}%
\bibitem [{\citenamefont {Marsiglio}(2020)}]{MARSIGLIO2020168102}%
  \BibitemOpen
  \bibfield  {author} {\bibinfo {author} {\bibfnamefont {F.}~\bibnamefont
  {Marsiglio}},\ }\href {\doibase 10.1016/j.aop.2020.168102} {\bibfield
  {journal} {\bibinfo  {journal} {Ann. Phys.}\ }\textbf {\bibinfo {volume}
  {417}},\ \bibinfo {pages} {168102} (\bibinfo {year} {2020})}\BibitemShut
  {NoStop}%
\bibitem [{\citenamefont {Greene}\ \emph {et~al.}(2020)\citenamefont {Greene},
  \citenamefont {Mandal}, \citenamefont {Poniatowski},\ and\ \citenamefont
  {Sarkar}}]{Greene2020}%
  \BibitemOpen
  \bibfield  {author} {\bibinfo {author} {\bibfnamefont {R.~L.}\ \bibnamefont
  {Greene}}, \bibinfo {author} {\bibfnamefont {P.~R.}\ \bibnamefont {Mandal}},
  \bibinfo {author} {\bibfnamefont {N.~R.}\ \bibnamefont {Poniatowski}}, \ and\
  \bibinfo {author} {\bibfnamefont {T.}~\bibnamefont {Sarkar}},\ }\href
  {\doibase 10.1146/annurev-conmatphys-031119-050558} {\bibfield  {journal}
  {\bibinfo  {journal} {Annu. Rev. Condens. Matter Phys.}\ }\textbf {\bibinfo
  {volume} {11}},\ \bibinfo {pages} {213–229} (\bibinfo {year}
  {2020})}\BibitemShut {NoStop}%
\bibitem [{\citenamefont {Shibauchi}\ \emph {et~al.}(2014)\citenamefont
  {Shibauchi}, \citenamefont {Carrington},\ and\ \citenamefont
  {Matsuda}}]{Shibauchi2014}%
  \BibitemOpen
  \bibfield  {author} {\bibinfo {author} {\bibfnamefont {T.}~\bibnamefont
  {Shibauchi}}, \bibinfo {author} {\bibfnamefont {A.}~\bibnamefont
  {Carrington}}, \ and\ \bibinfo {author} {\bibfnamefont {Y.}~\bibnamefont
  {Matsuda}},\ }\href {\doibase 10.1146/annurev-conmatphys-031113-133921}
  {\bibfield  {journal} {\bibinfo  {journal} {Annu. Rev. Condens. Matter
  Phys.}\ }\textbf {\bibinfo {volume} {5}},\ \bibinfo {pages} {113–135}
  (\bibinfo {year} {2014})}\BibitemShut {NoStop}%
\bibitem [{\citenamefont {Paschen}\ and\ \citenamefont
  {Si}(2020)}]{Paschen2020}%
  \BibitemOpen
  \bibfield  {author} {\bibinfo {author} {\bibfnamefont {S.}~\bibnamefont
  {Paschen}}\ and\ \bibinfo {author} {\bibfnamefont {Q.}~\bibnamefont {Si}},\
  }\href {\doibase 10.1038/s42254-020-00262-6} {\bibfield  {journal} {\bibinfo
  {journal} {Nat. Rev. Phys.}\ }\textbf {\bibinfo {volume} {3}},\ \bibinfo
  {pages} {9–26} (\bibinfo {year} {2020})}\BibitemShut {NoStop}%
\bibitem [{\citenamefont {Jaoui}\ \emph {et~al.}(2022)\citenamefont {Jaoui},
  \citenamefont {Das}, \citenamefont {Di~Battista}, \citenamefont
  {Díez-Mérida}, \citenamefont {Lu}, \citenamefont {Watanabe}, \citenamefont
  {Taniguchi}, \citenamefont {Ishizuka}, \citenamefont {Levitov},\ and\
  \citenamefont {Efetov}}]{Jaoui2022}%
  \BibitemOpen
  \bibfield  {author} {\bibinfo {author} {\bibfnamefont {A.}~\bibnamefont
  {Jaoui}}, \bibinfo {author} {\bibfnamefont {I.}~\bibnamefont {Das}}, \bibinfo
  {author} {\bibfnamefont {G.}~\bibnamefont {Di~Battista}}, \bibinfo {author}
  {\bibfnamefont {J.}~\bibnamefont {Díez-Mérida}}, \bibinfo {author}
  {\bibfnamefont {X.}~\bibnamefont {Lu}}, \bibinfo {author} {\bibfnamefont
  {K.}~\bibnamefont {Watanabe}}, \bibinfo {author} {\bibfnamefont
  {T.}~\bibnamefont {Taniguchi}}, \bibinfo {author} {\bibfnamefont
  {H.}~\bibnamefont {Ishizuka}}, \bibinfo {author} {\bibfnamefont
  {L.}~\bibnamefont {Levitov}}, \ and\ \bibinfo {author} {\bibfnamefont
  {D.~K.}\ \bibnamefont {Efetov}},\ }\href {\doibase
  10.1038/s41567-022-01556-5} {\bibfield  {journal} {\bibinfo  {journal} {Nat.
  Phys.}\ }\textbf {\bibinfo {volume} {18}},\ \bibinfo {pages} {633–638}
  (\bibinfo {year} {2022})}\BibitemShut {NoStop}%
\bibitem [{\citenamefont {Ghiotto}\ \emph {et~al.}(2021)\citenamefont
  {Ghiotto}, \citenamefont {Shih}, \citenamefont {Pereira}, \citenamefont
  {Rhodes}, \citenamefont {Kim}, \citenamefont {Zang}, \citenamefont {Millis},
  \citenamefont {Watanabe}, \citenamefont {Taniguchi}, \citenamefont {Hone},
  \citenamefont {Wang}, \citenamefont {Dean},\ and\ \citenamefont
  {Pasupathy}}]{Ghiotto2021}%
  \BibitemOpen
  \bibfield  {author} {\bibinfo {author} {\bibfnamefont {A.}~\bibnamefont
  {Ghiotto}}, \bibinfo {author} {\bibfnamefont {E.-M.}\ \bibnamefont {Shih}},
  \bibinfo {author} {\bibfnamefont {G.~S. S.~G.}\ \bibnamefont {Pereira}},
  \bibinfo {author} {\bibfnamefont {D.~A.}\ \bibnamefont {Rhodes}}, \bibinfo
  {author} {\bibfnamefont {B.}~\bibnamefont {Kim}}, \bibinfo {author}
  {\bibfnamefont {J.}~\bibnamefont {Zang}}, \bibinfo {author} {\bibfnamefont
  {A.~J.}\ \bibnamefont {Millis}}, \bibinfo {author} {\bibfnamefont
  {K.}~\bibnamefont {Watanabe}}, \bibinfo {author} {\bibfnamefont
  {T.}~\bibnamefont {Taniguchi}}, \bibinfo {author} {\bibfnamefont {J.~C.}\
  \bibnamefont {Hone}}, \bibinfo {author} {\bibfnamefont {L.}~\bibnamefont
  {Wang}}, \bibinfo {author} {\bibfnamefont {C.~R.}\ \bibnamefont {Dean}}, \
  and\ \bibinfo {author} {\bibfnamefont {A.~N.}\ \bibnamefont {Pasupathy}},\
  }\href {\doibase 10.1038/s41586-021-03815-6} {\bibfield  {journal} {\bibinfo
  {journal} {Nature}\ }\textbf {\bibinfo {volume} {597}},\ \bibinfo {pages}
  {345–349} (\bibinfo {year} {2021})}\BibitemShut {NoStop}%
\bibitem [{\citenamefont {Michon}\ \emph {et~al.}(2019)\citenamefont {Michon},
  \citenamefont {Girod}, \citenamefont {Badoux}, \citenamefont {Kačmarčík},
  \citenamefont {Ma}, \citenamefont {Dragomir}, \citenamefont {Dabkowska},
  \citenamefont {Gaulin}, \citenamefont {Zhou}, \citenamefont {Pyon},
  \citenamefont {Takayama}, \citenamefont {Takagi}, \citenamefont {Verret},
  \citenamefont {Doiron-Leyraud}, \citenamefont {Marcenat}, \citenamefont
  {Taillefer},\ and\ \citenamefont {Klein}}]{Michon2019}%
  \BibitemOpen
  \bibfield  {author} {\bibinfo {author} {\bibfnamefont {B.}~\bibnamefont
  {Michon}}, \bibinfo {author} {\bibfnamefont {C.}~\bibnamefont {Girod}},
  \bibinfo {author} {\bibfnamefont {S.}~\bibnamefont {Badoux}}, \bibinfo
  {author} {\bibfnamefont {J.}~\bibnamefont {Kačmarčík}}, \bibinfo {author}
  {\bibfnamefont {Q.}~\bibnamefont {Ma}}, \bibinfo {author} {\bibfnamefont
  {M.}~\bibnamefont {Dragomir}}, \bibinfo {author} {\bibfnamefont {H.~A.}\
  \bibnamefont {Dabkowska}}, \bibinfo {author} {\bibfnamefont {B.~D.}\
  \bibnamefont {Gaulin}}, \bibinfo {author} {\bibfnamefont {J.-S.}\
  \bibnamefont {Zhou}}, \bibinfo {author} {\bibfnamefont {S.}~\bibnamefont
  {Pyon}}, \bibinfo {author} {\bibfnamefont {T.}~\bibnamefont {Takayama}},
  \bibinfo {author} {\bibfnamefont {H.}~\bibnamefont {Takagi}}, \bibinfo
  {author} {\bibfnamefont {S.}~\bibnamefont {Verret}}, \bibinfo {author}
  {\bibfnamefont {N.}~\bibnamefont {Doiron-Leyraud}}, \bibinfo {author}
  {\bibfnamefont {C.}~\bibnamefont {Marcenat}}, \bibinfo {author}
  {\bibfnamefont {L.}~\bibnamefont {Taillefer}}, \ and\ \bibinfo {author}
  {\bibfnamefont {T.}~\bibnamefont {Klein}},\ }\href {\doibase
  10.1038/s41586-019-0932-x} {\bibfield  {journal} {\bibinfo  {journal}
  {Nature}\ }\textbf {\bibinfo {volume} {567}},\ \bibinfo {pages} {218–222}
  (\bibinfo {year} {2019})}\BibitemShut {NoStop}%
\bibitem [{\citenamefont {Grissonnanche}\ \emph {et~al.}(2021)\citenamefont
  {Grissonnanche}, \citenamefont {Fang}, \citenamefont {Legros}, \citenamefont
  {Verret}, \citenamefont {Laliberté}, \citenamefont {Collignon},
  \citenamefont {Zhou}, \citenamefont {Graf}, \citenamefont {Goddard},
  \citenamefont {Taillefer},\ and\ \citenamefont
  {Ramshaw}}]{Grissonnanche2021}%
  \BibitemOpen
  \bibfield  {author} {\bibinfo {author} {\bibfnamefont {G.}~\bibnamefont
  {Grissonnanche}}, \bibinfo {author} {\bibfnamefont {Y.}~\bibnamefont {Fang}},
  \bibinfo {author} {\bibfnamefont {A.}~\bibnamefont {Legros}}, \bibinfo
  {author} {\bibfnamefont {S.}~\bibnamefont {Verret}}, \bibinfo {author}
  {\bibfnamefont {F.}~\bibnamefont {Laliberté}}, \bibinfo {author}
  {\bibfnamefont {C.}~\bibnamefont {Collignon}}, \bibinfo {author}
  {\bibfnamefont {J.}~\bibnamefont {Zhou}}, \bibinfo {author} {\bibfnamefont
  {D.}~\bibnamefont {Graf}}, \bibinfo {author} {\bibfnamefont {P.~A.}\
  \bibnamefont {Goddard}}, \bibinfo {author} {\bibfnamefont {L.}~\bibnamefont
  {Taillefer}}, \ and\ \bibinfo {author} {\bibfnamefont {B.~J.}\ \bibnamefont
  {Ramshaw}},\ }\href {\doibase 10.1038/s41586-021-03697-8} {\bibfield
  {journal} {\bibinfo  {journal} {Nature}\ }\textbf {\bibinfo {volume} {595}},\
  \bibinfo {pages} {667–672} (\bibinfo {year} {2021})}\BibitemShut {NoStop}%
\bibitem [{\citenamefont {Zhu}\ \emph {et~al.}(2022)\citenamefont {Zhu},
  \citenamefont {Voneshen}, \citenamefont {Raymond}, \citenamefont {Lipscombe},
  \citenamefont {Tam},\ and\ \citenamefont {Hayden}}]{Zhu2022}%
  \BibitemOpen
  \bibfield  {author} {\bibinfo {author} {\bibfnamefont {M.}~\bibnamefont
  {Zhu}}, \bibinfo {author} {\bibfnamefont {D.~J.}\ \bibnamefont {Voneshen}},
  \bibinfo {author} {\bibfnamefont {S.}~\bibnamefont {Raymond}}, \bibinfo
  {author} {\bibfnamefont {O.~J.}\ \bibnamefont {Lipscombe}}, \bibinfo {author}
  {\bibfnamefont {C.~C.}\ \bibnamefont {Tam}}, \ and\ \bibinfo {author}
  {\bibfnamefont {S.~M.}\ \bibnamefont {Hayden}},\ }\href {\doibase
  10.1038/s41567-022-01825-3} {\bibfield  {journal} {\bibinfo  {journal} {Nat.
  Phys.}\ }\textbf {\bibinfo {volume} {19}},\ \bibinfo {pages} {99–105}
  (\bibinfo {year} {2022})}\BibitemShut {NoStop}%
\bibitem [{\citenamefont {Benhabib}\ \emph {et~al.}(2015)\citenamefont
  {Benhabib}, \citenamefont {Sacuto}, \citenamefont {Civelli}, \citenamefont
  {Paul}, \citenamefont {Cazayous}, \citenamefont {Gallais}, \citenamefont
  {M\'easson}, \citenamefont {Zhong}, \citenamefont {Schneeloch}, \citenamefont
  {Gu}, \citenamefont {Colson},\ and\ \citenamefont {Forget}}]{Benhabib2015}%
  \BibitemOpen
  \bibfield  {author} {\bibinfo {author} {\bibfnamefont {S.}~\bibnamefont
  {Benhabib}}, \bibinfo {author} {\bibfnamefont {A.}~\bibnamefont {Sacuto}},
  \bibinfo {author} {\bibfnamefont {M.}~\bibnamefont {Civelli}}, \bibinfo
  {author} {\bibfnamefont {I.}~\bibnamefont {Paul}}, \bibinfo {author}
  {\bibfnamefont {M.}~\bibnamefont {Cazayous}}, \bibinfo {author}
  {\bibfnamefont {Y.}~\bibnamefont {Gallais}}, \bibinfo {author} {\bibfnamefont
  {M.-A.}\ \bibnamefont {M\'easson}}, \bibinfo {author} {\bibfnamefont {R.~D.}\
  \bibnamefont {Zhong}}, \bibinfo {author} {\bibfnamefont {J.}~\bibnamefont
  {Schneeloch}}, \bibinfo {author} {\bibfnamefont {G.~D.}\ \bibnamefont {Gu}},
  \bibinfo {author} {\bibfnamefont {D.}~\bibnamefont {Colson}}, \ and\ \bibinfo
  {author} {\bibfnamefont {A.}~\bibnamefont {Forget}},\ }\href {\doibase
  10.1103/PhysRevLett.114.147001} {\bibfield  {journal} {\bibinfo  {journal}
  {Phys. Rev. Lett.}\ }\textbf {\bibinfo {volume} {114}},\ \bibinfo {pages}
  {147001} (\bibinfo {year} {2015})}\BibitemShut {NoStop}%
\bibitem [{\citenamefont {Volovik}(2017)}]{Volovik2017}%
  \BibitemOpen
  \bibfield  {author} {\bibinfo {author} {\bibfnamefont {G.~E.}\ \bibnamefont
  {Volovik}},\ }\href {\doibase 10.1063/1.4974185} {\bibfield  {journal}
  {\bibinfo  {journal} {Low Temp. Phys.}\ }\textbf {\bibinfo {volume} {43}},\
  \bibinfo {pages} {47–55} (\bibinfo {year} {2017})}\BibitemShut {NoStop}%
\bibitem [{\citenamefont {Doiron-Leyraud}\ \emph {et~al.}(2017)\citenamefont
  {Doiron-Leyraud}, \citenamefont {Cyr-Choinière}, \citenamefont {Badoux},
  \citenamefont {Ataei}, \citenamefont {Collignon}, \citenamefont {Gourgout},
  \citenamefont {Dufour-Beauséjour}, \citenamefont {Tafti}, \citenamefont
  {Laliberté}, \citenamefont {Boulanger}, \citenamefont {Matusiak},
  \citenamefont {Graf}, \citenamefont {Kim}, \citenamefont {Zhou},
  \citenamefont {Momono}, \citenamefont {Kurosawa}, \citenamefont {Takagi},\
  and\ \citenamefont {Taillefer}}]{DoironLeyraud2017}%
  \BibitemOpen
  \bibfield  {author} {\bibinfo {author} {\bibfnamefont {N.}~\bibnamefont
  {Doiron-Leyraud}}, \bibinfo {author} {\bibfnamefont {O.}~\bibnamefont
  {Cyr-Choinière}}, \bibinfo {author} {\bibfnamefont {S.}~\bibnamefont
  {Badoux}}, \bibinfo {author} {\bibfnamefont {A.}~\bibnamefont {Ataei}},
  \bibinfo {author} {\bibfnamefont {C.}~\bibnamefont {Collignon}}, \bibinfo
  {author} {\bibfnamefont {A.}~\bibnamefont {Gourgout}}, \bibinfo {author}
  {\bibfnamefont {S.}~\bibnamefont {Dufour-Beauséjour}}, \bibinfo {author}
  {\bibfnamefont {F.~F.}\ \bibnamefont {Tafti}}, \bibinfo {author}
  {\bibfnamefont {F.}~\bibnamefont {Laliberté}}, \bibinfo {author}
  {\bibfnamefont {M.-E.}\ \bibnamefont {Boulanger}}, \bibinfo {author}
  {\bibfnamefont {M.}~\bibnamefont {Matusiak}}, \bibinfo {author}
  {\bibfnamefont {D.}~\bibnamefont {Graf}}, \bibinfo {author} {\bibfnamefont
  {M.}~\bibnamefont {Kim}}, \bibinfo {author} {\bibfnamefont {J.-S.}\
  \bibnamefont {Zhou}}, \bibinfo {author} {\bibfnamefont {N.}~\bibnamefont
  {Momono}}, \bibinfo {author} {\bibfnamefont {T.}~\bibnamefont {Kurosawa}},
  \bibinfo {author} {\bibfnamefont {H.}~\bibnamefont {Takagi}}, \ and\ \bibinfo
  {author} {\bibfnamefont {L.}~\bibnamefont {Taillefer}},\ }\href {\doibase
  10.1038/s41467-017-02122-x} {\bibfield  {journal} {\bibinfo  {journal} {Nat.
  Commun.}\ }\textbf {\bibinfo {volume} {8}} (\bibinfo {year} {2017}),\
  10.1038/s41467-017-02122-x}\BibitemShut {NoStop}%
\bibitem [{\citenamefont {Wu}\ \emph {et~al.}(2018)\citenamefont {Wu},
  \citenamefont {Scheurer}, \citenamefont {Chatterjee}, \citenamefont
  {Sachdev}, \citenamefont {Georges},\ and\ \citenamefont
  {Ferrero}}]{PhysRevX.8.021048}%
  \BibitemOpen
  \bibfield  {author} {\bibinfo {author} {\bibfnamefont {W.}~\bibnamefont
  {Wu}}, \bibinfo {author} {\bibfnamefont {M.~S.}\ \bibnamefont {Scheurer}},
  \bibinfo {author} {\bibfnamefont {S.}~\bibnamefont {Chatterjee}}, \bibinfo
  {author} {\bibfnamefont {S.}~\bibnamefont {Sachdev}}, \bibinfo {author}
  {\bibfnamefont {A.}~\bibnamefont {Georges}}, \ and\ \bibinfo {author}
  {\bibfnamefont {M.}~\bibnamefont {Ferrero}},\ }\href {\doibase
  10.1103/PhysRevX.8.021048} {\bibfield  {journal} {\bibinfo  {journal} {Phys.
  Rev. X}\ }\textbf {\bibinfo {volume} {8}},\ \bibinfo {pages} {021048}
  (\bibinfo {year} {2018})}\BibitemShut {NoStop}%
\bibitem [{\citenamefont {Kim}\ \emph {et~al.}(2016)\citenamefont {Kim},
  \citenamefont {Herlinger}, \citenamefont {Moon}, \citenamefont {Koshino},
  \citenamefont {Taniguchi}, \citenamefont {Watanabe},\ and\ \citenamefont
  {Smet}}]{Kim2016}%
  \BibitemOpen
  \bibfield  {author} {\bibinfo {author} {\bibfnamefont {Y.}~\bibnamefont
  {Kim}}, \bibinfo {author} {\bibfnamefont {P.}~\bibnamefont {Herlinger}},
  \bibinfo {author} {\bibfnamefont {P.}~\bibnamefont {Moon}}, \bibinfo {author}
  {\bibfnamefont {M.}~\bibnamefont {Koshino}}, \bibinfo {author} {\bibfnamefont
  {T.}~\bibnamefont {Taniguchi}}, \bibinfo {author} {\bibfnamefont
  {K.}~\bibnamefont {Watanabe}}, \ and\ \bibinfo {author} {\bibfnamefont
  {J.~H.}\ \bibnamefont {Smet}},\ }\href {\doibase
  10.1021/acs.nanolett.6b01906} {\bibfield  {journal} {\bibinfo  {journal}
  {Nano Lett.}\ }\textbf {\bibinfo {volume} {16}},\ \bibinfo {pages}
  {5053–5059} (\bibinfo {year} {2016})}\BibitemShut {NoStop}%
\bibitem [{\citenamefont {Lee}(2021)}]{PALee2021}%
  \BibitemOpen
  \bibfield  {author} {\bibinfo {author} {\bibfnamefont {P.~A.}\ \bibnamefont
  {Lee}},\ }\href {\doibase 10.1103/PhysRevB.104.035140} {\bibfield  {journal}
  {\bibinfo  {journal} {Phys. Rev. B}\ }\textbf {\bibinfo {volume} {104}},\
  \bibinfo {pages} {035140} (\bibinfo {year} {2021})}\BibitemShut {NoStop}%
\bibitem [{\citenamefont {Ma}\ \emph {et~al.}(2023)\citenamefont {Ma},
  \citenamefont {Zeng}, \citenamefont {Cao},\ and\ \citenamefont
  {Feng}}]{PhysRevB.108.134502}%
  \BibitemOpen
  \bibfield  {author} {\bibinfo {author} {\bibfnamefont {X.}~\bibnamefont
  {Ma}}, \bibinfo {author} {\bibfnamefont {M.}~\bibnamefont {Zeng}}, \bibinfo
  {author} {\bibfnamefont {Z.}~\bibnamefont {Cao}}, \ and\ \bibinfo {author}
  {\bibfnamefont {S.}~\bibnamefont {Feng}},\ }\href {\doibase
  10.1103/PhysRevB.108.134502} {\bibfield  {journal} {\bibinfo  {journal}
  {Phys. Rev. B}\ }\textbf {\bibinfo {volume} {108}},\ \bibinfo {pages}
  {134502} (\bibinfo {year} {2023})}\BibitemShut {NoStop}%
\bibitem [{\citenamefont {Ma}\ \emph {et~al.}(2024)\citenamefont {Ma},
  \citenamefont {Zeng}, \citenamefont {Guo},\ and\ \citenamefont
  {Feng}}]{PhysRevB.110.094520}%
  \BibitemOpen
  \bibfield  {author} {\bibinfo {author} {\bibfnamefont {X.}~\bibnamefont
  {Ma}}, \bibinfo {author} {\bibfnamefont {M.}~\bibnamefont {Zeng}}, \bibinfo
  {author} {\bibfnamefont {H.}~\bibnamefont {Guo}}, \ and\ \bibinfo {author}
  {\bibfnamefont {S.}~\bibnamefont {Feng}},\ }\href {\doibase
  10.1103/PhysRevB.110.094520} {\bibfield  {journal} {\bibinfo  {journal}
  {Phys. Rev. B}\ }\textbf {\bibinfo {volume} {110}},\ \bibinfo {pages}
  {094520} (\bibinfo {year} {2024})}\BibitemShut {NoStop}%
\bibitem [{\citenamefont {Esterlis}\ and\ \citenamefont
  {Schmalian}(2019)}]{Esterlis2019}%
  \BibitemOpen
  \bibfield  {author} {\bibinfo {author} {\bibfnamefont {I.}~\bibnamefont
  {Esterlis}}\ and\ \bibinfo {author} {\bibfnamefont {J.}~\bibnamefont
  {Schmalian}},\ }\href {\doibase 10.1103/PhysRevB.100.115132} {\bibfield
  {journal} {\bibinfo  {journal} {Phys. Rev. B}\ }\textbf {\bibinfo {volume}
  {100}},\ \bibinfo {pages} {115132} (\bibinfo {year} {2019})}\BibitemShut
  {NoStop}%
\bibitem [{\citenamefont {Auvray}\ \emph {et~al.}(2019)\citenamefont {Auvray},
  \citenamefont {Loret}, \citenamefont {Benhabib}, \citenamefont {Cazayous},
  \citenamefont {Zhong}, \citenamefont {Schneeloch}, \citenamefont {Gu},
  \citenamefont {Forget}, \citenamefont {Colson}, \citenamefont {Paul},
  \citenamefont {Sacuto},\ and\ \citenamefont {Gallais}}]{Auvray2019}%
  \BibitemOpen
  \bibfield  {author} {\bibinfo {author} {\bibfnamefont {N.}~\bibnamefont
  {Auvray}}, \bibinfo {author} {\bibfnamefont {B.}~\bibnamefont {Loret}},
  \bibinfo {author} {\bibfnamefont {S.}~\bibnamefont {Benhabib}}, \bibinfo
  {author} {\bibfnamefont {M.}~\bibnamefont {Cazayous}}, \bibinfo {author}
  {\bibfnamefont {R.~D.}\ \bibnamefont {Zhong}}, \bibinfo {author}
  {\bibfnamefont {J.}~\bibnamefont {Schneeloch}}, \bibinfo {author}
  {\bibfnamefont {G.~D.}\ \bibnamefont {Gu}}, \bibinfo {author} {\bibfnamefont
  {A.}~\bibnamefont {Forget}}, \bibinfo {author} {\bibfnamefont
  {D.}~\bibnamefont {Colson}}, \bibinfo {author} {\bibfnamefont
  {I.}~\bibnamefont {Paul}}, \bibinfo {author} {\bibfnamefont {A.}~\bibnamefont
  {Sacuto}}, \ and\ \bibinfo {author} {\bibfnamefont {Y.}~\bibnamefont
  {Gallais}},\ }\href {\doibase 10.1038/s41467-019-12940-w} {\bibfield
  {journal} {\bibinfo  {journal} {Nat. Commun.}\ }\textbf {\bibinfo {volume}
  {10}},\ \bibinfo {pages} {5209} (\bibinfo {year} {2019})}\BibitemShut
  {NoStop}%
\bibitem [{\citenamefont {Metlitski}\ and\ \citenamefont
  {Sachdev}(2010)}]{Metlitski2010}%
  \BibitemOpen
  \bibfield  {author} {\bibinfo {author} {\bibfnamefont {M.~A.}\ \bibnamefont
  {Metlitski}}\ and\ \bibinfo {author} {\bibfnamefont {S.}~\bibnamefont
  {Sachdev}},\ }\href {\doibase 10.1103/PhysRevB.82.075127} {\bibfield
  {journal} {\bibinfo  {journal} {Phys. Rev. B}\ }\textbf {\bibinfo {volume}
  {82}},\ \bibinfo {pages} {075127} (\bibinfo {year} {2010})}\BibitemShut
  {NoStop}%
\bibitem [{\citenamefont {Zhong}\ \emph {et~al.}(2022)\citenamefont {Zhong},
  \citenamefont {Chen}, \citenamefont {Chen}, \citenamefont {Xu}, \citenamefont
  {Hashimoto}, \citenamefont {He}, \citenamefont {ichi Uchida}, \citenamefont
  {Lu}, \citenamefont {Mo},\ and\ \citenamefont {Shen}}]{Shen2022}%
  \BibitemOpen
  \bibfield  {author} {\bibinfo {author} {\bibfnamefont {Y.}~\bibnamefont
  {Zhong}}, \bibinfo {author} {\bibfnamefont {Z.}~\bibnamefont {Chen}},
  \bibinfo {author} {\bibfnamefont {S.-D.}\ \bibnamefont {Chen}}, \bibinfo
  {author} {\bibfnamefont {K.-J.}\ \bibnamefont {Xu}}, \bibinfo {author}
  {\bibfnamefont {M.}~\bibnamefont {Hashimoto}}, \bibinfo {author}
  {\bibfnamefont {Y.}~\bibnamefont {He}}, \bibinfo {author} {\bibfnamefont
  {S.}~\bibnamefont {ichi Uchida}}, \bibinfo {author} {\bibfnamefont
  {D.}~\bibnamefont {Lu}}, \bibinfo {author} {\bibfnamefont {S.-K.}\
  \bibnamefont {Mo}}, \ and\ \bibinfo {author} {\bibfnamefont {Z.-X.}\
  \bibnamefont {Shen}},\ }\href {\doibase 10.1073/pnas.2204630119} {\bibfield
  {journal} {\bibinfo  {journal} {Proc. Natl. Acad. Sci.}\ }\textbf {\bibinfo
  {volume} {119}},\ \bibinfo {pages} {e2204630119} (\bibinfo {year}
  {2022})}\BibitemShut {NoStop}%
\bibitem [{\citenamefont {Maslov}\ and\ \citenamefont
  {Chubukov}(2016)}]{Maslov2016}%
  \BibitemOpen
  \bibfield  {author} {\bibinfo {author} {\bibfnamefont {D.~L.}\ \bibnamefont
  {Maslov}}\ and\ \bibinfo {author} {\bibfnamefont {A.~V.}\ \bibnamefont
  {Chubukov}},\ }\href {\doibase 10.1088/1361-6633/80/2/026503} {\bibfield
  {journal} {\bibinfo  {journal} {Rep. Prog. Phys.}\ }\textbf {\bibinfo
  {volume} {80}},\ \bibinfo {pages} {026503} (\bibinfo {year}
  {2016})}\BibitemShut {NoStop}%
\bibitem [{\citenamefont {Kim}\ \emph {et~al.}(1994)\citenamefont {Kim},
  \citenamefont {Furusaki}, \citenamefont {Wen},\ and\ \citenamefont
  {Lee}}]{YBKim1994}%
  \BibitemOpen
  \bibfield  {author} {\bibinfo {author} {\bibfnamefont {Y.~B.}\ \bibnamefont
  {Kim}}, \bibinfo {author} {\bibfnamefont {A.}~\bibnamefont {Furusaki}},
  \bibinfo {author} {\bibfnamefont {X.-G.}\ \bibnamefont {Wen}}, \ and\
  \bibinfo {author} {\bibfnamefont {P.~A.}\ \bibnamefont {Lee}},\ }\href
  {\doibase 10.1103/PhysRevB.50.17917} {\bibfield  {journal} {\bibinfo
  {journal} {Phys. Rev. B}\ }\textbf {\bibinfo {volume} {50}},\ \bibinfo
  {pages} {17917} (\bibinfo {year} {1994})}\BibitemShut {NoStop}%
\bibitem [{\citenamefont {Maslov}\ \emph {et~al.}(2011)\citenamefont {Maslov},
  \citenamefont {Yudson},\ and\ \citenamefont {Chubukov}}]{Chubukov2011}%
  \BibitemOpen
  \bibfield  {author} {\bibinfo {author} {\bibfnamefont {D.~L.}\ \bibnamefont
  {Maslov}}, \bibinfo {author} {\bibfnamefont {V.~I.}\ \bibnamefont {Yudson}},
  \ and\ \bibinfo {author} {\bibfnamefont {A.~V.}\ \bibnamefont {Chubukov}},\
  }\href {\doibase 10.1103/PhysRevLett.106.106403} {\bibfield  {journal}
  {\bibinfo  {journal} {Phys. Rev. Lett.}\ }\textbf {\bibinfo {volume} {106}},\
  \bibinfo {pages} {106403} (\bibinfo {year} {2011})}\BibitemShut {NoStop}%
\bibitem [{\citenamefont {Li}\ \emph {et~al.}(2023)\citenamefont {Li},
  \citenamefont {Sharma}, \citenamefont {Levchenko},\ and\ \citenamefont
  {Maslov}}]{Songci2023}%
  \BibitemOpen
  \bibfield  {author} {\bibinfo {author} {\bibfnamefont {S.}~\bibnamefont
  {Li}}, \bibinfo {author} {\bibfnamefont {P.}~\bibnamefont {Sharma}}, \bibinfo
  {author} {\bibfnamefont {A.}~\bibnamefont {Levchenko}}, \ and\ \bibinfo
  {author} {\bibfnamefont {D.~L.}\ \bibnamefont {Maslov}},\ }\href {\doibase
  10.1103/PhysRevB.108.235125} {\bibfield  {journal} {\bibinfo  {journal}
  {Phys. Rev. B}\ }\textbf {\bibinfo {volume} {108}},\ \bibinfo {pages}
  {235125} (\bibinfo {year} {2023})}\BibitemShut {NoStop}%
\bibitem [{\citenamefont {Hlubina}\ and\ \citenamefont
  {Rice}(1995)}]{Hlubina1995}%
  \BibitemOpen
  \bibfield  {author} {\bibinfo {author} {\bibfnamefont {R.}~\bibnamefont
  {Hlubina}}\ and\ \bibinfo {author} {\bibfnamefont {T.~M.}\ \bibnamefont
  {Rice}},\ }\href {\doibase 10.1103/PhysRevB.51.9253} {\bibfield  {journal}
  {\bibinfo  {journal} {Phys. Rev. B}\ }\textbf {\bibinfo {volume} {51}},\
  \bibinfo {pages} {9253} (\bibinfo {year} {1995})}\BibitemShut {NoStop}%
\bibitem [{\citenamefont {Rosch}(1999)}]{Rosch1999}%
  \BibitemOpen
  \bibfield  {author} {\bibinfo {author} {\bibfnamefont {A.}~\bibnamefont
  {Rosch}},\ }\href {\doibase 10.1103/PhysRevLett.82.4280} {\bibfield
  {journal} {\bibinfo  {journal} {Phys. Rev. Lett.}\ }\textbf {\bibinfo
  {volume} {82}},\ \bibinfo {pages} {4280} (\bibinfo {year}
  {1999})}\BibitemShut {NoStop}%
\bibitem [{\citenamefont {Lee}\ and\ \citenamefont
  {Nagaosa}(1992)}]{LeeNagaosa1992}%
  \BibitemOpen
  \bibfield  {author} {\bibinfo {author} {\bibfnamefont {P.~A.}\ \bibnamefont
  {Lee}}\ and\ \bibinfo {author} {\bibfnamefont {N.}~\bibnamefont {Nagaosa}},\
  }\href {\doibase 10.1103/PhysRevB.46.5621} {\bibfield  {journal} {\bibinfo
  {journal} {Phys. Rev. B}\ }\textbf {\bibinfo {volume} {46}},\ \bibinfo
  {pages} {5621} (\bibinfo {year} {1992})}\BibitemShut {NoStop}%
\bibitem [{\citenamefont {Guo}(2023)}]{Haoyu2023}%
  \BibitemOpen
  \bibfield  {author} {\bibinfo {author} {\bibfnamefont {H.}~\bibnamefont
  {Guo}},\ }\href@noop {} {\enquote {\bibinfo {title} {{Is the
  Migdal-Eliashberg Theory for 2+1D Critical Fermi Surface Stable?}}}\ }
  (\bibinfo {year} {2023}),\ \Eprint {http://arxiv.org/abs/2311.03455}
  {arXiv:2311.03455 [cond-mat.str-el]} \BibitemShut {NoStop}%
\bibitem [{\citenamefont {Guo}(2024)}]{Haoyu2024}%
  \BibitemOpen
  \bibfield  {author} {\bibinfo {author} {\bibfnamefont {H.}~\bibnamefont
  {Guo}},\ }\href {\doibase 10.1103/PhysRevB.110.155130} {\bibfield  {journal}
  {\bibinfo  {journal} {Phys. Rev. B}\ }\textbf {\bibinfo {volume} {110}},\
  \bibinfo {pages} {155130} (\bibinfo {year} {2024})}\BibitemShut {NoStop}%
\bibitem [{\citenamefont {Legros}\ \emph {et~al.}(2022)\citenamefont {Legros},
  \citenamefont {Post}, \citenamefont {Chauhan}, \citenamefont {Rickel},
  \citenamefont {He}, \citenamefont {Xu}, \citenamefont {Shi}, \citenamefont
  {Bo\ifmmode \check{z}\else \v{z}\fi{}ovi\ifmmode~\acute{c}\else \'{c}\fi{}},
  \citenamefont {Crooker},\ and\ \citenamefont {Armitage}}]{Legros2022}%
  \BibitemOpen
  \bibfield  {author} {\bibinfo {author} {\bibfnamefont {A.}~\bibnamefont
  {Legros}}, \bibinfo {author} {\bibfnamefont {K.~W.}\ \bibnamefont {Post}},
  \bibinfo {author} {\bibfnamefont {P.}~\bibnamefont {Chauhan}}, \bibinfo
  {author} {\bibfnamefont {D.~G.}\ \bibnamefont {Rickel}}, \bibinfo {author}
  {\bibfnamefont {X.}~\bibnamefont {He}}, \bibinfo {author} {\bibfnamefont
  {X.}~\bibnamefont {Xu}}, \bibinfo {author} {\bibfnamefont {X.}~\bibnamefont
  {Shi}}, \bibinfo {author} {\bibfnamefont {I.}~\bibnamefont {Bo\ifmmode
  \check{z}\else \v{z}\fi{}ovi\ifmmode~\acute{c}\else \'{c}\fi{}}}, \bibinfo
  {author} {\bibfnamefont {S.~A.}\ \bibnamefont {Crooker}}, \ and\ \bibinfo
  {author} {\bibfnamefont {N.~P.}\ \bibnamefont {Armitage}},\ }\href {\doibase
  10.1103/PhysRevB.106.195110} {\bibfield  {journal} {\bibinfo  {journal}
  {Phys. Rev. B}\ }\textbf {\bibinfo {volume} {106}},\ \bibinfo {pages}
  {195110} (\bibinfo {year} {2022})}\BibitemShut {NoStop}%
\bibitem [{\citenamefont {Guo}\ \emph {et~al.}(2024)\citenamefont {Guo},
  \citenamefont {Valentinis}, \citenamefont {Schmalian}, \citenamefont
  {Sachdev},\ and\ \citenamefont {Patel}}]{Guo2024}%
  \BibitemOpen
  \bibfield  {author} {\bibinfo {author} {\bibfnamefont {H.}~\bibnamefont
  {Guo}}, \bibinfo {author} {\bibfnamefont {D.}~\bibnamefont {Valentinis}},
  \bibinfo {author} {\bibfnamefont {J.}~\bibnamefont {Schmalian}}, \bibinfo
  {author} {\bibfnamefont {S.}~\bibnamefont {Sachdev}}, \ and\ \bibinfo
  {author} {\bibfnamefont {A.~A.}\ \bibnamefont {Patel}},\ }\href {\doibase
  10.1103/PhysRevB.109.075162} {\bibfield  {journal} {\bibinfo  {journal}
  {Phys. Rev. B}\ }\textbf {\bibinfo {volume} {109}},\ \bibinfo {pages}
  {075162} (\bibinfo {year} {2024})}\BibitemShut {NoStop}%
\bibitem [{\citenamefont {Irkhin}\ \emph {et~al.}(2001)\citenamefont {Irkhin},
  \citenamefont {Katanin},\ and\ \citenamefont
  {Katsnelson}}]{PhysRevB.64.165107}%
  \BibitemOpen
  \bibfield  {author} {\bibinfo {author} {\bibfnamefont {V.~Y.}\ \bibnamefont
  {Irkhin}}, \bibinfo {author} {\bibfnamefont {A.~A.}\ \bibnamefont {Katanin}},
  \ and\ \bibinfo {author} {\bibfnamefont {M.~I.}\ \bibnamefont {Katsnelson}},\
  }\href {\doibase 10.1103/PhysRevB.64.165107} {\bibfield  {journal} {\bibinfo
  {journal} {Phys. Rev. B}\ }\textbf {\bibinfo {volume} {64}},\ \bibinfo
  {pages} {165107} (\bibinfo {year} {2001})}\BibitemShut {NoStop}%
\bibitem [{\citenamefont {Gindikin}\ and\ \citenamefont
  {Chubukov}(2024)}]{Chubukov2024}%
  \BibitemOpen
  \bibfield  {author} {\bibinfo {author} {\bibfnamefont {Y.}~\bibnamefont
  {Gindikin}}\ and\ \bibinfo {author} {\bibfnamefont {A.~V.}\ \bibnamefont
  {Chubukov}},\ }\href {\doibase 10.1103/PhysRevB.109.115156} {\bibfield
  {journal} {\bibinfo  {journal} {Phys. Rev. B}\ }\textbf {\bibinfo {volume}
  {109}},\ \bibinfo {pages} {115156} (\bibinfo {year} {2024})}\BibitemShut
  {NoStop}%
\bibitem [{\citenamefont {Prange}\ and\ \citenamefont
  {Kadanoff}(1964)}]{PhysRev.134.A566}%
  \BibitemOpen
  \bibfield  {author} {\bibinfo {author} {\bibfnamefont {R.~E.}\ \bibnamefont
  {Prange}}\ and\ \bibinfo {author} {\bibfnamefont {L.~P.}\ \bibnamefont
  {Kadanoff}},\ }\href {\doibase 10.1103/PhysRev.134.A566} {\bibfield
  {journal} {\bibinfo  {journal} {Phys. Rev.}\ }\textbf {\bibinfo {volume}
  {134}},\ \bibinfo {pages} {A566} (\bibinfo {year} {1964})}\BibitemShut
  {NoStop}%
\bibitem [{\citenamefont {Hartnoll}\ \emph {et~al.}(2014)\citenamefont
  {Hartnoll}, \citenamefont {Mahajan}, \citenamefont {Punk},\ and\
  \citenamefont {Sachdev}}]{Hartnoll2014}%
  \BibitemOpen
  \bibfield  {author} {\bibinfo {author} {\bibfnamefont {S.~A.}\ \bibnamefont
  {Hartnoll}}, \bibinfo {author} {\bibfnamefont {R.}~\bibnamefont {Mahajan}},
  \bibinfo {author} {\bibfnamefont {M.}~\bibnamefont {Punk}}, \ and\ \bibinfo
  {author} {\bibfnamefont {S.}~\bibnamefont {Sachdev}},\ }\href {\doibase
  10.1103/PhysRevB.89.155130} {\bibfield  {journal} {\bibinfo  {journal} {Phys.
  Rev. B}\ }\textbf {\bibinfo {volume} {89}},\ \bibinfo {pages} {155130}
  (\bibinfo {year} {2014})}\BibitemShut {NoStop}%
\end{thebibliography}%

\begin{appendix}

\section{A. Self-consistent solutions of the Eliashberg-type equations}
For the case $J\ne 0, J^\prime = 0$ and $w=0$,
the NFL boson and fermion self-energy corrections are derived as [see Supplemental Material Sec.~1.A.1 for details]
\begin{equation}
	\begin{aligned}
		&\Pi_J(i\Omega_m,\bm{q})\simeq  -\tilde{J}^2 \frac{|\Omega_m|}{|\epsilon_{\bm q}|},\\
		&\Sigma_J(i\omega_n,{\bm k}_v)\simeq -i|J|\sqrt{\frac{N'}{N}}\mathrm{sgn}(\omega_n)|\omega_n|^{1/2}.
	\end{aligned}
	\label{sef_J}
\end{equation}
Note that the clean system yields the same fermion self-energy as Eq.~(\ref{fesecl})
and the boson self-energy is consistent with Eq.~(\ref{bo-se-J}) with a modest impurity scattering rate $\Gamma\ll |\epsilon_{\bm q}|$.
The one-loop results in Eq.(\ref{sef_J}) is in fact self-consistent in the sense that they satisfy the Migdal-Eliashberg equations.
This is verified by substituting the renormalized NFL Green function 
$G^{-1}(i\omega_n,{\bm k})\simeq  -\epsilon_{\bm k}-\Sigma_{J}(i\omega_n,{\bm k})$ back to evaluate the polarization bubble.
We end up with the same exponent $1/2$ for the fermion self-energy near the VHS as illustrated in Fig.~\ref{fig2}(b,c).

Conversely, for $J=0, J^\prime \ne 0$ and $w=0$, we obtain the well-known MFL with a Ohmic damped boson
\begin{equation}
	\begin{aligned}
		&\Pi_{J^\prime}(i\Omega_m,\bm{q})\simeq  - \tilde{J}'^2 \Lambda_\theta^2|\Omega_m|,\\
		&\Sigma_{J^\prime}(i\omega_n,{\bm k})\simeq -i|J'|^2\Lambda_\theta \omega_n\ln\frac{\Lambda_{\rm U}^2}{\tilde{J}'^2|\omega_n|}.
	\end{aligned}
	\label{sef_JJ}
\end{equation}

\begin{figure}
	\centering
	\includegraphics[width=0.3\linewidth]{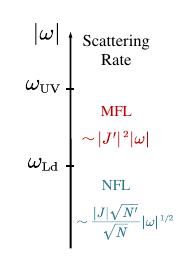}
	\caption{\textbf{Scattering rates without impurity $w=0$.}
		Schematic crossover diagram. The NFL is low-frequencies $|\omega|<\omega_{\rm Ld}$ and the MFL is in the finite frequency regime $\omega_{\rm Ld}<|\omega|\ll \omega_{\rm UV}$. The corresponding scattering rates are shown in blue and red, respectively.
		}
	\label{fig5}
\end{figure}

When both interactions are turned on $J\ne 0, J^\prime \ne 0$, 
it is necessary to take both damping mechanism into account for the critical bosons.
We show that the interplay between the spatially uniform and random interactions $J, J'({\bm x})$ 
results in two distinct NFL phases in a clean system with $w = 0$, 
corresponding to the low-frequency and finite-frequency regimes, as illustrated in Fig.~\ref{fig5}.
We present the analytic results below and refer length calculations to the Supplementary Materials Sec.~S5.
Namely, we explicitly work with the damped boson taking a form
\begin{equation}
	\begin{aligned}
		&\Pi(i\Omega_m,\bm{q})= \Pi_J(i\Omega_m,\bm{q}) + \Pi_{J^\prime}(i\Omega_m,\bm{q}),
	\end{aligned}
	\label{sef-bo}
\end{equation}
As depicted in Fig.~\ref{fig5}, we obtain the NFL at low-frequencies $|\omega|<\omega_{\rm Ld}$
which then crossovers to the MFL in the finite frequency regime $\omega_{\rm Ld}<|\omega|\ll \omega_{\rm UV}$.
The NFL and MFL are nothing but the solutions in Eq.~(\ref{sef_J}) and Eq.~(\ref{sef_JJ}) 
which are derived for $J\ne 0, J=0$ and $J=0, J^\prime \ne 0$, respectively.
And, the crossover energy scale is ${\omega_{\rm Ld}}\equiv  N^\prime |{J}|^2/(N |{J}'|^4\Lambda_\theta^4)$.
Heuristically, ${\omega_{\rm Ld}}$ is determined by comparing 
the NFL and MFL self energies.
Notably, when $J \neq 0$ and $J'= 0$, the crossover energy scale ${\omega_{\rm Ld}}$ diverges.
This indicates that the low-energy regime is enlarged and the system remains to be the NFL at all frequencies, 
which is consistent with Eq.~(\ref{fesecl}).
Eventually, in the high-energy region $|\omega|\gg \omega_{\rm UV}\equiv N^\prime \Lambda_{\rm U}^2/N|{J}'|^2 \exp{(-1/|J'|^2\Lambda_\theta)}$, the bare dynamic term $i\omega$ in Eq.~(\ref{eff}) overwrites all NFL fermion self-energies, 
ultimately exhibiting Fermi liquid behavior.

As illustrated in Table~\ref{tab1}, the dirty system with $w\ne 0$ exhibits the same MFL behavior ${\rm Im}\Sigma_{{\cal J}} \sim {\cal J}^2 |\omega|$ in the presence of either interaction ${\cal J}=J,J^\prime$.
Therefore, we conclude that when both types of interactions are present, the system is still a MFL.\\

\section{B. Diagrammatic calculation of optical conductivity}
We use the Kubo formula 
\begin{equation}
\mathrm{Re}[\sigma(\omega,T)]=\frac{\mathrm{Im}[\Xi(i\omega_n\rightarrow \omega+i0^+,0)]}{\omega},
\label{KF}
\end{equation}
to evaluate the conductivity perturbatively in terms of spatially uniform Yukawa interaction $J$. 
For a generic momentum-dependent form factor in the Yukawa interaction, the
current operator can be decomposed into a normal part, determined by the
fermion group velocity, and an anomalous part arising from the momentum
derivative of the form factor. 
At leading order in $\mathcal{O}(|J|^2)$, the
normal-current correlation function $\Xi$ receives the self-energy, vertex,
and AL contributions shown in
Fig.~\ref{fig3}(d-g), respectively. We will show that their sum is constrained
by momentum conservation.
The correlations involving the anomalous current have been discussed in
Ref.~\cite{Chubukov2024,Songci2023}. 
For the Pomeranchuk form factor considered here,
$f_{\bm k}=\cos k_x-\cos k_y$, we have explicitly verified that both the
mixed normal-anomalous correlation and the anomalous-anomalous correlation
are parametrically subleading at the order of interest. 
We therefore neglect the anomalous-current contribution in what follows.

First, the self-energy diagram in Fig.~\ref{fig3}(d) gives
\begin{equation}
\begin{aligned}
	\Xi_{\rm SE}(i\omega_n)=&-N\int_k v_{\bm k}^2 G(i\omega,\bm{k})[G(i\omega+i\omega_n,\bm{k})]^2\Sigma_{J}(i\omega+i\omega_n,\bm{k})\\
	&
	+(\omega_n\rightarrow -\omega_n).
	\label{XiSE}
	\end{aligned}
\end{equation}
where $\int_k \equiv T\sum_\omega\int \frac{d^2\bm{k}}{(2\pi)^2}$ 
is the integral over the three-momentum which is denoted as $k = (i\omega, {\bm k})$.
The vertex diagram in Fig.~\ref{fig3}(e) gives,
\begin{equation}
\begin{aligned}
	\Xi_{\rm V}(i\omega_n)=&-N|J|^2\int_{k,q} v_{{\bm k}}v_{{\bm k}+{\bm q}} G(i\omega,\bm{k})G(i\omega+i\omega_n,\bm{k})\\
	& \times D(i\Omega,\bm{q})G(i\omega+i\Omega+i\omega_n,\bm{k}+\bm{q})G(i\omega+i\Omega,\bm{k}+\bm{q}).\\
	\end{aligned}
	\label{XiV}
\end{equation}

And, the two AL diagrams in Fig.~\ref{fig3}(f,g) give a seemingly higher order term
\begin{equation}
	\begin{aligned}
		&\Xi_{\rm AL}(i\omega_n)=\Xi_{\rm AL_1}(i\omega_n)+\Xi_{\rm AL_2}(i\omega_n)\\
		=&|J|^4\frac{N^2}{N'}\int_q D(i\Omega-i\omega_n/2,\bm{q})D(i\Omega+i\omega_n/2,\bm{q}) \\
		& \quad \times  \int_{k}v_{\bm k}G(i\omega-i\omega_n/2,\bm{k}) G(i\omega+i\omega_n/2,\bm{k})G(i\omega+i\Omega,\bm{k}+\bm{q})\\
 		& \quad \times \int_{k^\prime} v_{\bm k'} G(i\omega'-i\omega_n/2,\bm{k}')G(i\omega'+i\omega_n/2,\bm{k}')\\
 		&\quad \times [G(i\omega'+i\Omega,\bm{k}'+\bm{q})+G(i\omega'-i\Omega,\bm{k}'-\bm{q})].\\
	\end{aligned}
\end{equation}
Here $G(i\omega,\bm{k})$ is the electron Green's function, and $D(i\Omega,\bm{q})$ is the damped boson Green's function.
The product of two electron Green functions can be approximated in the limit $|\Sigma_J|\ll \Gamma$ as
\begin{equation}
\begin{aligned}
	&G(i\omega+i\omega_n/2,\bm{k})G(i\omega-i\omega_n/2,\bm{k})\\
	\simeq &\frac{\nabla G_{k}(i\omega_n)}{i\omega_n+i\Gamma[\mathrm{sgn}(\omega+\omega_n/2)-\mathrm{sgn}(\omega-\omega_n/2)]}.
\end{aligned}
\end{equation}
We abbreviate the functional difference as $\nabla {\cal O}_k(i\omega_n)={\cal O}(i\omega-i\omega_n/2,\bm{k})-{\cal O}(i\omega+i\omega_n/2,\bm{k})$.
Then, by repeatedly using this modified Ward identity, 
we incorporate the AL diagrams into a unified expression at leading order, which reads
\begin{equation}
	\begin{aligned}
		 &\Xi_{\rm AL+SE+V}(i\omega_n)\\
		 &=\frac{N\tilde{J}^2|J|^2}{4(|\omega_n|+2\Gamma)^2} \int_q D(i\Omega+i\omega_n/2,{\bm q}) D(i\Omega-i\omega_n/2,{\bm q}) \\
 &\times \int_k G(i\omega+i\Omega,\bm{k}+\bm{q})\nabla G_{k}(i\omega_n) \int_{k^\prime} G(i\omega',\bm{k}')\nabla G_{k'-q}(i\omega_n)(\Delta v)^2,\\
	\end{aligned}
	\label{SE+V+ALv}
\end{equation}
where $\Delta v = v_{\bm{k}'-\bm{q}}-v_{\bm{k}'}+v_{\bm{k}+\bm{q}}-v_{\bm{k}}$ is the change of group velocity, which appears frequently in the Boltzmann equation \cite{Chubukov2011,Maslov2016}. 

Let's define a weighted integral 
\begin{equation}
\int^*_k\equiv T\sum_\omega \int \frac{d^2{\bm k}}{(2\pi)^2}G(i\omega+i\Omega,\bm{k}+\bm{q})\nabla G_{k}(i\omega_n)
\label{SI_int}
\end{equation}
and an external three-momentum $k_0=(i\omega_n,0)$, then Eq.(\ref{SE+V+ALv}) transforms into the abbreviated form in Eq.~(\ref{totxi}) of the maintext.

We then restore the energy scale of the electronic dispersion that was left implicit in the main text and analyze the dimensional consistency of the resulting \emph{ee}- and impurity-induced Drude weights.
In the low-frequency limit, the relevant energy-momentum integral in the
retarded current-current correlation function is linear in the external
frequency. It can therefore be parametrized as
\begin{equation}
    \left[
    \tilde{J}^{2}|J|^{2}
    \int_q D_+(q)D_-(q)
    \int_{k}^{\ast}\int_{k'-q}^{\ast}
    (\Delta v)^2
    \right]_{i\omega_n\rightarrow \omega+i0^+}
    \equiv
    \frac{i\omega D_{ee}}{\tau_{\mathrm{tr}}},
\end{equation}
where \(1/\tau_{\mathrm{tr}}\) is the transport scattering rate determined by
temperature and interaction effects. Using the Kubo formula \eqref{KF}, the
contribution of \emph{ee} scattering in Eq.~\eqref{SE+V+ALv} to the real part
of the optical conductivity in the low-frequency limit is then
\begin{equation}
    \operatorname{Re}\sigma_{ee}(\omega)
    \approx  
    \frac{ND_{ee}}{\Gamma^2 \tau_{\mathrm{tr}}}
    \frac{1}{1+\omega^2/\Gamma^2}.
    \label{eeee}
\end{equation}
This has the same Lorentzian form as the Drude contribution from impurity
scattering,
\begin{equation}
    \operatorname{Re}\sigma_{ie}(\omega)
    \approx 
    \frac{ND_{ie}}{\Gamma}
    \frac{1}{1+\omega^2/\Gamma^2}.
\end{equation}
In both cases, the impurity scattering rate \(\Gamma\) sets the width of the
Drude peak. The difference lies in the overall spectral weight: the
\emph{ee}-induced contribution is suppressed relative to the impurity Drude
weight by the dimensionless factor
\begin{equation}
    \frac{D_{ee}}{D_{ie}}\frac{1}{\Gamma\tau_{\mathrm{tr}}}.
\end{equation}
This point is also confirmed by the fact that the impurity- and
\emph{ee}-scattering contributions to the resistivity in
Eq.~\eqref{rho_J} have the same physical dimensions.

For an ordinary convex Fermi surface, the Drude weight scales as
\begin{equation}
    D_{ie} \sim \int_{\bm k} v_F^2 \delta(\epsilon_{\bm k}-\mu)
    \sim m v_F^2
    \sim \mu
    \sim \frac{n_0}{m},
\end{equation}
which gives the familiar Drude form of a nearly free-electron system. 
By contrast, in the low-energy effective theory in Eq.~\eqref{eff}, the hopping
scale has been absorbed into the saddle-point dispersion near the VHS.
Restoring this scale explicitly, one may write
\begin{equation}
    \epsilon_{\bm k}
    =
    \alpha k_x^2-\beta k_y^2-\mu ,
\end{equation}
where \(\alpha,\beta\sim t a_0^2\), with \(t\) the microscopic hopping amplitude
and \(a_0\) the lattice spacing. The corresponding Drude weight then scales as
\begin{equation}
    D_{ie}
    \sim
    \alpha^2
    \int_{\bm k}
    k^2
    \delta(\alpha k_x^2-\beta k_y^2)
    \sim
    \alpha \Lambda_{\rm U}^2
    \sim
    t a_0^2\Lambda_{\rm U}^2
    \sim
    t ,
\end{equation}
up to dimensionless factors, where \(\Lambda_{\rm U}\sim a_0^{-1}\) is the
ultraviolet momentum cutoff. Thus, as in the case of a conventional convex
Fermi surface, the Drude weight carries the dimension of energy. Equivalently,
whenever the hopping scale is restored, the cutoff-dependent factors appearing
in the main text should be understood schematically as
\begin{equation}
    \Lambda_{\rm U}^2
    \;\longrightarrow\;
    t a_0^2 \Lambda_{\rm U}^2 .
\end{equation}

For the VHS saddle-point dispersion considered here, the prefactor
of the \emph{ee} contribution in Eq.~\eqref{eeee} scales as
\begin{equation}
    \frac{D_{ee}}{\Gamma\tau_{\mathrm{tr}}}
    \sim
    \frac{|J|^2T}{\Gamma}.
\end{equation}
Since \(T\) and \(\Gamma\) both have the dimension of energy in units
\(\hbar=k_B=1\), the ratio \(T/\Gamma\) is dimensionless. After restoring the
hopping scale, the current-vertex factor contains the same microscopic energy
scale as the Drude weight,
$
    |J|^2 \sim t a_0^2\Lambda_{\rm U}^2 \sim t .
$
Therefore,
$
    D_{ee}/\Gamma\tau_{\mathrm{tr}}
$
also carries the dimension of energy, as required for the optical spectral
weight. Thus the \emph{ee}-induced Drude weight is dimensionally consistent
with the VHS effective theory.

Finally, we analyze $\Delta v$ for the normal  \emph{ee} scattering channels under different types of dispersions:
For a generic 2D convex Fermi surface as depicted in Fig.~\ref{fig4}(a),
there is only one intersection point that is inequivalent under inversion transformation
meaning that the injecting ${\bm k}, {\bm k}^\prime$ are at most swapped.
The generic Fermi point on a convex Fermi surface disperse linearly as ${\epsilon}_{\bm k}\simeq v_F k_x$,
which leads to a constant group velocity $v_x={\rm const}$ 
and then a vanishing $\Delta v_x=0$.
Then, the subleading-order terms lead to $\sigma(\omega\gg T,\Gamma=0) \sim |\omega|^{2/3}$\cite{Guo2022,Chubukov2024}.
For the saddle-point dispersion $\epsilon_{\bm k}=k_x^2-a k_y^2$ at the VHS, as shown in Fig.~\ref{fig4}(b), 
there are two intersection points that are inequivalent under inversion transformation
with ${\bm k}$ situated near the VHS and ${\bm k}^\prime$ on the convex portion of the Fermi surface. 
By exchanging a small-${\bm q}$, the outgoing momentum ${\bm k}+{\bm q}$ exceeds the Brillouin zone
and is folded back to the same area by the reciprocal lattice vector ${\bm G}$.
Therefore, the scattering processes can be classified into two distinct types.
(i) For the swap channel at the VHS, both ${\bm k}$ and ${\bm k'}$ in Eqs.~(\ref{SE+V+ALv}) are from the vicinity of the same VHS, which leads to $\Delta v_x = 0$. 
Despite that the Galilean invariance is absence,
the swapping between the inversion equivalent intersection points gives rise to a vanishing result.
(ii) For the extra channel, given that ${\bm k}$ is from the VHS vicinity, ${\bm k'}$ can be from a generic convex Fermi surface or another VHS, which leads to $\Delta v_x\ne 0$. Thus, the scattering in the extra channel contributes to a nonzero conductivity. 
We provide a step-by-step derivation in Supplemental Material Sec.~S3.A and Sec.~S4.B.\\

\begin{table*}
\centering
\caption{Scattering rate and optical conductivity at $T=0$.}
\label{tab1}%
\begin{tabular*}{\textwidth}{ccccc}
\toprule
 & ${-\rm Im}[\Sigma_{J}^{\rm R}(\omega,{\bm k})]$ & ${\rm Re}[\sigma^{J}(\omega,T=0)]$& ${-\rm Im}[\Sigma_{J'}^{\rm R}(\omega,{\bm k})]$ & ${\rm Re}[\sigma^{J'}(\omega,T=0)]$ \\                
$\Gamma\ll |\tilde{J}||\omega|^{1/2}$     & $|\tilde{J}| |\omega|^{1/2}$   & $\frac{N|J|^2|\omega|}{{\rm max}(\Gamma^2,\omega^2)}$ &\raisebox{-1.50ex}{$|J'|^2|\omega|$}& \raisebox{-1.50ex}{$\frac{N\Lambda_{\rm U}^2|J'|^2|\omega|}{{\rm max}(\Gamma^2,\omega^2)}$} \\
$\Gamma\gg |\tilde{J}||\omega|^{1/2}$   & $\frac{|J|^2}{\Gamma}|\omega|$ & $\frac{N|J|^2|\omega|^2}{\Gamma^3}$&&\\
\end{tabular*}
\end{table*}

\section{C. Optical conductivity at zero temperature}
In contrast to the \emph{dc} conductivity, which is infinite in absence of the momentum relaxation mechanism (\emph{e.g.} the umklapp scattering or impurities), 
the optical conductivity is free from the constraint imposed by momentum conservation.
It sensitively depends on both the Galilean invariance of the low-energy dispersion\cite{Chubukov2011} 
and the geometry of the 2D Fermi surface\cite{Songci2023}.

Then, we consider a relative stronger impurity scattering $|\omega|\ll \Gamma$
(or equivalently, in the low-frequency regime). The diagrams in Fig.~\ref{fig3}(d-g) lead to
\begin{equation}
\begin{aligned}
\Xi_{\rm SE+V+AL}&(i\omega_n,T=0) \sim -N|J|^2 \Big(\frac{|\omega_n|}{|\omega_n|+\Gamma}\Big)^2\ln\frac{\tilde{J}^2|\omega_n|+\Gamma^2}{\Lambda_{\rm U}^4},
\end{aligned}
\label{Xi}
\end{equation}
where $\Lambda_{\rm U}$ is the UV cutoff for momentum. In the clean limit, the optical conductivity comes from the imaginary part of the logarithmic function and differs from the weak scattering result in Eq.(\ref{sig_T0_Gm0}) by a factor [see Supplemental Material Sec.~S4.B.1]
\begin{equation}
\begin{aligned}
	  &\mathrm{Re}\big[\sigma^{J}\big(\Gamma^2/\tilde{J}^2 \ll |\omega|\ll \Gamma,T=0\big)\big]\\
	 & \simeq -\mathrm{Re}[\sigma^{J}(\Gamma^2/\tilde{J}^2 \ll  |\omega| ,\Gamma\ll |\omega|,T=0)]\times \frac{\omega^2}{\Gamma^2}
	  \sim -\frac{N|J|^2}{\Gamma^2}|\omega|. \\
\end{aligned}
	\label{clean_T0}
\end{equation}
The scaling factor $\omega^2/\Gamma^2$ converts the $|\omega|^{-1}$ tail in Eq.(\ref{sig_T0_Gm0}) 
into a linear-in-$\omega$ leading order term.
Furthermore, in the opposite dirty limit $\Gamma^2 \gg \tilde{J}^2 |\omega|$, the leading-order imaginary part of the retarded current-current correlation comes from power-law function in Eq.(\ref{Xi}). And, this leads to the optical conductivity 
\begin{equation}
{\rm Re}\big[\sigma^{J}\big(|\omega|\ll \Gamma^2/\tilde{J}^2,|\omega|\ll \Gamma, T=0 \big)\big]\sim -\frac{N|J|^2}{\Gamma^3}|\omega|^2\ln\frac{\Lambda_{\rm U}^2}{\Gamma}.
\label{dirty_T0}
\end{equation}
Note that this result is an order in $|\omega|/\Gamma$ smaller than Eq.(\ref{clean_T0}).
Conclusively, Eq.~(\ref{clean_T0},\ref{dirty_T0}) are the main results for the optical conductivity at $T=0$
which are summarized in Table.~\ref{tab1}.

In addition, due to the breaking of momentum conservation under impurity scattering, 
the zeroth order diagram $\Xi_0$ in Fig.~\ref{fig3}(c) gives rise to a finite \emph{dc} conductivity
\begin{equation}
	\mathrm{Re}[\sigma^0(|\omega|\ll \Gamma,T=0)]\sim \frac{N\Lambda_{\rm U}^2}{\Gamma},
\end{equation}
And, the total transport scattering rate is given by
\begin{equation}
	N\mathrm{Re}\Big[\frac{1}{\sigma\big(\Gamma^2/\tilde{J}^2 \ll |\omega|\ll \Gamma,T=0\big)}\Big] \simeq 
	\frac{\Gamma}{\Lambda_{\rm U}^2}+\frac{|J|^2}{\Lambda_{\rm U}^4} |\omega|.
\end{equation}
The optical conductivity arises at VHS is converted into a linear-in-$\omega$ scaling in $T=0$ with a residual \emph{dc} conductivity by introducing a modest amount of impurity. The leading order terms in $o(|J|^2)$ is rendered finite by the extra \emph{ee} scattering channel at the Lifshitz transition. And, the coefficient of \emph{ee} scattering rate is independent piece of information apart from the impurity scattering rate $\Gamma$ and the two scattering rates are additive satisfying the Matthiessen rule\cite{Maslov2016}.
Conventionally, the Matthiessen rule dictates that \emph{ee} scattering rate is much smaller that the impurity one.
This sets up a criteria $|\omega|\ll \Gamma \Lambda_{\rm U}^2/|J|^2$ which is readily satisfied by $\Lambda_{\rm U}/|J| \gg 1$.  \\
\end{appendix}

\clearpage


\makeatletter
\renewcommand{\theequation}{S\arabic{equation}}
\setcounter{equation}{0}
\renewcommand{\thefigure}{S\arabic{figure}}
\setcounter{figure}{0}
\renewcommand{\thesection}{S\arabic{section}}

\onecolumngrid

\section{Self-energy at the van Hove singularity}
\label{NFLs}

In this section, we separately consider the self-energy contributions resulting from two types of Yukawa interactions.

\subsection{Spatially uniform Yukawa interactions $J$}
\label{J}
In this subsection, we derive the analytic expressions for the non-Fermi liquid (NFL)
in the low-ferquency limit.
We start with the low-energy effective theory in Eq.~(1) of the main text,
which is rewritten here as,
 \begin{equation}
		\mathcal{L}=\sum_{m}\psi^\dagger_m(\partial_\tau-\partial_x^2+a\partial_y^2)\psi_m+\sum_l\phi_l^*(\Delta -\partial_\tau^2 - \bm{\nabla}^2 )\phi_l +\frac{1}{\sqrt{N}}\sum_{mn}w_{mn}(\bm{x})\psi_m^\dagger\psi_n+\frac{1}{\sqrt{NN'}}\sum_{mnl}J_{mnl}\psi^\dagger_m\psi_n\phi_l+h.c..
		\label{eff1}
\end{equation}
where $m,n=1, 2, . . ., N$ are the flavors of fermion field and $l=1, 2, ..., N'$ is the flavors of the boson field. The space uniform coupling $J_{mnl}$ is random in the space of flavors with Gaussian distribution:
  \begin{equation}
  	\overline{J_{mnl}}=0,\ \ \ \  \overline{J_{mnl}J^*_{m'n'l'}}=|J|^2\delta_{mm'}\delta_{nn'}\delta_{ll'}.
  \end{equation}
  The space dependent potential disorder $w_{mn}(\bm x)$ is random in flavor and coordinate spaces obeying the Gaussian distribution:
  \begin{equation}
  	\overline{w_{mn}(\bm{x})}=0,\ \ \ \  \overline{w_{mn}(\bm{x})w^*_{m'n'}(\bm{x}')}=|w|^2\delta_{mm'}\delta_{nn'}\delta(\bm{x}-\bm{x}').
  \end{equation}
  
The NFL are induced near the VHS with spatially uniform Yukawa interactions $J$.
In the following parts, we first consider the case without potential disorder and then examine the case with it.

\subsubsection{potential disorder $|w|=0$}
\label{Jw=0}
In the calculations of this subsection, we only consider the case of $w=0$.
We perform the calculations within the Eliashberg framework, starting with the computation of the boson self-energy, followed by the fermion self-energy.

We now derive the bosonic self-energy generated by the fermionic
particle-hole excitations near the van Hove saddle points. The one-loop
polarization is
\begin{equation}
\Pi_{\rm J}(i\Omega_m,\bm q)
=
-|J|^2\frac{N}{N'}
T\sum_n
\int\frac{d^2\bm k}{(2\pi)^2}
\frac{1}{i\omega_n-\epsilon_{\bm k}}
\frac{1}{i\omega_n+i\Omega_m-\epsilon_{\bm{k+q}}},
\label{eq:PiJ_start}
\end{equation}
$
\epsilon_{\bm k}=k_x^2-a k_y^2,
\ a>0.
$
After performing the Matsubara-frequency summation, we obtain
\begin{equation}
\Pi_{\rm J}(i\Omega_m,\bm q)
=
-|J|^2\frac{N}{N'}
\int\frac{d^2\bm k}{(2\pi)^2}
\frac{
n_{\rm F}(\epsilon_{\bm k})
-n_{\rm F}(\epsilon_{\bm{k+q}})
}{
i\Omega_m+\epsilon_{\bm k}-\epsilon_{\bm{k+q}}
}.
\label{eq:PiJ_after_sum}
\end{equation}
At zero temperature,
\(n_{\rm F}(\epsilon_{\bm k})=\theta(-\epsilon_{\bm k})\).
Making the changes of variables
\(\bm k\rightarrow\bm k-\bm q\) and
\(\bm k\rightarrow-\bm k\) in the second term gives
\begin{equation}
\begin{aligned}
\Pi_{\rm J}(i\Omega_m,\bm q)
={}&
-|J|^2\frac{N}{N'}
\int\frac{d^2\bm k}{(2\pi)^2}
\left[
\frac{1}{
i\Omega_m+\epsilon_{\bm k}-\epsilon_{\bm{k+q}}
}
-
\frac{1}{
i\Omega_m+\epsilon_{\bm{k+q}}-\epsilon_{\bm k}
}
\right]
\theta(-\epsilon_{\bm k})
\\
={}&
|J|^2\frac{N}{N'}
\int^{*}\frac{d^2\bm k}{(2\pi)^2}
\frac{
2\left(
\epsilon_{\bm q}+2k_xq_x-2ak_yq_y
\right)
}{
\left(
\epsilon_{\bm q}+2k_xq_x-2ak_yq_y
\right)^2+\Omega_m^2
},
\label{eq:PiJ_occupied}
\end{aligned}
\end{equation}
where the asterisk denotes the occupied region
$
-\sqrt a\,|k_y|<k_x<\sqrt a\,|k_y|.
$
The integration over \(k_x\) then yields
\begin{equation}
\begin{aligned}
\Pi_{\rm J}(i\Omega_m,\bm q)
={}&
|J|^2\frac{N}{8\pi^2N'q_x}
\int_{-\infty}^{\infty}dk_y
\Big\{
\ln\!\left[
\left(
\epsilon_{\bm q}
+2\sqrt a\,|k_y|q_x
-2ak_yq_y
\right)^2+\Omega_m^2
\right]
\\
&\hspace{3.5cm}
-
\ln\!\left[
\left(
\epsilon_{\bm q}
-2\sqrt a\,|k_y|q_x
-2ak_yq_y
\right)^2+\Omega_m^2
\right]
\Big\}.
\label{eq:PiJ_log_integral}
\end{aligned}
\end{equation}

To regularize the momentum integral, we impose the ultraviolet cutoff
\(|k_y|<\Lambda_{\rm U}\). After separating the \(k_y>0\) and
\(k_y<0\) regions, the required integrals can be written in terms of
\begin{equation}
I(c)
=
\int_0^{\Lambda_{\rm U}}dx\,
\ln\!\left[\Omega_m^2+(b+cx)^2\right],
\label{eq:Ic_def}
\end{equation}
where
\begin{equation}
b=\epsilon_{\bm q},
\qquad
c_{\pm}=2\sqrt a\left(q_x\pm\sqrt a\,q_y\right).
\end{equation}
The exact integral is
\begin{equation}
\begin{aligned}
I(c)
={}&
-2\Lambda_{\rm U}
+
\frac{2|\Omega_m|}{c}
\left[
\arctan\left(
\frac{b+c\Lambda_{\rm U}}{|\Omega_m|}
\right)
-
\arctan\left(
\frac{b}{|\Omega_m|}
\right)
\right]
\\
&+
\frac{b}{c}
\ln
\frac{
\Omega_m^2+(b+c\Lambda_{\rm U})^2
}{
\Omega_m^2+b^2
}
+
\Lambda_{\rm U}
\ln\!\left[
\Omega_m^2+(b+c\Lambda_{\rm U})^2
\right].
\label{eq:Ic_exact}
\end{aligned}
\end{equation}
Only the part odd under \(c\rightarrow-c\) survives in
Eq.~\eqref{eq:PiJ_log_integral}. We first discuss the finite low-energy
part and then return to the cutoff-sensitive logarithmic contribution.

In the patch regime
\begin{equation}
|c_{\pm}|\Lambda_{\rm U}\gg
\max\left(|b|,|\Omega_m|\right),
\label{eq:patch_regime}
\end{equation}
the cutoff-independent odd contribution to Eq.~\eqref{eq:Ic_exact} is
\begin{equation}
I_{\rm fin}^{\rm odd}(c)
=
-\frac{2|\Omega_m|}{c}
\arctan\left(\frac{b}{|\Omega_m|}\right).
\label{eq:Ic_finite}
\end{equation}
Substituting this term into Eq.~\eqref{eq:PiJ_log_integral}, we obtain
\begin{equation}
\begin{aligned}
\Pi_{{\rm J}}(i\Omega_m,\bm q)
={}&
|J|^2\frac{N}{N'}\frac{1}{4\pi^2}
\Bigg[
\frac{|\Omega_m|}
{q_x(-aq_y-\sqrt a\,q_x)}
\arctan\left(
\frac{\epsilon_{\bm q}}{|\Omega_m|}
\right)-
\frac{|\Omega_m|}
{q_x(-aq_y+\sqrt a\,q_x)}
\arctan\left(
\frac{\epsilon_{\bm q}}{|\Omega_m|}
\right)
\Bigg]
\\
={}&
-|J|^2\frac{N}{N'}
\frac{1}{2\pi^2\sqrt a}
\frac{|\Omega_m|}{\epsilon_{\bm q}}
\arctan\left(
\frac{\epsilon_{\bm q}}{|\Omega_m|}
\right).
\label{eq:PiJ_dynamic_full}
\end{aligned}
\end{equation}
Eq.~\eqref{eq:PiJ_dynamic_full} is valid for nonzero momentum within
the patch regime \eqref{eq:patch_regime}.

In the static regime,
\begin{equation}
|\epsilon_{\bm q}|\gg|\Omega_m|,
\end{equation}
we use
\(
\arctan(\epsilon_{\bm q}/|\Omega_m|)
\simeq
(\pi/2)\operatorname{sgn}(\epsilon_{\bm q})
\)
and obtain the Landau-damping contribution
\begin{equation}
\Pi_{{\rm J}}(i\Omega_m,\bm q)
\simeq
-|J|^2\frac{N}{N'}
\frac{1}{4\pi\sqrt a}
\frac{|\Omega_m|}{|\epsilon_{\bm q}|}.
\label{eq:PiJ_static_limit}
\end{equation}
This is the damping kernel used in the subsequent low-energy
calculations.

In the intermediate dynamical regime,
\begin{equation}
|\epsilon_{\bm q}|\ll|\Omega_m|,
\qquad
|q_x\pm\sqrt a\,q_y|
\gg
\frac{|\Omega_m|}{\Lambda_{\rm U}},
\end{equation}
Eq.~\eqref{eq:PiJ_dynamic_full} instead gives
\begin{equation}
\Pi_{{\rm J}}(i\Omega_m,\bm q)
\simeq
-|J|^2\frac{N}{N'}
\frac{1}{2\pi^2\sqrt a}.
\label{eq:PiJ_intermediate_dynamic}
\end{equation}
This constant does not, however, describe the strict
\(\bm q\rightarrow0\) limit at fixed nonzero \(\Omega_m\). For a finite
cutoff, when
\begin{equation}
|q_x\pm\sqrt a\,q_y|
\ll
\frac{|\Omega_m|}{\Lambda_{\rm U}},
\end{equation}
the boundary terms retained in Eq.~\eqref{eq:Ic_exact} cancel the
constant in Eq.~\eqref{eq:PiJ_intermediate_dynamic}, and the polarization
vanishes as \(q^2/\Omega_m^2\). Thus,
\begin{equation}
\Pi_{\rm J}(i\Omega_m,\bm 0)=0,
\qquad \Omega_m\neq0,
\end{equation}
as required by the Ward identity. The different regimes of the finite
dynamical contribution can therefore be summarized as
\begin{equation}
\Pi_{{\rm J}}(i\Omega_m,\bm q)
\simeq
\begin{cases}
-\displaystyle
|J|^2\frac{N}{N'}
\frac{1}{4\pi\sqrt a}
\frac{|\Omega_m|}{|\epsilon_{\bm q}|},
&
|\epsilon_{\bm q}|\gg|\Omega_m|,
\\[9pt]
-\displaystyle
|J|^2\frac{N}{N'}
\frac{1}{2\pi^2\sqrt a},
&
|\epsilon_{\bm q}|\ll|\Omega_m|,
\quad
|q_x\pm\sqrt a\,q_y|
\gg
\dfrac{|\Omega_m|}{\Lambda_{\rm U}},
\\[10pt]
0,
&
|\epsilon_{\bm q}|\ll|\Omega_m|,
\quad
|q_x\pm\sqrt a\,q_y|
\ll
\dfrac{|\Omega_m|}{\Lambda_{\rm U}}.
\end{cases}
\label{eq:PiJ_three_regimes}
\end{equation}

We next consider the second odd term in Eq.~\eqref{eq:Ic_exact},
\begin{equation}
I_{\rm div}^{\rm odd}(c)
=
\frac{b}{c}
\ln
\frac{
\Omega_m^2+c^2\Lambda_{\rm U}^2
}{
\Omega_m^2+b^2
},
\label{eq:Ic_divergent}
\end{equation}
where we have used
\(|c|\Lambda_{\rm U}\gg|b|\).
In the static regime
\(|\epsilon_{\bm q}|\gg|\Omega_m|\), this term becomes independent of
frequency and gives
\begin{equation}
\Pi_{{\rm J},{\rm UV}}(0,\bm q)
=
\frac{N|J|^2}{8\pi^2N'}
\left[
\frac{1}{\sqrt a}
\ln\frac{16a^2\Lambda_{\rm U}^4}{\epsilon_{\bm q}^2}
+
\frac{q_y}{q_x}
\ln\left(
\frac{q_x-\sqrt a\,q_y}
{q_x+\sqrt a\,q_y}
\right)^2
\right].
\label{eq:PiJ_divergent}
\end{equation}
The first term contains the logarithmic dependence on both the cutoff and
\(\epsilon_{\bm q}\). The second term approaches a
direction-dependent constant when the origin is approached along
\(q_x=d\sqrt a\,q_y\), with \(d\neq1\). Its precise form therefore
depends on the cutoff geometry and on the subtraction convention used
to define \(\Pi_{\rm J}(0,\bm q\rightarrow0)\).
The same separation can be seen more transparently in hyperbolic
coordinates. Up to cutoff-scheme-dependent constants and
direction-dependent boundary terms, one finds, for
\(\epsilon_{\bm q}\neq0\),
\begin{equation}
\Pi_{\rm J}(i\Omega_m,\epsilon_{\bm q})
=
-\frac{|J|^2N}{2\pi^2N'\sqrt a}
\left[
\frac{|\Omega_m|}{\epsilon_{\bm q}}
\arctan\left(
\frac{\epsilon_{\bm q}}{|\Omega_m|}
\right)
-\frac{1}{2}
\ln
\frac{
|\epsilon_{\bm q}|\Lambda_{\rm U}^2
}{
\epsilon_{\bm q}^2+\Omega_m^2
}
-1
\right].
\label{eq:PiJ_hyperbolic}
\end{equation}
The first term in Eq.~\eqref{eq:PiJ_hyperbolic} is precisely the finite
dynamical contribution in Eq.~\eqref{eq:PiJ_dynamic_full}, whereas the
second term represents the logarithmically cutoff-sensitive part. The
remaining constant depends on the subtraction convention and can be
absorbed into the definition of the bosonic mass.

The treatment of the logarithmic term here is quite interesting. 
The simplest approach would be to follow the effective low-energy prescriptions employed in Refs.\cite{PhysRevLett.130.083603,PhysRevB.90.161106}: we retain only the contribution from fermionic states close to the Fermi surface. The cutoff-sensitive terms in Eq.\eqref{eq:PiJ_divergent}, which also receive contributions from momenta near the boundary of the saddle-point patch, are interpreted as a renormalization of the bare bosonic dispersion and are therefore not included explicitly in the Landau-damped propagator. However, here we would like to discuss this issue more carefully:
We introduce a low-energy cutoff
\begin{equation}
	E_{\rm IR}\sim
\max \left\{
T,\ \Gamma,\ t_z,\ E_L
\right\},
\end{equation}
which respectively represent the temperature, the fermionic broadening, the finite \(k_z\) dispersion, and the finite-size energy scale. Any effect that smears the VHS regularizes the singularity; consequently, the logarithmically divergent term becomes:
\begin{equation}
	\frac{N|J|^2}{8\pi^2N'}
\frac{1}{\sqrt a}
\ln\frac{\Lambda_{\rm U}^2}{\sqrt{\epsilon_{\bm q}^2+E_{\rm IR}^2}}
\end{equation}
Upon substituting this result into the bosonic propagator (need to subtract $\Pi_{\rm J}(0,0)$), the momentum-integration region in hyperbolic coordinates becomes:
\begin{equation}
	\int_{|\Omega|}^{E_{\rm IR}}dq^2 \frac{1}{q^2\cosh(2\theta)+\tilde{J}^2\frac{|\Omega_m|}{q^2}}\times ... +\int_{E_{\rm IR}}^{\Lambda_U^2}dq^2 \frac{1}{q^2\cosh(2\theta)+\tilde{J}^2\ln \frac{q^2}{E_{\rm IR}}}\times ...-\int_{E_{\rm IR}}^{\Lambda_U^2}dq^2 \frac{\tilde{J}^2\frac{|\Omega_m|}{q^2}}{(q^2\cosh(2\theta)+\tilde{J}^2\ln \frac{q^2}{E_{\rm IR}})^2}\times ...
\end{equation}
Here, the ellipsis denotes the remaining parts of the integral. As long as the momentum integration does not generate a power-law dependence on the ultraviolet cutoff, \((\Lambda_U)^{\alpha>0}\), the logarithmic static correction does not modify the leading nonanalytic low-frequency scaling obtained by neglecting it. In particular, even when the calculation exhibits a power-law ultraviolet dependence, \((\Lambda_U)^{\alpha>0}\), the leading frequency-independent contribution from the logarithmic bosonic self-energy does not enter certain dissipative observables defined through the imaginary part of an analytically continued retarded correlation function.
Since all subsequent results exhibit at most logarithmic ultraviolet divergences, the logarithmic static correction leaves all low-frequency exponents unchanged and only modifies nonuniversal prefactors.
Note that all these results are only applicable at the low-frequency limit; above that, crossover behavior will occur.

The NFL self-energy provides an intrinsic regularization of the static
VHS singularity. Within the momentum-independent approximation of
Eq.~\eqref{sic-fer-5}, write
\begin{equation}
 \Sigma_{\rm J}(i\omega,0)-\Sigma_{\rm J}(0,0)
 =-i\gamma_{\rm J}\operatorname{sgn}(\omega)\sqrt{|\omega|},
 \qquad
 \gamma_{\rm J}=\frac{|J|}{4\sqrt{\pi}}
 \sqrt{\frac{N'}{N}}\left(a^{1/4}+a^{-1/4}\right).
\end{equation}
Retaining the VHS contribution, the dressed static bubble gives
\begin{equation}
 \begin{aligned}
 \Pi_{\rm J}(0,\bm q)
 &\simeq\frac{\tilde{J}^2}{\pi}
 \int_0^1dx\int_0^{\Lambda_\omega}d\omega\,
 \frac{\omega+\gamma_{\rm J}\sqrt{\omega}}
 {(\omega+\gamma_{\rm J}\sqrt{\omega})^2
   +[x(1-x)\epsilon_{\bm q}]^2},\\
 \Pi_{\rm J}(0,0)
 &\simeq\frac{\tilde{J}^2}{\pi}
 \int_0^{\Lambda_\omega}\frac{d\omega}
 {\omega+\gamma_{\rm J}\sqrt{\omega}}
 =\frac{2\tilde{J}^2}{\pi}
 \ln\left(1+\frac{\sqrt{\Lambda_\omega}}{\gamma_{\rm J}}\right)<\infty,
 \end{aligned}
\end{equation}
where $\Lambda_\omega$ is a frequency cutoff; nonsingular band and patch-boundary
contributions are omitted. The low-frequency integrand is
$\propto\omega^{-1/2}$ and is integrable. For
$|\epsilon_{\bm q}|\ll\gamma_{\rm J}^2$, the leading low-energy momentum
correction is
\begin{equation}
 \left[\Pi_{\rm J}(0,\bm q)-\Pi_{\rm J}(0,0)\right]_{\rm low}
 \simeq-\frac{\tilde{J}^2}{\gamma_{\rm J}^2}
 \int_0^1dx\,|x(1-x)\epsilon_{\bm q}|
 =-\frac{\tilde{J}^2}{6\gamma_{\rm J}^2}|\epsilon_{\bm q}|
 =O(q^2).
\end{equation}
Thus the finite uniform part renormalizes the boson mass, while the
remaining static term has the same momentum scaling as the bare spatial
kernel, supporting the self-consistency of the normal-state scaling solution.

Accordingly, the effective bosonic kernel used below is determined by
Eq.~\eqref{eq:PiJ_static_limit}.
All our calculations of the fermion self-energy and transport properties are performed under the condition that Landau damping exists. Accordingly, the boson momentum $|\epsilon_{\bm q}|$ integral should, in principle, be taken from $|\Omega|$ to $\infty$. However, we have verified that extending the integration limits from $0$ to $\infty$ does not lead to any significant difference in the low frequency limit.
 
We note that the result in the static limit mimics the Landau damping term for the case of generic convex Fermi surface: $|\Omega_m|/v_{\rm F}|q|$. 
We take the static limit and the Landau damping (at VHS) in the following calculations. To be more concrete, the boson self energy can be expressed in a simpler form by dividing the momentum space into different regions:
 \begin{equation}
 	\Pi_{\rm J}(i\Omega_m,\bm{q})=\left\{\begin{matrix}
\tilde{J}^2\frac{|\Omega_m|}{q_x^2-a q_y^2} & (-\sqrt{a}q_y<q_x<\sqrt{a}q_y) \equiv \mathbb{A} \\
 -\tilde{J}^2\frac{|\Omega_m|}{q_x^2-a q_y^2} & (-\frac{|q_x|}{\sqrt{a}}< q_y<\frac{|q_x|}{\sqrt{a}}) \equiv \mathbb{B}
\end{matrix}\right.,
\label{sic-bos-1}
 \end{equation}		
where the two different regions are labelled as $\mathbb{A}$ (within the Fermi surface) and $\mathbb{B}$ (outside the Fermi surface). For convenience, we set $\tilde{J}^2 \equiv |J|^2 \frac{N}{N’} \frac{1}{4\pi \sqrt{a}}$.\\

At the quantum critical point (QCP), where $\Delta_{\rm eff} \equiv \Delta - \Pi(0,0) = 0$, the bosonic modes are strongly overdamped at small frequencies and near momenta $(0,0)$, where the bosons condense. This can, in turn, destroy the electronic quasiparticles. The scattering rate can be read off from the fermion self energy, which is defined as
 \begin{equation}
 	\begin{aligned}
 		\Sigma_{\rm J}(i\omega_n,0)=&|J|^2T\sum_m[\int_\mathbb{A} \frac{d^2 \bm{q}}{(2\pi)^2}\frac{1}{\bm{q}^2-\tilde{J}^2\frac{|\Omega_m|}{q_x^2-a q_y^2}}\frac{1}{i\omega_n+i\Omega_m-q_x^2+a q_y^2}+\int_\mathbb{B}\frac{d^2 \bm{q}}{(2\pi)^2}\frac{1}{\bm{q}^2+\tilde{J}^2\frac{|\Omega_m|}{q_x^2-a q_y^2}}\frac{1}{i\omega_n+i\Omega_m-q_x^2+a q_y^2}\\
 	\end{aligned}
 \end{equation}
  	
We provide a step-by-step derivation for integration procedures in region $\mathbb{A}$, while the integration over region $\mathbb{B}$ is similar and thus omitted.
\begin{equation}
		\int_\mathbb{A} \frac{d^2 \bm{q}}{(2\pi)^2}\frac{1}{\bm{q}^2-\tilde{J}^2\frac{|\Omega_m|}{q_x^2-a q_y^2}}\frac{1}{i\omega_n+i\Omega_m-q_x^2+a q_y^2}
		\simeq \frac{1}{\pi^2}\int_0^\infty d q_y\frac{1}{q_y^2+\tilde{J}^2\frac{|\Omega_m|}{a q_y^2}}\frac{\mathrm{ArcTanh}(\frac{\sqrt{a}q_y}{\sqrt{i\omega_n+i\Omega_m+a q_y^2}})}{\sqrt{i\omega_n+i\Omega_m+a q_y^2}},
	\label{sic-fer-4}	
\end{equation} 	
where we neglect the $q_x$ component of the Landau-damped boson propagator in region $\mathbb{A}$, because the main contribution of the integral is concentrated near the positive $q_y$-axis. As the integration region approaches the boundary of $\mathbb{A}$, the fermion propagator diverges in the static limit and plays a dominant role. In the static limit $|\omega_n + \Omega_m| \ll a q_y^2$, we define a branch cut along the negative $x$-axis to calculate the square root on the denominator which is approximated as
 $\sqrt{i\omega_n + i\Omega_m + a q_y^2} \simeq \sqrt{a} q_y \mathrm{e}^{i\mathrm{sgn}(\omega_n + \Omega_m) \frac{\eta}{2}}$ with $\eta \equiv |\omega_n + \Omega_m| / (a q_y^2) \rightarrow 0^+$. The final result is
\begin{equation}
\sim	\frac{1}{16|\tilde{J}||\Omega_m|^\frac{1}{2}}[\ln \gamma-i\mathrm{sgn}(\omega_n+\Omega_m)].
	\label{sic-fer-1}
\end{equation}
where the log term reads $\ln \gamma \simeq \ln \left(\frac{4 a \Lambda_{\rm U}^2}{|\omega_n + \Omega_m|}\right)$. Similarly, the integration over the domain $\mathbb{B}$ yields:
\begin{equation}
	\sim -\frac{1}{16|\tilde{J}||\Omega_m|^\frac{1}{2}\sqrt{a}}[\mathrm{sgn}(\omega_n+\Omega_m)i+\ln \gamma].
	\label{sic-fer-2}
\end{equation}
Finally, we arrive at the fermion self-energy at zero temperature, which takes a form
\begin{equation}
	\Sigma_{\rm J}(i\omega_n,0)-\Sigma_{\rm J}(0,0)\simeq -i\frac{|J|}{4\sqrt{\pi}}\sqrt{\frac{N'}{N}}[a^{1/4}+a^{-1/4}]\mathrm{sgn}(\omega_n)|\omega_n|^{1/2}.
		\label{sic-fer-5}
 \end{equation}
We have used the minimal subtraction renormalization scheme to eliminate the divergence in the real part. Specifically, when $a = 1$, the real part of the Fermi self-energy cancels out automatically.
Eq.~(\ref{sic-fer-5}), together with Eq.~(\ref{sic-bos-1}), is the main result of this subsection.\\

We obtain a NFL. Fig.~2b-c in the main text presents the frequency dependence of the fermion self-energy, verified through numerical self-consistent calculations. Specifically, we use Eq.~(\ref{sic-fer-5}) to compute the renormalized Fermi Green’s function, which is then substituted back into the self-energy calculation. It is found that, within the Eliashberg framework, the fermion self-energy in the vicinity of the VHS region is indeed proportional to $|\omega_n|^{1/2}$.
All these calculations requires that the Migdal approximation is satisfied, namely $|\epsilon_{\bm q}|=|q_x^2-aq_y^2| \gg |\Omega|$.

We close this subsection by remarking that the expression $\mathrm{Im}\Sigma(i\omega_n,0)\sim |\omega_n|^{1/2}$ is consistent with the scaling analysis: We assume the scaling dimension of space to be $1$ and that of time to be $z$, respectively. The deviation from Fermi liquid occurs near the critical fixed point, where we expect the interaction to be marginal. When the spatially uniform interaction dominates, resulting in a dynamical critical exponent $z=4$. It leads to the fermion self-energy $\sim |\omega_n|^{1/2}$ and is consistent with the analytical results in Eq.(\ref{sic-fer-5}).

\subsubsection{potential disorder $|w|\ne 0$}
In this subsection, we consider the effects of potential disorder $w$ in Eq.~(\ref{eff1}), namely the impact of impurities, which manifests in the fermion self-energy as the impurity-induced scattering rate:
\begin{equation}
	\Sigma_{\rm w}(i\omega_n)=|w|^2\int\frac{d^2\bm{k}}{(2\pi)^2}\frac{1}{i\omega-\epsilon_{\bm k}-\Sigma(i\omega_n)} =-i\frac{\Lambda_\theta}{2\pi}|w|^2\mathrm{sgn}(\omega_n),
	\label{pose}
\end{equation}
where we divide the momentum space into two regions for calculation: the region outside the Fermi surface and the region inside the Fermi surface. For the region outside the Fermi surface, we apply the transformation 
\begin{equation}
	k_x=k\cosh(\theta), \ \ \ k_y=k\sinh(\theta),
\end{equation}
to simplify the calculation in hyperbolic coordinates. Similarly, for the region inside the Fermi surface, we use the transformation 
\begin{equation}
	k_x=k\sinh(\theta), \ \ \ k_y=k\cosh(\theta),
\end{equation}
for the computation.
Here the $\Lambda_\theta$ is the dimensionless cutoff for angular $\theta$-variable.
For direct integration over the momentum in the $k_x-k_y$ coordinates, we find $\Lambda_\theta\sim \ln \Lambda_{\rm U}^2$, where $\Lambda_{\rm U}$ is the UV cutoff of the momentum.
Note that $\mathrm{sgn}\{\omega_n-\mathrm{Im}[\Sigma(i\omega_n)]\}=\mathrm{sgn}(\omega_n)$. For convenience, we set the impurity scattering rate
\begin{equation}
	\Gamma \equiv \frac{\Lambda_\theta}{2\pi}|w|^2.
\end{equation}
So the electron Green’s function can be written as  $G_{\rm w}(i\omega_n,\bm{k})= 1/(i\omega_n+i\Gamma\mathrm{sgn}(\omega_n)-\epsilon_{\bm k})$. We work at the low-frequency limit ($|\omega|\ll \Gamma$), at zero temperature, the electron Green’s function can be approximated as $G_{\rm w}(i\omega_n,\bm{k})\approx 1/(i\Gamma\mathrm{sgn}(\omega_n)-\epsilon_{\bm k})$.

We find that in the dirty limit $\tilde{J}^2|\omega|\ll \Gamma^2$, the fermion self-energy exhibits characteristics of a marginal Fermi liquid (MFL).
While in the clean limit $\tilde{J}^2|\omega|\gg \Gamma^2$, the system still exhibits the z = 4 NFL behavior as shown in Eq.~(\ref{sic-fer-5}).
 We first perform calculations in the limit $\Gamma\ll |\epsilon_{\bm q}|$ and later demonstrate that our results do not depend on this limit. 

When $\Gamma\ll |\epsilon_{\bm q}|$, the width of the bosonic spectral function peak, $|\mathrm{Im}\Pi^{\rm R}(\Omega,\bm{q})|\sim \frac{|\Omega|}{|\epsilon_{\bm q}|}\sim |\epsilon_{\bm q}|$, is much greater than the width $|\mathrm{Im}\Sigma^{\rm R}(\omega,\bm{k})|\sim \Gamma$ of the fermionic spectral function peak. Therefore, the fermionic spectral function can be safely approximated by a delta function:
\begin{equation}
	G_{\rm w}(i\omega_n,\bm{k})=\int \frac{dq}{\pi}\frac{\mathrm{Im}G_{\rm w}^{\rm R}(q,\bm{k})}{q-i\omega_n}\approx -i \frac{\Gamma}{\epsilon_{\bm k}^2+\Gamma^2}\mathrm{sgn}(\omega_n)\approx -i\pi\mathrm{sgn}(\omega_n)\delta(\epsilon_{\bm k}).
	\label{vapp}
\end{equation}
Thus, the boson self-energy is given by:
 \begin{equation}
 \begin{aligned}
  	\Pi_{\rm J}(i\Omega_m,\bm{q})=&-|J|^2\frac{N}{N'}T\sum_n\int\frac{d^2\bm{k}}{(2\pi)^2}G_{\rm w}(i\omega_n,\bm{k})G_{\rm w}(i\omega_n+i\Omega_m,\bm{k}+\bm{q})\\
  	=&\frac{|J|^2N\pi^2}{N'}\int_{-\Lambda_\omega}^{{\Lambda_\omega}}\frac{d\omega}{2\pi}\mathrm{sgn}(\omega)\mathrm{sgn}(\omega+\Omega_m) \int \frac{d^2\bm{k}}{4\pi^2}\frac{\Gamma/\pi}{\epsilon_{\bm k}^2+\Gamma^2}\frac{\Gamma/\pi}{\epsilon_{\bm{k}+\bm{q}}^2+\Gamma^2}\\
  	\approx &\frac{|J|^2N\pi^2}{N'}\int_{-\Lambda_\omega}^{{\Lambda_\omega}}\frac{d\omega}{2\pi}\mathrm{sgn}(\omega)\mathrm{sgn}(\omega+\Omega_m) \int \frac{d^2\bm{k}}{4\pi^2\sqrt{a}}\delta(\epsilon_{\bm k})\delta(\epsilon_{\bm{k}+\bm{q}})\\
  	=&\frac{|J|^2N}{4\pi\sqrt{a}N'|\epsilon_{\bm q}|}(\Lambda_\omega-|\Omega_m|).
  \end{aligned}
   \label{bsej}
 \end{equation}
 Here, we introduced an ultraviolet cutoff $\Lambda_\omega$ to prevent divergence in the frequency integral. Similarly, we are only concerned with the low-energy physics, so this divergent part is omitted. Specifically, when $\Gamma\ll |\epsilon_{\bm q}|$, $\frac{\Lambda_\omega}{|\epsilon_{\bm q}|}<\frac{\Gamma}{|\epsilon_{\bm q}|}\ll 1$ itself is a small quantity and can be directly ignored.  Thus, the frequency-dependent boson self-energy is NOT different from the case without potential disorder which is written in Eq.~(\ref{sic-bos-1}). As promised, the approximation (\ref{vapp}) is not strictly necessary. By performing the actual integration over $\bm{k}$, in Eq.~(\ref{bsej}) without relying on the delta-function approximation, one arrives at the same result:
\begin{equation}
\begin{aligned}	
&\int \frac{d^2\bm{k}}{4\pi^2}\frac{\Gamma/\pi}{\epsilon_{\bm k}^2+\Gamma^2}\frac{\Gamma/\pi}{\epsilon_{\bm{k}+\bm{q}}^2+\Gamma^2}\\
=&\int \frac{dk_+dk_-}{8\pi^2\sqrt{a}}\frac{\Gamma/\pi}{k_+^2k_-^2+\Gamma^2}\frac{\Gamma/\pi}{(k_++q_+)^2(k_-+q_-)^2+\Gamma^2}\\
=&\frac{\Gamma^2}{8\pi^4\sqrt{a}|\epsilon_{\bm q}|^3}\int dxdy \frac{1}{x^2y^2+\Gamma^2/\epsilon_{\bm q}^2} \frac{1}{(x+1)^2(y+1)^2+\Gamma^2/\epsilon_{\bm q}^2}\\
=&\frac{\Gamma^2}{8\pi^3\sqrt{a}|\epsilon_{\bm q}|^3}\int dy\frac{\mathrm{Abs}(1+y)+\mathrm{Abs}(y)}{\frac{\Gamma}{|\epsilon_{\bm q}|}y^2(1+y)^2+\frac{\Gamma^3}{|\epsilon_{\bm q}|^3}\{1+2y(1+y)+2\mathrm{Abs}[y(1+y)]\}}\\
=&\frac{\Gamma}{4\pi^3\sqrt{a}|\epsilon_{\bm q}|^2}\int dy\frac{\mathrm{Abs}(1+y)}{y^2(1+y)^2+\frac{\Gamma^2}{|\epsilon_{\bm q}|^2}\{1+2y(1+y)+2\mathrm{Abs}[y(1+y)]\}}\\
\equiv &\frac{1}{4\pi^2\sqrt{a}|\epsilon_{\bm q}|} f(\frac{\Gamma}{|\epsilon_{\bm q}|})\\
\approx &\frac{1}{4\pi^2\sqrt{a}|\epsilon_{\bm q}|}+\mathcal{O}(\Gamma/|\epsilon_{\bm q}|^2)
\end{aligned}
\label{gam=del}
\end{equation}
In the first equality, we performed the following variable substitution $k_+=k_x-k_y, k_-=k_x-k_y$, and at the second equality, we applied the dimensionless transformation $x=k_+/q_+, y=k_-/q_-$.

\begin{figure}[htbp]
	\centering
	\includegraphics[width=\textwidth]{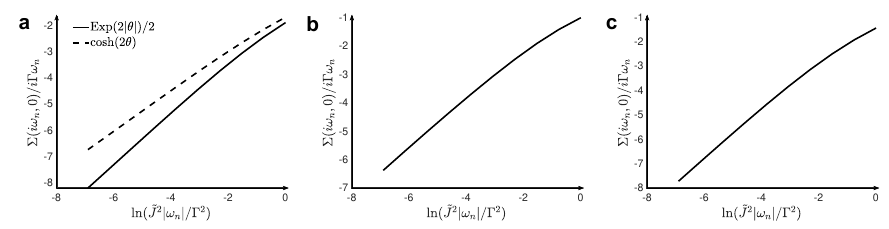}
	\caption{\textbf{The fermion self-energy under different frequencies and impurity scattering rates.} \textbf{a} When $\Gamma\ll |\epsilon_{\bm q}|$ and  $\tilde{J}^2|\omega|/\Gamma^2\in (10^{-3},10^{0})$. The dashed and solid lines represent the numerical results obtained without and with using approximation $\cosh(2\theta)\approx e^{2|\theta|}/2$, respectively. \textbf{b} When $\Gamma\gg |\epsilon_{\bm q}|$ and $\tilde{J}^2|\omega|/\Gamma^2\in (10^{-3},10^{0})$. \textbf{c} When incorporating the precise bosonic damping (\ref{gam=del}) and $\tilde{J}^2|\omega|/\Gamma^2\in (10^{-3},10^{0})$. } 
	\label{sen}
\end{figure}

However, when calculating the fermion self-energy $\Sigma_{\rm J}(i\omega_n,0)$, directly applying the approximation (\ref{vapp}) is unreasonable, as this would lead to a vanishing boson self-energy. Here, we provide a more careful calculation for the fermion self-energy at the VHS.
\begin{equation}
\begin{aligned}	
\Sigma_{\rm J}(i\omega_n,0)=&|J|^2T\sum_m\int \frac{d^2 \bm{q}}{(2\pi)^2}\frac{1}{\bm{q}^2+\tilde{J}^2\frac{|\Omega_m|}{|\epsilon_{\bm q}|}}G_{\rm w}(i\omega_n+i\Omega_m,\bm{q})\\
=& -i|J|^2T\sum_m\int \frac{d^2 \bm{q}}{(2\pi)^2}\frac{1}{\bm{q}^2+\tilde{J}^2\frac{|\Omega_m|}{|\epsilon_{\bm q}|}}\frac{\Gamma}{\epsilon_{\bm q}^2+\Gamma^2}\mathrm{sgn}(\omega_n+\Omega_m)\\
=&\frac{-i|J|^2}{\pi\tilde{J}^2}\mathrm{sgn}(\omega_n)\int \frac{d^2 \bm{q}}{(2\pi)^2}\frac{\Gamma}{\epsilon_{\bm q}^2+\Gamma^2}|\epsilon_{\bm q}|\ln[1+\frac{|\omega_n|\tilde{J}^2}{\bm{q}^2|\epsilon_{\bm q}|}]\\
\approx &\frac{i|J|^2\Gamma}{2\pi^3\tilde{J}^2}\mathrm{sgn}(\omega_n)\int_0^\infty dq^2 \mathrm{Ploylog}[2,-2\frac{|\omega_n|\tilde{J}^2}{q^4}]\frac{q^2}{\Gamma^2+q^4}\\
=&\frac{i|J|^2\Gamma}{2\pi^3\tilde{J}^2}\mathrm{sgn}(\omega_n)\{-\frac{1}{12}\ln(
   \frac{2\tilde{J}^2|\omega_n|}{\Gamma^2} ) [2 \pi^2 + \ln(2)^2 + \ln(\frac{\tilde{J}^2|\omega_n|}{\Gamma^2}) \ln(\frac{4\tilde{J}^2|\omega_n|}{\Gamma^2}) + 
    6 \mathrm{PolyLog}(2, \frac{\Gamma^2}{2\tilde{J}^2|\omega_n|})]\\
    & - \mathrm{PolyLog}(3, \frac{\Gamma^2}{2\tilde{J}^2|\omega_n|})\}\\
\approx &\left\{\begin{matrix}
 -i\frac{|J|^2}{2\pi^3\Gamma}\omega_n\ln\frac{\Gamma^2e^2}{2\tilde{J}^2|\omega_n|} &\tilde{J}^2|\omega_n|\ll \Gamma^2 \\
  -\frac{i|J|^2\Gamma}{24\pi^3\tilde{J}^2}\mathrm{sgn}(\omega_n)\ln(\frac{2\tilde{J}^2|\omega_n|}{\Gamma^2})[\ln^2(\frac{2\tilde{J}^2|\omega_n|}{\Gamma^2})+2\pi^2]&\tilde{J}^2|\omega_n|\gg \Gamma^2
\end{matrix}\right.
\end{aligned},
\label{tem6}
\end{equation}
where we performed the calculations in hyperbolic coordinates and utilized the approximation $\cosh(2\theta)\approx e^{2|\theta|}/2$, the PolyLog denotes the polylogarithm function. Fig.~\ref{sen}a presents numerical results that obtained without using the approximation $\cosh(2\theta)\approx e^{2|\theta|}/2$, yet still exhibit an MFL-type frequency dependence when $\tilde{J}^2|\omega|\ll \Gamma^2$. 

The results of the self-energy calculations in this subsection are self-consistent because, in the limit $\tilde{J}^2|\omega|\ll \Gamma^2$, $|\Sigma_{\rm J}(i\omega_n,0)|\ll \Gamma $, and the full propagator of the electrons is just $G_{\rm w}(i\omega_n,\bm{k})\approx 1/(i\Gamma\mathrm{sgn}(\omega_n)-\epsilon_{\bm k})$. In the clean limit $\tilde{J}^2|\omega|\gg \Gamma^2$, the result in Eq.~(\ref{tem6}) is not self-consistent; the true self-energy should be Eq.~(\ref{sic-fer-5}). In this case, impurity scattering is negligible, and the electronic Green’s function can be written as $G(i\omega_n,\bm{k})\approx 1/(i|J|\mathrm{sgn}(\omega_n)|\omega|^{1/2}-\epsilon_{\bm k})$, which has already been shown to be self-consistent in Fig.~2b of the main text.


We examine the opposite limit $\Gamma\gg |\epsilon_{\bm q}|$, and it is found that when  $\tilde{J}^2|\omega|\ll \Gamma^2$, the fermion self-energy still exhibits the characteristics of an MFL. 
The boson self-energy (\ref{gam=del}) needs to be modified as:
\begin{equation}
\begin{aligned}	
\int \frac{d^2\bm{k}}{4\pi^2}\frac{\Gamma/\pi}{\epsilon_{\bm k}^2+\Gamma^2}\frac{\Gamma/\pi}{\epsilon_{\bm{k}+\bm{q}}^2+\Gamma^2}
\approx \frac{1}{8\pi^3\sqrt{a}\Gamma}\ln\frac{4e\Gamma}{|\epsilon_{\bm q}|}+\mathcal{O}(\frac{|\epsilon_{\bm q}|}{\Gamma^2}).
\end{aligned}
\label{tem5}
\end{equation}
Similarly, the fermion self-energy needs to be modified to
\begin{equation}
\begin{aligned}	
\Sigma_{\rm J}(i\omega_n,0)=&|J|^2T\sum_m\int \frac{d^2 \bm{q}}{(2\pi)^2}\frac{1}{\bm{q}^2+\tilde{J}^2\frac{|\Omega_m|\ln\frac{4e\Gamma}{|\epsilon_{\bm q}|}}{\Gamma}}G_{\rm w}(i\omega_n+i\Omega_m,\bm{q})\\
\approx &\frac{i|J|^2\Gamma^2}{2\pi^3\tilde{J}^2}\mathrm{sgn}(\omega_n)\int_0^\Gamma dq^2 \mathrm{Ploylog}[2,-2\frac{|\omega_n|\tilde{J}^2}{q^2\Gamma}\ln (4e/q^2)]\frac{1}{(\Gamma^2+q^4)\ln (4e/q^2)}\\
\approx &-i\frac{|J|^2}{2\pi^3\Gamma}\omega_n\ln\frac{e^4\Gamma^4}{8\tilde{J}^4|\omega_n|^2} \ \ \ \mathrm{when}\   \tilde{J}^2|\omega|\ll \Gamma^2,
\end{aligned}
\label{tem7}
\end{equation}
and exhibits the characteristics of MFL. The results of the numerical verification are shown in Fig.~\ref{sen}b, where the system indeed exhibits MFL behavior when  $\tilde{J}^2|\omega|\ll \Gamma^2$.

Fig.~\ref{sen}c presents the fermion self-energy obtained by substituting the precise bosonic propagator $\tilde{J}^2\frac{|\Omega|}{|\epsilon_{\bm q}|} f(\frac{\Gamma}{|\epsilon_{\bm q}|})$ (\ref{gam=del}) without any approximations. In the dirty limit $\tilde{J}^2|\omega|\ll \Gamma^2$, it always exhibits the characteristics of MFL. In the clean limit $\tilde{J}^2|\omega|\gg \Gamma^2$, the impurity scattering rate can be neglected, and the system exhibits the NFL behavior described by Eq.~(\ref{sic-fer-5}).\\

\subsection{Spatially disordered Yukawa interactions $J'$}
\label{J'}
In this subsection, we derive the analytic expressions for the MFL
in the low-ferquency limit.
We start with the low-energy effective theory 
 \begin{equation}
		\mathcal{L}=\sum_{m}\psi^\dagger_m(\partial_\tau-\partial_x^2+a\partial_y^2)\psi_m+\sum_l\phi_l^*(\Delta -\partial_\tau^2 - \bm{\nabla}^2 )\phi_l +\frac{1}{\sqrt{N}}\sum_{mn}w_{mn}(\bm{x})\psi_m^\dagger\psi_n+\frac{1}{\sqrt{NN'}}\sum_{mnl}J_{mnl}'(\bm{x})\psi^\dagger_m\psi_n\phi_l+h.c..
	\label{eff2}
\end{equation}
where $m,n=1, 2, . . ., N$ are the flavors of fermion field and $l=1, 2, ..., N'$ is the flavors of the boson field. The space dependent coupling $J_{mnl}'(\bm{x})$ is random in the space of flavor and coordinate with Gaussian distribution:
  \begin{equation}
  	\overline{J_{mnl}'(\bm{x})}=0,\ \ \ \  \overline{J_{mnl}'(\bm{x})J^{'*}_{m'n'l'}(\bm{x}')}=|J'|^2\delta_{mm'}\delta_{nn'}\delta_{ll'}\delta(\bm{x}-\bm{x}').
  \end{equation}
  The space dependent potential disorder $w_{mn}$ is random in flavor and coordinate spaces obeying the Gaussian distribution:
  \begin{equation}
  	\overline{w_{mn}(\bm{x})}=0,\ \ \ \  \overline{w_{mn}(\bm{x})w^*_{m'n'}(\bm{x}')}=|w|^2\delta_{mm'}\delta_{nn'}\delta(\bm{x}-\bm{x}').
  \end{equation}
  
The MFL are induced near the VHS with spatially random Yukawa interactions $J'$.
In the following parts, we first consider the case without potential disorder and then examine the case with it.

\subsubsection{potential disorder $|w|=0$}
The boson self-energy correction $\Pi_{\rm J'}$ from the fermionic particle-hole excitation in the vicinity of the VHS saddle points is given by,
\begin{equation}
	\begin{aligned}
		\Pi_{\rm J'}(i\Omega_m)-\Pi_{\rm J'}(0)&=-|J'|^2\frac{N}{N'}T\sum_n \int \frac{d^2\bm{q}}{(2\pi)^2}\frac{1}{i\omega_n-q_x^2+a q_y^2-\Sigma(i\omega_n)} \\
		&\ \quad \times\int \frac{d^2\bm{k}}{(2\pi)^2}[\frac{1}{i\omega_n+i\Omega_m-k_x^2+a k_y^2-\Sigma(i\omega_n+i\Omega_m)}-\frac{1}{i\omega_n-k_x^2+a k_y^2-\Sigma(i\omega_n)}]\\
		&= -\frac{|J'|^2\Lambda_\theta^2N}{4a\pi^3N'}|\Omega_m|,
	\end{aligned}
	\label{sdc-bos}
\end{equation}
Similar to previous section, we have $\Pi_{\rm J'}(0)=\Delta$ at zero temperature. For convenience, we set $\tilde{J}'^2\equiv |J'|^2N(\Lambda_\theta)^2/4a\pi^3 N'$. The decoupling of momentum integrals in ${\bm q},{\bm k}$ reduces the computational complexity.\\

The fermion self-energy is defined as,
\begin{equation}
\begin{aligned}
	\Sigma_{\rm J'}(i\omega_n)=&|J'|^2T\sum_m \int \frac{d^2 \bm{k}}{(2\pi)^2}\frac{1}{\bm{k}^2+\tilde{J}'^2|\Omega_m|}\int \frac{d^2\bm{q}}{(2\pi)^2}\frac{1}{i\omega_n+i\Omega_m-q_x^2+a q_y^2-\Sigma(i\omega_n+i\Omega_m)}.
		\label{sdc-fer-1}
\end{aligned}
\end{equation}
The self-energy correction is derived as,
\begin{equation}
	\Sigma_{\rm J'}(i\omega_n)-\Sigma_{\rm J'}(0)=-i\frac{|J'|^2\Lambda_\theta}{8\pi^3\sqrt{a}}\omega_n\ln(\frac{e\Lambda_{\rm U}^2}{\tilde{J}'^2|\omega_n|}),
	\label{sdc-fer-2}
\end{equation}
where $\Lambda_{\rm U}$ is nothing but the UV cutoff introduced for the first momentum integral in Eq.~(\ref{sdc-fer-1}).
We remark that the analytical calculation is within the framework of self-consistent Elishberg theory.
And, the result in Eq.~(\ref{sdc-fer-2}) corresponds to the famous phenomenological theory known as the MFL.

\subsubsection{potential disorder $|w|\ne 0$}
In this subsection, we consider the effects of potential disorder. The integrals in Eqs.~(\ref{sdc-bos}) and (\ref{sdc-fer-1}) do not depend on the specific form of the self-energy $\Sigma$. Therefore, even if $\Sigma$ is replaced with the disorder self-energy $\Sigma_{\rm w}$ (\ref{pose}), the resulting outcome remains the same as Eq.~(\ref{sdc-fer-2}).

\newpage
\section{The specific heat}
In this section, we derive the temperature dependence of the bosonic contribution
to the specific heat. We show that impurity scattering cuts off the singular
momentum dependence of the Landau damping, thereby driving the low-energy
bosonic dynamics into an effectively local regime. Consequently, the singular
clean-limit behavior, $C_v^{\rm boson}\sim \sqrt{T}$, is replaced in the dirty
limit by a logarithmically enhanced linear-in-$T$ specific heat,
$C_v^{\rm boson}\sim T\ln T$.

We first consider the renormalization of the boson dynamics due to its
coupling to the fermions. After integrating out the fermions, the effective
Gaussian action for the bosonic mode takes the form
\begin{equation}
    S_{\phi}
    =
    T\sum_{m}
    \int\frac{d^2{\bm q}}{(2\pi)^2}\,
    \phi^*(i\Omega_m,{\bm q})
    D^{-1}(i\Omega_m,{\bm q})
    \phi(i\Omega_m,{\bm q}) ,
\end{equation}
where $D(i\Omega_m,{\bm q})$ is the renormalized boson propagator. At the
quantum critical point, we write
\begin{equation}
    D^{-1}(i\Omega_m,{\bm q})
    =
    {\bm q}^2-\Pi(i\Omega_m,{\bm q}),
\end{equation}
with $\Pi$ the fermion-induced boson self-energy.
The corresponding contribution to the free-energy density is
\begin{equation}
    F
    =
    T\sum_m
    \int\frac{d^2{\bm q}}{(2\pi)^2}
    \ln D^{-1}(i\Omega_m,{\bm q}) .
\end{equation}
Using the standard contour representation of the Matsubara sum, this can be
written as
\begin{equation}
    F
    =
    \int_0^\infty \frac{d\omega}{\pi}
    \int\frac{d^2{\bm q}}{(2\pi)^2}
    \coth\frac{\omega}{2T}\,
    \tan^{-1}
    \left[
    \frac{{\rm Im}\Pi^R(\omega,{\bm q})}{{\bm q}^2}
    \right] .
\end{equation}

For the clean system, the imaginary part of the retarded boson self-energy is
\begin{equation}
    {\rm Im}\Pi^R(\omega,{\bm q})
    =
    \frac{\tilde J^2 \omega}{|\epsilon_{\bm q}|}.
\end{equation}
Substituting this expression into the bosonic free-energy density, the specific
heat is obtained as
\begin{equation}
    C_v
    =
    -T\frac{\partial^2 F}{\partial T^2}
    =
    T\int_0^\infty \frac{d\omega}{\pi}
    \int\frac{d^2{\bm q}}{(2\pi)^2}
    \left[
    \frac{\partial^2}{\partial T^2}
    \coth\left(\frac{\omega}{2T}\right)
    \right]
    \tan^{-1}
    \left[
    \frac{{\rm Im}\Pi^R(\omega,{\bm q})}{{\bm q}^2}
    \right] .
\end{equation}
Explicitly,
\begin{equation}
    \frac{\partial^2}{\partial T^2}
    \coth\left(\frac{\omega}{2T}\right)
    =
    \frac{\omega}{T^3}
    {\rm csch}^2\left(\frac{\omega}{2T}\right)
    \left[
    \frac{\omega}{2T}
    \coth\left(\frac{\omega}{2T}\right)
    -1
    \right] .
\end{equation}
Therefore,
\begin{equation}
    C_v
    \sim 
    \frac{1}{T^2}\int_0^{T}d\omega \omega
    \int d^2{\bm q}
    \tan^{-1}
    \left[
    \frac{\tilde J^2\omega}
    {{\bm q}^2|\epsilon_{\bm q}|}
    \right]\sim \tilde{J} \sqrt{T} .
    \label{ssh1}
\end{equation}
  
In the dirty limit, the fermion-induced damping of the bosonic mode is modified
to
\begin{equation}
    {\rm Im}\Pi^R(\omega,{\bm q})
    =
    \frac{\tilde J^2\omega}{|\epsilon_{\bm q}|}
    f\left(\frac{\Gamma}{|\epsilon_{\bm q}|}\right),
\end{equation}
Keeping only the scaling form, the bosonic contribution to the specific heat
can be written as
\begin{equation}
\begin{aligned}
    C_v
   & \sim
    \frac{1}{T^2}
    \int_0^T d\omega\, \omega
    \int_0^\infty d\theta
    \left\{
    \int_0^\Gamma d(q^2)\,
    \tan^{-1}
    \left[
    \frac{\tilde J^2\omega}
    {q^2\Gamma\cosh(2\theta)}
    \right]
    +
    \int_\Gamma^\infty d(q^2)\,
    \tan^{-1}
    \left[
    \frac{\tilde J^2\omega}
    {q^4\cosh(2\theta)}
    \right]
    \right\}\\
    & \sim \frac{\tilde{J}^2}{\Gamma^2 T^2}
    \int_0^T d\omega\, \omega^2 \ln \frac{\Gamma^2}{\tilde{J}^2\omega}\sim \frac{\tilde{J}^2T}{\Gamma^2}
    \ln \frac{\Gamma^2}{\tilde{J}^2T},
    \end{aligned}
    \label{ssh2}
\end{equation}
where we have assumed $\tilde{J}^2T\ll \Gamma^2$. This result is consistent with the connection established in previous studies
between the marginal-Fermi-liquid behavior and the electronic specific heat
$C_v\sim T\ln T$. In the impurity-dominated regime, the bosonic damping becomes
effectively independent of momentum, because impurity scattering cuts off the
singular momentum dependence inherited from the clean fermionic spectrum.
Consequently, the bosonic fluctuations acquire a local, marginal form, which
naturally gives rise to the logarithmically enhanced linear specific heat found
above.

The specific heat of the full system contains several contributions: the bare electronic background, the bare bosonic background, and the renormalization induced by the electron-boson coupling. In principle, the interaction-induced contribution can be evaluated either from the fermionic self-energy or from the bosonic self-energy. The resulting temperature dependences should be identical, since the two approaches describe the same interaction-induced correction to the thermodynamics. In our analysis, we adopt the bosonic perspective mainly for convenience, as the singular low-energy contribution is most transparently captured by the renormalized bosonic propagator.
From the fermionic perspective, the interaction contribution to the free energy can be written as
\begin{equation}
F=
-\int_0^\infty \frac{d\omega}{\pi}
\int\frac{d^2{\bm k}}{(2\pi)^2}
\tanh\left(\frac{\omega}{2T}\right)
\tan^{-1}\left[
\frac{{\rm Im}\Sigma^R(\omega)}
{\epsilon_{\bm k}+{\rm Re}\Sigma^R(\omega)}
\right].	
\end{equation}
The corresponding specific heat therefore scales as
\begin{equation}
C_v\sim \frac{1}{T^2}\int_0^T d\omega\,\omega
\int d^2{\bm k}\,
\tan^{-1}\left[
\frac{{\rm Im}\Sigma^R(\omega)}
{\epsilon_{\bm k}+{\rm Re}\Sigma^R(\omega)}
\right]\sim \frac{1}{T^2}\int_0^T d\omega\,\omega\,
{\rm Re}\Sigma^R(\omega),
\end{equation}
which leads to the same conclusions as Eqs. \eqref{ssh1} and \eqref{ssh2}.

\newpage
\section{Transport property with potential disorder $|w|=0$}\label{v0con}

In this section, we use the Kubo formula to calculate the optical conductivity and resistivity. The real part of the conductivity can be expressed as
\begin{equation}
	\mathrm{Re}[\sigma (\Omega,T)]=-\frac{\mathrm{Im}\Xi^{\rm R}(\Omega,T)}{\Omega},
\label{Re_OC}
\end{equation}
where $\Xi^{\rm R}(\Omega,T)$ is the retarded dynamic current-current correction function. The current can be obtained from the low energy effective theory through minimal coupling:
\begin{equation}
	(j_{\bm{k},x},j_{\bm{k},y})\equiv (\partial H/\partial A_x,\partial H/\partial A_y)|_{\bm{A}=0} =\sum_{\bm{p},m} (2p_x+k_x,-2ap_y-ak_y)\psi^\dagger_{\bm{p}+\bm{k},m}\psi_{\bm{p},m}\equiv \sum_{\bm{p},m} v_{{\bm p}+{\bm k}/2}\psi^\dagger_{\bm{p}+\bm{k},m}\psi_{\bm{p},m}.
\end{equation}

In the following subsections, we first calculate the optical conductivity at zero temperature.
Then, we invoke the spectral representation to calculate the optical conductivity at finite temperatures, i.e. $|\Omega| \ll T$.

\subsection{$\sigma^{\rm J}(\Omega,T)$ from spatially uniform interaction}
\begin{figure}[htbp]
		\centering
		\includegraphics[width=0.7\textwidth]{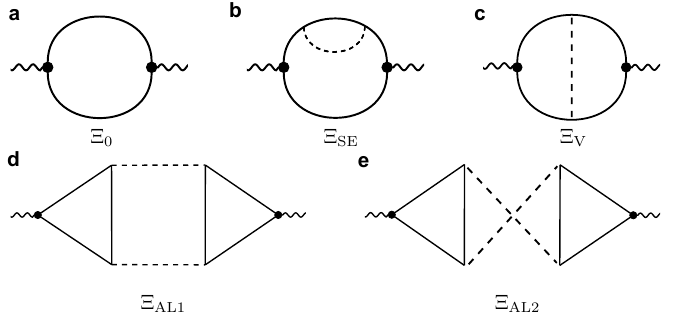}
		\caption{\textbf{Diagrams contributing to conductivity in the limit of large Fermi energy.} Lines represent fermion propagators, wave lines represent current operators, dashed lines represent full boson propagators, dots represent group velocity. We use $\Xi_0$ to represent the zeroth-order diagram \textbf{a}, $\Xi_{\rm SE}$ for the self-energy diagram \textbf{b}, $\Xi_{\rm V}$ for the vertex diagram \textbf{c}, and $\Xi_{\rm AL1}$ and $\Xi_{\rm AL2}$ for the Aslamazov-Larkin diagrams \textbf{d} and \textbf{e}, respectively.} 
		\label{figs2}
\end{figure}

We first consider the contribution to the conductivity arising from the interaction $J$, where the bosons are Landau-damped as described by Eq.(\ref{sic-bos-1}).
All electron-boson interaction vertices in Fig.~\ref{figs2} are from spatially uniform interaction $J$ and all calculations in this section are performed in the static limit. To put the conclusion first, only Feynman diagrams in Fig.~\ref{figs2}b-e contribute to the optical conductivity at $o(N|J|^2)$. \\  

The one-loop diagram shown in Fig.~\ref{figs2}a has no contribution to the optical conductivity in the low-frequency limit. We define $G^0(i\omega_n,\bm{k})=1/(i\omega_n-\epsilon_{\bm k})$ for the bare electron propagator in Fermi liquid.
The subscript is to distinguish between the electron propagators in the opposite limit $\Gamma\gg |\omega|$ in the next section.\\

\subsubsection{Derivation on the $T=0$ optical conductivity} 
\label{sigmaJ0}

The self-energy diagram in Fig.~\ref{figs2}b gives us
\begin{equation}
\begin{aligned}
	\Xi_{\rm SE,J}(i\Omega_m,0)=&-NT\sum_n\int \frac{d^2\bm{k}}{(2\pi)^2} v_{\bm k}^2 G ^0(i\omega_n,\bm{k})[G ^0(i\Omega_m+i\omega_n,\bm{k})]^2\Sigma_{\rm J}(i\omega_n+i\Omega_m,\bm{k})+(\Omega_m\rightarrow -\Omega_m)\\
	=& \frac{N}{i\Omega_m}T\sum_n\int \frac{d^2\bm{k}}{(2\pi)^2} v_{\bm k}^2 G ^0(i\omega_n,\bm{k})G ^0(i\Omega_m+i\omega_n,\bm{k})[\Sigma_{\rm J}(i\omega_n,\bm{k})-\Sigma_{\rm J}(i\omega_n+i\Omega_m,\bm{k})].
	\label{sic-sed}
	\end{aligned}
\end{equation}
where, in the second term, we have used transformation $\omega_n\rightarrow \omega_n+\Omega_m$ and the following relation:
\begin{equation}
	G^0(i\omega_n+i\Omega_m,\bm{k})G ^0(i\omega_n,\bm{k})=\frac{G ^0(i\omega_n,\bm{k})-G ^0(i\omega_n+i\Omega_m,\bm{k})}{i\Omega_m}.
	\label{con-for}
\end{equation}
This can be viewed as a bare version of the so-called Ward identity, which will be frequently used below.\\

The vertex diagram in Fig.~\ref{figs2}c gives us
\begin{equation}
\begin{aligned}
	\Xi_{\rm V,J}(i\Omega_m,0)=&-N|J|^2T\sum_l T\sum_n\int \frac{d^2\bm{k_2}}{(2\pi)^2} \int \frac{d^2\bm{k_1}}{(2\pi)^2} v_{{\bm k}_1}v_{{\bm k}_1+{\bm k}_2} G ^0(i\omega_n,\bm{k_1})G ^0(i\Omega_m+i\omega_n,\bm{k_1})D(i\Omega_l,\bm{k_2})\\
	&\times G ^0(i\omega_n+i\Omega_l+i\Omega_m,\bm{k_1}+\bm{k_2})G ^0(i\omega_n+i\Omega_l,\bm{k_1}+\bm{k_2})f^2_{\bm{k}_1+\bm{k}_2/2}\\
	=&-|J|^2NT\sum_l T\sum_n\int \frac{d^2\bm{k_2}}{(2\pi)^2} \int \frac{d^2\bm{k_1}}{(2\pi)^2} v_{{\bm k}_1}^2 G ^0(i\omega_n,\bm{k_1})G ^0(i\Omega_m+i\omega_n,\bm{k_1})D(i\Omega_l,\bm{k_2})\\
	&\quad \times G ^0(i\omega_n+i\Omega_l+i\Omega_m,\bm{k_1}+\bm{k_2})G ^0(i\omega_n+i\Omega_l,\bm{k_1}+\bm{k_2})f^2_{\bm{k}_1+\bm{k}_2/2}\\
	&-|J|^2NT\sum_l T\sum_n\int \frac{d^2\bm{k_2}}{(2\pi)^2} \int \frac{d^2\bm{k_1}}{(2\pi)^2} v_{{\bm k}_1}(v_{{\bm k}_1+{\bm{k}_2}}-v_{{\bm k}_1}) G ^0(i\omega_n,\bm{k_1})G ^0(i\Omega_m+i\omega_n,\bm{k_1})D(i\Omega_l,\bm{k_2})\\
	&\quad \times G ^0(i\omega_n+i\Omega_l+i\Omega_m,\bm{k_1}+\bm{k_2})G ^0(i\omega_n+i\Omega_l,\bm{k_1}+\bm{k_2})f^2_{\bm{k}_1+\bm{k}_2/2},\\
	\end{aligned}
	\label{XiV}
\end{equation}
 where $f^2_{\bm{k}_1+\bm{k}_2/2}$ is the form factor of the Yukawa interaction, which is set to 1 in this article.

By applying the Ward identity in Eq.~(\ref{con-for}), 
we note that the first term in Eq.~(\ref{XiV}) cancels exactly with Eq.~(\ref{sic-sed}) from
the self-energy diagram and the remaining terms are transformed as,
\begin{equation}
	\begin{aligned}
		\Xi_{\rm SE+V,J}(i\Omega_m,0)=&\frac{N|J|^2}{\Omega_m^2}T\sum_{\Omega_1}T\sum_{\omega}\int \frac{d^2\bm{k_2}}{(2\pi)^2} \int \frac{d^2\bm{k_1}}{(2\pi)^2} v_{{\bm k}_1}(v_{{\bm k}_1+{\bm{k}_2}}-v_{{\bm k}_1}) D(i\Omega_1,\bm{k_2})f^2_{\bm{k}_1+\bm{k}_2/2}[G ^0(i\omega,\bm{k_1})\\
	&-G ^0(i\Omega_m+i\omega,\bm{k_1})] [G ^0(i\Omega_1+i\omega,\bm{k_1}+\bm{k_2})-G ^0(i\Omega_1+i\omega+i\Omega_m,\bm{k_1}+\bm{k_2})].
	\end{aligned}
	\label{SE+V}
\end{equation}
It is worth noting that the result obtained so far in Eq.(\ref{SE+V}) is exact.
Namely, the above cancellation between the self energy and vertex diagram
occurs irrespective of specific form of the fermion and boson Green functions. 
To proceed further, we make approximation to facilitate the analytic calculations.
We consider the static limit which leads to $G ^0(i\omega+i\Omega_m,\bm{k_1})-G ^0(i\omega-i\Omega_m,\bm{k_1})\simeq -i\pi [\mathrm{sgn}(\omega+\Omega_m)-\mathrm{sgn}(\omega-\Omega_m)]\delta(\epsilon_{\bm{k}_1}) $, also known as the Prange-Kadanoff approximation \cite{PhysRev.134.A566}. We consider the bosonic propagators to be full propagators, i.e., $D(i\Omega,\bm{p})=1/(\bm{p}^2+\tilde{J}^2|\Omega_m|/|\epsilon_{\bm p}|)$. Substituting this into Eq.~(\ref{SE+V}), we get
\begin{equation}
	\begin{aligned}
		\Xi_{\rm SE+V,J}(i\Omega_m,0)(i\Omega_m,0)=&-\frac{N|J|^2\mathrm{sgn}(\Omega_m)}{2\Omega_m^2}\int d\Omega \int^{|\Omega_m|/2}_{-|\Omega_m|/2} d\omega \int \frac{d^2\bm{k_2}}{(2\pi)^2} \int \frac{d^2\bm{k_1}}{(2\pi)^2} 4k_{1x}k_{2x}D(i\Omega_1,\bm{k_2})\delta(\epsilon_{\bm{k}_1})\delta(\epsilon_{\bm{k}_1+\bm{k}_2})\\
		&\times [\mathrm{sgn}(\omega+\Omega+\Omega_m/2)-\mathrm{sgn}(\omega+\Omega-\Omega_m/2)]	\\
		=&\frac{|J|^2N\mathrm{sgn}(\Omega_m)}{4\pi^2\sqrt{a}\Omega_m^2}\int d\Omega \int_{-|\Omega_m|/2+\Omega}^{|\Omega_m|/2+\Omega}d\omega [\mathrm{sgn}(\omega+\Omega_m/2)-\mathrm{sgn}(\omega-\Omega_m/2)]
 \int \frac{d^2\bm{p}}{(2\pi)^2}\frac{p_x^2}{|\epsilon_{\bm p}|}D(i\Omega,\bm{p}) \\
 \approx &\frac{|J|^2N \Lambda_\theta}{4\pi^4\sqrt{a}\Omega_m^2}\int_0^{|\Omega_m|} d\Omega (|\Omega_m|-\Omega)\big[\ln \frac{\Lambda_{\rm U}^4\cosh(2\Lambda_\theta)}{\tilde{J}^2|\Omega|}-\Lambda_\theta\big]\\
 =&\frac{|J|^2N|\Lambda_\theta}{8\pi^4\sqrt{a}}[\ln \frac{e^{3/2}\Lambda_{\rm U}^4\cosh(2\Lambda_\theta)}{\tilde{J}^2|\Omega_m|}-\Lambda_\theta]
		\end{aligned}
	\label{xiJ}
\end{equation}

Next, we turn to the calculation of the three-loop Aslamazov-Larkin (AL) diagrams in Fig.~\ref{figs2}d and Fig.~\ref{figs2}e, which reads
\begin{equation}
	\begin{aligned}
		 \Xi_{\rm AL_1,J}+\Xi_{\rm AL_2, J}=&|J|^4\frac{N^2}{N'}T\sum_{\omega_1}T\sum_{\omega_2}T\sum_{\Omega}\int \frac{d^2\bm{k}_1}{(2\pi)^2}\int \frac{d^2\bm{k}_2}{(2\pi)^2}\int \frac{d^2\bm{p}}{(2\pi)^2}v_{\bm{k}_1}v_{\bm{k}_2}G ^0(i\omega_1-i\Omega_m/2,\bm{k}_1)G ^0(i\omega_1+i\Omega_m/2,\bm{k}_1)\\
 &\times G ^0(i\omega_1+i\Omega,\bm{k}_1+\bm{p})f^2_{\bm{k}_1+\bm{p}/2}D(i\Omega-i\Omega_m/2,\bm{p})D(i\Omega+i\Omega_m/2,\bm{p})G ^0(i\omega_2-i\Omega_m/2,\bm{k}_2)\\
 &\times G ^0(i\omega_2+i\Omega_m/2,\bm{k}_2)[G ^0(i\omega_2+i\Omega,\bm{k}_2+\bm{p})f^2_{\bm{k}_2+\bm{p}/2}+G ^0(i\omega_2-i\Omega,\bm{k}_2-\bm{p})f^2_{\bm{k}_2-\bm{p}/2}].\\
	\end{aligned}
\end{equation}
Using equation
\begin{equation}
	\begin{aligned}
		&D(i\Omega-i\Omega_m/2,\bm{p})D(i\Omega+i\Omega_m/2,\bm{p})\\
		=&\frac{D(i\Omega-i\Omega_m/2,\bm{p})-D(i\Omega+i\Omega_m/2,\bm{p})}{D^{-1}(i\Omega+i\Omega_m/2,\bm{p})-D^{-1}(i\Omega-i\Omega_m/2,\bm{p})}\\
		=&\frac{D(i\Omega-i\Omega_m/2,\bm{p})-D(i\Omega+i\Omega_m/2,\bm{p})}{\Pi_{\rm J}(i\Omega-i\Omega_m/2,\bm{p})-\Pi_{\rm J}(i\Omega+i\Omega_m/2,\bm{p})}\\
		=&-\frac{N'}{|J|^2N}\frac{D(i\Omega-i\Omega_m/2,\bm{p})-D(i\Omega+i\Omega_m/2,\bm{p})}{T\sum_\omega \int \frac{d^2\bm{k}}{(2\pi)^2}G ^0(i\omega-i\Omega,\bm{k}-\bm{p})f^2_{\bm{k}-\bm{p}/2}[G ^0(i\omega-i\Omega_m/2,\bm{k})-G ^0(i\omega+i\Omega_m/2,\bm{k})]},
	\end{aligned}
	\label{DD=D-D}
\end{equation}
we can further simplify the current-current correlation corresponding to the AL diagrams as:
\begin{equation}
	\begin{aligned}
		 \Xi_{\rm AL_1,J}+\Xi_{\rm AL_2, J}(i\Omega_m)=&\frac{|J|^2N}{i\Omega_m} T\sum_{\omega_1}T\sum_{\Omega}\int \frac{d^2\bm{k}_1}{(2\pi)^2}\int \frac{d^2\bm{p}}{(2\pi)^2}v_{\bm{k}_1}G ^0(i\omega_1-i\Omega_m/2,\bm{k}_1)G ^0(i\omega_1+i\Omega_m/2,\bm{k}_1)\\
 &\times G ^0(i\omega_1+i\Omega,\bm{k}_1+\bm{p})f^2_{\bm{k}_1+\bm{p}/2}[D(i\Omega-i\Omega_m/2,\bm{p})-D(i\Omega+i\Omega_m/2,\bm{p})]\\
 &\times [\bar{v}_p(i\Omega_m)-\bar{v}_{-p}(i\Omega_m)].\\
	\end{aligned}
	\label{AL}
\end{equation}
Here, the average group velocity $\bar{v}$ is defined as:
\begin{equation}
	\bar{v}_{p}(i\Omega_m)\equiv \frac{T\sum_\omega \int \frac{d^2\bm{k}}{(2\pi)^2}G ^0(i\omega+i\Omega,\bm{k}+\bm{p})f^2_{\bm{k}+\bm{p}/2}[G ^0(i\omega-i\Omega_m/2,\bm{k})-G ^0(i\omega+i\Omega_m/2,\bm{k})]v_{\bm k}}{T\sum_\omega \int \frac{d^2\bm{k}}{(2\pi)^2}G ^0(i\omega+i\Omega,\bm{k}+\bm{p})f^2_{\bm{k}+\bm{p}/2}[G ^0(i\omega-i\Omega_m/2,\bm{k})-G ^0(i\omega+i\Omega_m/2,\bm{k})]}.
	\label{agv}
\end{equation}
It is worth noting that the result obtained so far in Eq.~(\ref{AL}) is exact. Specifically, for a constant group velocity, the contributions of both the AL diagrams and the SE+V diagrams vanish, respectively.

Collecting terms from the SE+V diagrams in Eq.~(\ref{SE+V}), the total contribution to the conductivity is obtained as:
\begin{equation}
	\begin{aligned}
		 \Xi_{\rm AL+SE+V,J}(i\Omega_m)=&-\frac{|J|^2N}{\Omega_m^2} T\sum_{\omega_1}T\sum_{\Omega}\int \frac{d^2\bm{k}_1}{(2\pi)^2}\int \frac{d^2\bm{p}}{(2\pi)^2}v_{1x}\overline{\Delta v}[G ^0(i\omega_1-i\Omega_m/2,\bm{k}_1)-G ^0(i\omega_1+i\Omega_m/2,\bm{k}_1)]\\
 &\times G ^0(i\omega_1+i\Omega,\bm{k}_1+\bm{p})f^2_{\bm{k}_1+\bm{p}/2}[D(i\Omega-i\Omega_m/2,\bm{p})-D(i\Omega+i\Omega_m/2,\bm{p})].\\
	\end{aligned}
	\label{SE+V+AL}
\end{equation}
We will base all subsequent analytical calculations of the conductivity on this expression. However, in order to facilitate comparison with the Boltzmann equation, we proceed to further simplify it. Here, we define
\begin{equation}
	\begin{aligned}
	&\overline{\Delta v}\\
	\equiv & \bar{v}_{p}(i\Omega_m)-\bar{v}_{-p}(i\Omega_m)+v_{\bm{k}_1+\bm{p}}-v_{\bm{k}_1}\\
	=&\frac{T\sum_{\omega_2} \int \frac{d^2{\bm k }_2}{(2\pi)^2}G^0(i\omega_2+i\Omega,{\bm k}_2+\bm{p})f^2_{{\bm k}_2+\bm{p}/2}[G ^0(i\omega_2-i\Omega_m/2,{\bm k}_2)-G^0(i\omega_2+i\Omega_m/2,{\bm k}_2)](v_{{\bm k}_2}+v_{\bm{k}_1+\bm{p}}-v_{\bm{k}_1})}{T\sum_{\omega_2} \int \frac{d^2{\bm k}_2}{(2\pi)^2}G ^0(i\omega_2+i\Omega,{\bm k}_2+\bm{p})
	f^2_{{\bm k}_2+\bm{p}/2}[G^0(i\omega_2-i\Omega_m/2,{\bm k}_2)-G^0(i\omega_2+i\Omega_m/2,{\bm k}_2)]}\\
	&+\frac{T\sum_{\omega_2} \int \frac{d^2{\bm k }_2}{(2\pi)^2}G^0(i\omega_2-i\Omega,{\bm k}_2-\bm{p})f^2_{{\bm k}_2-\bm{p}/2}[G ^0(i\omega_2-i\Omega_m/2,{\bm k}_2)-G^0(i\omega_2+i\Omega_m/2,{\bm k}_2)]v_{{\bm k}_2}}{T\sum_{\omega_2} \int \frac{d^2{\bm k}_2}{(2\pi)^2}G ^0(i\omega_2-i\Omega,{\bm k}_2-\bm{p})
	f^2_{{\bm k}_2-\bm{p}/2}[G^0(i\omega_2-i\Omega_m/2,{\bm k}_2)-G^0(i\omega_2+i\Omega_m/2,{\bm k}_2)]}\\
	=&\frac{T\sum_{\omega_2} \int \frac{d^2{\bm k }_2}{(2\pi)^2}G^0(i\omega_2+i\Omega,{\bm k}_2+\bm{p})f^2_{{\bm k}_2+\bm{p}/2}[G ^0(i\omega_2-i\Omega_m/2,{\bm k}_2)-G^0(i\omega_2+i\Omega_m/2,{\bm k}_2)](v_{{\bm k}_2}-v_{{\bm k}_2+{\bm p}}+v_{\bm{k}_1+\bm{p}}-v_{\bm{k}_1})}{T\sum_{\omega_2} \int \frac{d^2{\bm k}_2}{(2\pi)^2}G ^0(i\omega_2+i\Omega,{\bm k}_2+\bm{p})
	f^2_{{\bm k}_2+\bm{p}/2}[G^0(i\omega_2-i\Omega_m/2,{\bm k}_2)-G^0(i\omega_2+i\Omega_m/2,{\bm k}_2)]}\\
	=&\frac{T\sum_{\omega_2} \int \frac{d^2{\bm k }_2}{(2\pi)^2}G^0(i\omega_2,{\bm k}_2)f^2_{{\bm k}_2-\bm{p}/2}[G^0(i\omega_2-i\Omega-i\Omega_m/2,{\bm k}_2-\bm{p})-G^0(i\omega_2-i\Omega+i\Omega_m/2,{\bm k}_2-{\bm p})]\Delta v}{T\sum_{\omega_2} \int \frac{d^2{\bm k }_2}{(2\pi)^2}G^0(i\omega_2,{\bm k}_2)f^2_{{\bm k}_2-\bm{p}/2}[G^0(i\omega_2-i\Omega-i\Omega_m/2,{\bm k}_2-\bm{p})-G^0(i\omega_2-i\Omega+i\Omega_m/2,{\bm k}_2-{\bm p})]}
	\label{Delv}
	\end{aligned}
\end{equation}
where $\Delta v=v_{{\bm k}_2-{\bm p}}-v_{{\bm k}_2}+v_{\bm{k}_1+\bm{p}}-v_{\bm{k}_1}$. The third equality is obtained by performing the transformation $(\omega_2,{\bm k}_2)\rightarrow (\omega_2+i\Omega\pm i\Omega_m/2,{\bm k}_2+\bm{p})$ on the second expression.
Using Eqs.~(\ref{DD=D-D}) and (\ref{Delv}), Eq.~(\ref{SE+V+AL}) can be simplified as:
\begin{equation}
	\begin{aligned}
		 \Xi_{\rm AL+SE+V,J}=&-\frac{|J|^4N^2}{N'\Omega_m^2} T\sum_{\omega_2}T\sum_{\omega_1}T\sum_{\Omega}\int \frac{d^2{\bm k }_2}{(2\pi)^2}\int \frac{d^2\bm{k}_1}{(2\pi)^2}\int \frac{d^2\bm{p}}{(2\pi)^2} v_{\bm{k}_1}[G ^0(i\omega_1-i\Omega_m/2,\bm{k}_1)-G ^0(i\omega_1+i\Omega_m/2,\bm{k}_1)]\\
 &\times G^0(i\omega_1+i\Omega,\bm{k}_1+\bm{p})f^2_{\bm{k}_1+\bm{p}/2}D(i\Omega-i\Omega_m/2,\bm{p})D(i\Omega+i\Omega_m/2,\bm{p})G^0(i\omega_2,{\bm k}_2)\\
 &\times f^2_{{\bm k}_2-\bm{p}/2}[G^0(i\omega_2-i\Omega-i\Omega_m/2,{\bm k}_2-\bm{p})-G^0(i\omega_2-i\Omega+i\Omega_m/2,{\bm k}_2-{\bm p})]\Delta v.\\
	\end{aligned}
\end{equation}
After the change of variables $(\omega_1,{\bm k}_1)\rightarrow (\omega_1-\Omega/2,{\bm k}_1-{\bm p}/2)$ and $(\omega_2,{\bm k}_2)\rightarrow (\omega_2+\Omega/2,{\bm k}_2+{\bm p}/2)$, the above expression can be rewritten as:
\begin{equation}
	\begin{aligned}
		 \Xi_{\rm AL+SE+V,J}=&-\frac{|J|^4N^2}{N'\Omega_m^2} T\sum_{\omega_2}T\sum_{\omega_1}T\sum_{\Omega}\int \frac{d^2{\bm k }_2}{(2\pi)^2}\int \frac{d^2\bm{k}_1}{(2\pi)^2}\int \frac{d^2\bm{p}}{(2\pi)^2} v_{{\bm k}_1-{\bm p}/2}G^0(i\omega_1+i\Omega/2,\bm{k}_1+\bm{p}/2)f^2_{\bm{k}_1}\\
 &\times [G ^0(i\omega_1-i\Omega/2-i\Omega_m/2,\bm{k}_1-{\bm p}/2)-G ^0(i\omega_1-i\Omega/2+i\Omega_m/2,\bm{k}_1-{\bm p}/2)]D(i\Omega-i\Omega_m/2,\bm{p})\\
 &\times D(i\Omega+i\Omega_m/2,\bm{p})G^0(i\omega_2+i\Omega/2,{\bm k}_2+{\bm p}/2)f^2_{{\bm k}_2}[G^0(i\omega_2-i\Omega/2-i\Omega_m/2,{\bm k}_2-\bm{p}/2)\\
 &-G^0(i\omega_2-i\Omega/2+i\Omega_m/2,{\bm k}_2-{\bm p}/2)](v_{{\bm k}_2-{\bm p}/2}-v_{{\bm k}_2+{\bm p}/2}+v_{\bm{k}_1+\bm{p}/2}-v_{\bm{k}_1-{\bm p}/2}).\\
	\end{aligned}
	\label{bolz1}
\end{equation}
Noting that
\begin{equation}
\begin{aligned}
&T\sum_{\omega_1}G^0(i\omega_1+i\Omega/2,\bm{k}_1+\bm{p}/2)f^2_{\bm{k}_1}	[G ^0(i\omega_1-i\Omega/2-i\Omega_m/2,\bm{k}_1-{\bm p}/2)-G ^0(i\omega_1-i\Omega/2+i\Omega_m/2,\bm{k}_1-{\bm p}/2)]\\
=& T\sum_{\omega_1}[G^0(i\omega_1+i\Omega/2+i\Omega_m/2,\bm{k}_1+\bm{p}/2)-G^0(i\omega_1+i\Omega/2-i\Omega_m/2,\bm{k}_1+\bm{p}/2)]f^2_{\bm{k}_1}G ^0(i\omega_1-i\Omega/2,\bm{k}_1-{\bm p}/2),
\end{aligned}
\end{equation}
we then perform the variable substitution $(\Omega,{\bm p})\rightarrow -(\Omega,{\bm p})$ (The boson Green’s function is a function of the absolute value of momentum or frequency.), so that the Eq.~(\ref{bolz1}) becomes
\begin{equation}
	\begin{aligned}
		 \Xi_{\rm AL+SE+V,J}=&\frac{|J|^4N^2}{N'\Omega_m^2} T\sum_{\omega_2}T\sum_{\omega_1}T\sum_{\Omega}\int \frac{d^2{\bm k }_2}{(2\pi)^2}\int \frac{d^2\bm{k}_1}{(2\pi)^2}\int \frac{d^2\bm{p}}{(2\pi)^2} v_{{\bm k}_1+{\bm p}/2}G^0(i\omega_1+i\Omega/2,\bm{k}_1+\bm{p}/2)f^2_{\bm{k}_1}\\
 &\times [G ^0(i\omega_1-i\Omega/2-i\Omega_m/2,\bm{k}_1-{\bm p}/2)-G ^0(i\omega_1-i\Omega/2+i\Omega_m/2,\bm{k}_1-{\bm p}/2)]D(i\Omega-i\Omega_m/2,\bm{p})\\
 &\times D(i\Omega+i\Omega_m/2,\bm{p})G^0(i\omega_2+i\Omega/2,{\bm k}_2+{\bm p}/2)f^2_{{\bm k}_2}[G^0(i\omega_2-i\Omega/2-i\Omega_m/2,{\bm k}_2-\bm{p}/2)\\
 &-G^0(i\omega_2-i\Omega/2+i\Omega_m/2,{\bm k}_2-{\bm p}/2)](v_{{\bm k}_2-{\bm p}/2}-v_{{\bm k}_2+{\bm p}/2}+v_{\bm{k}_1+\bm{p}/2}-v_{\bm{k}_1-{\bm p}/2}).\\
	\end{aligned}
	\label{bolz2}
\end{equation}
By performing substitution $(\omega_1,{\bm k}_1)\leftrightarrow (\omega_2,{\bm k}_2)$ in Eqs.~(\ref{bolz1}) and (\ref{bolz2}), and adding the results of Eq.~(\ref{bolz1}) and Eq.~(\ref{bolz2}), we obtain:
\begin{equation}
	\begin{aligned}
		 \Xi_{\rm AL+SE+V,J}=&\frac{|J|^4N^2}{4N'\Omega_m^2} T\sum_{\omega_2}T\sum_{\omega_1}T\sum_{\Omega}\int \frac{d^2{\bm k }_2}{(2\pi)^2}\int \frac{d^2\bm{k}_1}{(2\pi)^2}\int \frac{d^2\bm{p}}{(2\pi)^2} G^0(i\omega_1+i\Omega/2,\bm{k}_1+\bm{p}/2)f^2_{\bm{k}_1}\\
 &\times [G ^0(i\omega_1-i\Omega/2-i\Omega_m/2,\bm{k}_1-{\bm p}/2)-G ^0(i\omega_1-i\Omega/2+i\Omega_m/2,\bm{k}_1-{\bm p}/2)]D(i\Omega-i\Omega_m/2,\bm{p})\\
 &\times D(i\Omega+i\Omega_m/2,\bm{p})G^0(i\omega_2+i\Omega/2,{\bm k}_2+{\bm p}/2)f^2_{{\bm k}_2}[G^0(i\omega_2-i\Omega/2-i\Omega_m/2,{\bm k}_2-\bm{p}/2)\\
 &-G^0(i\omega_2-i\Omega/2+i\Omega_m/2,{\bm k}_2-{\bm p}/2)](v_{{\bm k}_2-{\bm p}/2}-v_{{\bm k}_2+{\bm p}/2}+v_{\bm{k}_1+\bm{p}/2}-v_{\bm{k}_1-{\bm p}/2})^2.\\
	\end{aligned}
\end{equation}
By performing the inverse variable substitutions $(\omega_1,{\bm k}_1)\rightarrow (\omega_1+\Omega/2,{\bm k}_1+{\bm p}/2)$ and $(\omega_2,{\bm k}_2)\rightarrow (\omega_2-\Omega/2,{\bm k}_2-{\bm p}/2)$, we recover Eq.~(b9) in the main text when impurities are absent ($\Gamma=0$), namely,
\begin{equation}
	\begin{aligned}
		 \Xi_{\rm AL+SE+V,J}=&\frac{|J|^4N^2}{4N'\Omega_m^2} T\sum_{\omega_2}T\sum_{\omega_1}T\sum_{\Omega}\int \frac{d^2{\bm k }_2}{(2\pi)^2}\int \frac{d^2\bm{k}_1}{(2\pi)^2}\int \frac{d^2\bm{p}}{(2\pi)^2} [G ^0(i\omega_1-i\Omega_m/2,\bm{k}_1)-G ^0(i\omega_1+i\Omega_m/2,\bm{k}_1)]\\
 &\times G^0(i\omega_1+i\Omega,\bm{k}_1+\bm{p})f^2_{\bm{k}_1+\bm{p}/2}D(i\Omega-i\Omega_m/2,\bm{p})D(i\Omega+i\Omega_m/2,\bm{p})G^0(i\omega_2,{\bm k}_2)\\
 &\times f^2_{{\bm k}_2-\bm{p}/2}[G^0(i\omega_2-i\Omega-i\Omega_m/2,{\bm k}_2-\bm{p})-G^0(i\omega_2-i\Omega+i\Omega_m/2,{\bm k}_2-{\bm p})](\Delta v)^2.\\
	\end{aligned}
	\label{bolz}
\end{equation}
This expression is exact, and the factor of $\Delta v$ also appears in the Boltzmann equation. Although Eq.~(\ref{bolz}) appears more concise in form, it is far less convenient for analytical calculations compared to Eq.~(\ref{SE+V+AL}). Therefore, we proceed with Eq.~(\ref{SE+V+AL}) for our analytical calculations, while using Eq.~(\ref{bolz}) for physical analysis.

For a Galilean-invariant system, which is spatially uniform and isotropic, the electron dispersion $\epsilon_{\bm k}$ can be expressed as a function of $|{\bm k}|$, ensuring that the Fermi surface is always convex.
For a momentum-conserving scattering process on the Fermi surface, a small-momentum transfer can only result in a single intersection point that is inequivalent under inversion symmetry as shown in Fig.~2b. 
At this intersection point, if the dispersion expands up to the first (second)-order momentum term $|{\bm k}| \ (|{\bm k}|^2)$, then $\Delta v= 0$, leading to a trivial conductivity \cite{Chubukov2024}. However, if the dispersion further expands to higher-order momentum terms, $\Delta v\ne 0$ leads to a nontrivial conductivity.

The saddle-point-type dispersion of VHS explicitly breaks Galilean invariance, leading to transport behavior entirely different from that of a convex Fermi surface. 
In the swap channel—i.e., the conventional scattering process—since all incoming and outgoing momenta lie near the VHS, we have $\Delta v_x =2k_{2x}-2p_x-2k_{2x}+2k_{1x}+2p_x-2k_{1x}=0$.
So the frequency dependence of the optical conductivity experiences an exact cancellation between the terms SE$+$V and AL1$+$AL2, which leads to a vanishing result
\begin{equation}
	\mathrm{Re}[\sigma^{\rm J}_{xx}(\Omega\gg T)]=-\frac{\mathrm{Im}[\Xi(i\Omega_m,0)]|_{i\Omega_m\rightarrow \Omega+i0^+}}{\Omega}\simeq  \frac{N|J|^2\Lambda_\theta}{16\pi^3\sqrt{a}|\Omega|}-\frac{N|J|^2\Lambda_\theta}{16\pi^3\sqrt{a}|\Omega|}=0.
	\label{condJ}
\end{equation}
However, for our case depicted in Fig~1b of the main text, the VHS saddle points emerge at the critical point where the Fermi surface changes from convex to concave. This provides an extra channel as illustrated in Fig.~4b. 
This scattering channel can be understood as the scattering process $\psi_{{\bm k}_1+{\bm p}}^\dagger \psi_{{\bm k}_1}\psi_{{\bm k}_2-{\bm p}}^\dagger \psi_{{\bm k}_2}$, where the fermion with momentum ${\bm k}_1$ is located at the VHS, while the fermion with momentum ${\bm k}_2$ is on the convex Fermi surface. This channel satisfies energy and momentum conservation and involves only the exchange of a small momentum ${\bm p}$.
To put the conclusion first, we show that the extra channel makes a non-zero contribution to the optical conductivity.\\

In order to introduce the extra channel depicted in Fig.~4b of the main text,
one has to derive the two inversion inequivalent intersecting points (black solid dots) 
between the shifted Fermi surfaces.
To this end, we first introduce expression for the electron dispersion of the entire Fermi surface
by including higher-order terms in addition to the saddle point, which reads
\begin{equation}
\epsilon_{\bm k}=k_x^2-a k_y^2+b k_y^4\ (b>0).
\label{feis}
\end{equation}
The optical conductivity arises solely from small-momentum scattering. This scattering process conserves both energy and momentum and is equivalent to a shift of the Fermi surface by $2\bm{p}$, as shown in Fig.~4a-b of the main text, i.e., $\epsilon_{{\bm k}-{\bm p}}=\epsilon_{{\bm k}+{\bm p}}=0$. For simplicity, we solve this equation in the case of $p_x = 0$, $p_y \neq 0$ (which does not affect the final conclusion), we obtain two inversion inequivalent solutions: 
\begin{equation}
\begin{aligned}
& {\bm k}_1 = (k_{1x},k_{1y})=(p_y\sqrt{a-b p_y^2},0) ,\\
& {\bm k}_2 = (k_{2x},k_{2y})\simeq (a/2\sqrt{b},\sqrt{a/2b-p_y^2}).\\
\end{aligned}
\label{scatter_12}
\end{equation}
We note the 1st type of solution ${\bm k}_1$ is at the VHS saddle point ${\bm k}_{\rm VHS}=(0,0)$
in the 0'th order approximation in terms of $o(p)$.
The subleading term gives rise to nothing but the saddle point dispersion,
\begin{equation}
\epsilon_{\bm{k}-\bm{k}_1}|_{\bm{p}=0} \simeq k_x^2 -ak_y^2.
\end{equation}
In addition, there's a 2nd type of solution $\bm{k}_2$ in Eq.(\ref{scatter_12}) which locates on the convex region of the Fermi surface.
And, the electron dispersion near this point is expanded as
\begin{equation}
 \epsilon_{\bm{k}-\bm{k}_2}|_{\bm{p}=0}=k_x^2-a(k_y-k_{2y})^2+b(k_y-k_{2y})^4|_{\bm{p}=0}\simeq -\frac{a}{\sqrt{b}}k_x +ak_y^2.
\end{equation}
We note that the linear and quadratic dispersion along the radical and tangential directions, respectively, is a feature of the convex Fermi surface. 
The small-momentum boson in the extra channel scattering process connects the fermions at the VHS and the convex Fermi surface. 

For the Feynman diagram in Fig.~\ref{figs2}, Eq.~(\ref{SE+V+AL}) or Eq.~(\ref{bolz}) indicates that the extra channel scattering in optical conductivity actually amounts to evaluating the SE$+$V diagrams using the 1st type of solution which has been given by Eq.~(\ref{xiJ}); While a part of fermion loop in AL$_1$$+$AL$_2$ using the 2nd type of solution. 

For the Fermi surface described by the form of Eq.~(\ref{feis}), the $\overline{\Delta v_x} =\overline{v_x}_{,p}(\Omega_m)-\overline{v_x}_{,-p}(\Omega_m)+v_{k_{1x}+p_x}-v_{k_{1x}}=2p_x$ for the extra channel is non-zero, resulting in a finite optical conductivity.
For Fermi surfaces similar in shape to those in copper-based superconductors, the extra channel occurs between two different VHS saddle points, the $\overline{\Delta v_x} =4p_x$ and the final optical conductivity is twice that of the Fermi surface described by Eq.~(\ref{feis}).

Finally, we arrive at the total optical conductivity by including the extra channel,
which takes a form
\begin{equation}
	\mathrm{Re}[\sigma^{\rm J}_{xx}(\Omega\gg T)]=-\frac{\mathrm{Im}[\Xi(i\Omega_m,0)]|_{i\Omega_m\rightarrow \Omega+i0^+}}{\Omega}\simeq  -\frac{N|J|^2\Lambda_\theta}{16\pi^3\sqrt{a}|\Omega|}.
	\label{sigmaJ}
\end{equation}
This corresponds to Eq.~(7) in the main text. Furthermore, the reflection symmetry ensures that the Hall conductivity $\mathrm{Re}[\sigma^{\rm J}_{xy}(\Omega\gg T)]=0$.

An important question is what form the boson damping should take.
Because the extra channel described by Eq.~(\ref{feis}) involves both the VHS and the convex part of the Fermi surface, the resulting Landau damping takes two distinct forms.
In all calculations of this section, we adopt the damping form at the VHS (\ref{sic-bos-1}). However, replacing the Landau damping in Eq.~(\ref{xiJ}) and (\ref{sigmaT1})  with the Landau damping $\sim |\Omega|/|p|$ at general convex Fermi surface does not affect our results. The frequency/temperature dependence of the conductvity remains the same as in Eq.~(\ref{xiJ}) and (\ref{sigmaT1}), except for a correction in the coefficients.
For a Fermi surface like that in cuprate superconductors, the extra channel occurs between two different VHS, so the Landau damping takes the same form given in Eq.~(\ref{sic-bos-1}).

\subsubsection{Boson thermal mass from interaction $J$}
\label{BTMJ}

\begin{figure}[htbp]
	\centering
	\includegraphics[width=0.7\textwidth]{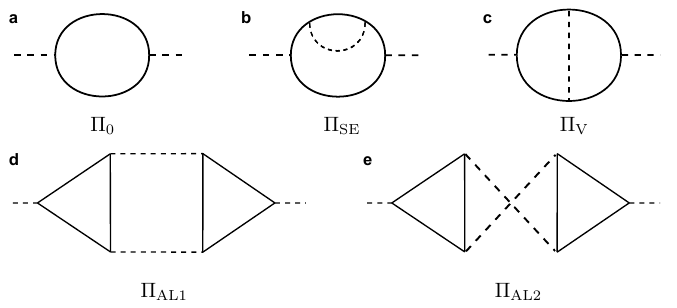}
	\caption{\textbf{Feynman diagrams of the boson mass in the limit of large Fermi energy.} Lines represent fermion propagators, dashed lines represent full boson propagators. We use \textbf{a} $\Pi_0$ to represent the zeroth-order diagram, \textbf{b} $\Pi_{\rm SE}$ for the self-energy diagram, \textbf{c} $\Pi_{\rm V}$ for the vertex diagram, and \textbf{d} $\Pi_{\rm AL1}$ and \textbf{e} $\Pi_{\rm AL2}$ for the Aslamazov-Larkin diagrams, respectively.} 
	\label{figs1}
\end{figure}

The original boson mass is $T$-independent determining the distance towards the QCP.
Considering the effect of higher order boson interactions,
the boson mass is renormalized and acquires a $T$-dependence.
The boson thermal mass is originally calculated by Moriya using the self-consistent renormalization method, which plays an important role in affecting the transport properties\cite{Hartnoll2014}.
The higher order boson interaction diagrams, illustrated in Fig.~\ref{figs1}, contributes to the quadratic boson term by contracting all other external boson legs.
Theses Feynman diagrams are nothing but the bare polarization $\Pi_0$,
self-energy diagram $\Pi_{\rm SE}$, vertex diagram $\Pi_{\rm V}$ and two AL diagrams $\Pi_{\rm AL1},\Pi_{\rm AL2}$.
The important difference between the diagrams in Fig.~\ref{figs2} and Fig.~\ref{figs1}
is that the boson polarization diagrams have constant vertices while the vertices (black solid dots) in current correlation diagrams are momentum-dependent.
The fermion Green function in principal has two origins at the Lifshitz transition,
namely near the VHS saddle points and on the convex part of the Fermi surface.\\

Let's start with the case of spatially uniform interaction $J$.
For the VHS, the one-loop diagram $\Pi_0$ in Fig.~\ref{figs1}a does not exhibit temperature dependence when performing the momentum integration in hyperbolic coordinates. This temperature independence constant can be absorbed into the effective mass $\Delta_{\rm eff}$.
Since the interaction vertex is a constant, the contributions from $\Pi_{\rm SE}+\Pi_{\rm V}$ and from $\Pi_{\rm AL1}+\Pi_{\rm AL2}$ (shown in Fig.~\ref{figs1}b-e are zero. Thus, up to the leading order at $o(|J|^4)$, the polarization diagrams evaluated using the VHS saddle point dispersion do not contribute to the effective thermal mass in the quantum critical region.
For the convex Fermi surface, Ref.~\cite{Hartnoll2014} demonstrates that $m^2(T)\simeq U T\ln T\sim \tilde{J}^4 N'T\ln (T)/ k_{\rm F}^4 $ when $\Gamma=0$. As a result, the total effective thermal mass takes a form, 
\begin{equation}
m^2(T)\simeq \frac{\tilde{J}^4N'}{k_{\rm F}^4} T\ln T.
\label{mTJ}
\end{equation}

However, if we directly integrate in the momentum $k_x-k_y$  coordinates, the leading-order one-loop diagram (Fig.~\ref{figs1}a) gives a logarithmic temperature dependence, i.e., $\Pi_0(0,0)\sim \tilde{J}^2\ln \frac{\Lambda^2_U}{T}$. Then the effective thermal mass takes a form different from that of Eq.~(\ref{mTJ}). But we would like to point out that regardless of the temperature dependence of the effective thermal mass, in the quantum critical regime $m^2(T)\ll \tilde{J}^2$, the conductivity (\ref{restj}) is still proportional to temperature. The difference arises in the region $m^2(T)\gg \tilde{J}^2$, where the different effective thermal masses will significantly affect the conductivity. If the effective thermal mass $m^2(T)\sim \ln T$, the conductivity remains linearly proportional to temperature.

\subsubsection{The finite temperature ($|\Omega| \ll T$) optical conductivity}

 At finite temperature, we start from the definition for the real part of the optical conductivity in Eq.(\ref{Re_OC}) and conduct the calculation in the spectral representation.


In this subsection, we calculate the optical conductivity in the finite-temperature regime $T\gg |\Omega|$.

We follow similar routines as in Sec.~\ref{sigmaJ0}
and start with Eq.~(\ref{SE+V+AL}) and (\ref{Delv}), which yields
\begin{equation}
	\begin{aligned}
		\Xi_{\rm{SE+V+AL}, J}(i\Omega_m,T)=&\frac{N|J|^2}{\Omega_m^2} T\sum_{\Omega_1}T\sum_{\omega}\int \frac{d^2\bm{k_2}}{(2\pi)^2} \int \frac{d^2\bm{k_1}}{(2\pi)^2} v_{{\bm k}_1}\overline{\Delta v} D(i\Omega_1,\bm{k_2})\\
	&\times G ^0(i\Omega_1+i\omega,\bm{k}_1+\bm{k}_2)\Big\{ G^0 (i\omega-i\Omega_m,\bm{k}_1) +G ^0(i\omega+i\Omega_m,\bm{k}_1) -2G^0 (i\omega,\bm{k}_1) \Big\} .
	\end{aligned}
\end{equation}
By using the spectral representation,
\begin{equation}
	D(i\Omega_1,\bm{k}) = \int \frac{d q}{2\pi} \frac{2{\rm Im}D(q+i\eta,\bm{k})}{q-i\Omega_1},
\end{equation}
we carry out the summation over the Matsubara frequency $\Omega_1$
and obtain
\begin{equation}
	\begin{aligned}
		& \Xi_{\rm{SE+V+AL}, J}(i\Omega_m,T)	= -\frac{N|J|^2}{\Omega_m^2} T\sum_{\omega}\int \frac{d^2\bm{k_2}}{(2\pi)^2} \int \frac{d^2\bm{k_1}}{(2\pi)^2} 4k_{1x}k_{2x} \int \frac{d q}{\pi}\mathrm{Im}D(q+i\eta,\bm{k_2}) \\
	&\quad\quad\quad\quad\quad\quad \times \Big\{ n_{\rm B}(q)G ^0(q+i\omega,\bm{k_1}+\bm{k_2})G^0_{\rm }(i\omega-i\Omega_m,\bm{k}_1) \\
	&\quad\quad\quad \quad\quad\quad\quad -\frac{n_{\rm F}(\epsilon_{\bm{k}_1+\bm{k}_2})}{\epsilon_{\bm{k}_1+\bm{k}_2}-i\omega-q}G^0 (i\omega-i\Omega_m,\bm{k}_1)+(\Omega_m\rightarrow -\Omega_m)-2(\Omega_m\rightarrow 0)\Big\}.
	\end{aligned}
\end{equation}

Then, summing over the Matsubara frequency $\omega$, we obtain:
\begin{equation}
	\begin{aligned}
		\Xi_{\rm{SE+V+AL}, J}(i\Omega_m,T)=&-\frac{N|J|^2}{\Omega_m^2}\int \frac{d^2\bm{k_2}}{(2\pi)^2} \int \frac{d^2\bm{k_1}}{(2\pi)^2} 4k_{1x}k_{2x} \int \frac{d q}{\pi}\mathrm{Im}D(q+i\eta,\bm{k_2})\\
	&\times \Big\{ \frac{[n_{\rm B}(q)+n_{\rm F}(\epsilon_{\bm{k}_1+\bm{k}_2})][n_{\rm F}(\epsilon_{\bm{k}_1+\bm{k}_2}-q)-n_{\rm F}(\epsilon_{\bm{k}_1})]}{\epsilon_{\bm{k}_1+\bm{k}_2}-q-\epsilon_{\bm{k}_1}-i\Omega_m}+(\Omega_m\rightarrow -\Omega_m)-2(\Omega_m\rightarrow 0)\Big\}.
	\end{aligned}
\end{equation}
where $n_{\rm F}$ and $n_{\rm B}$ are the Fermi-Dirac and Bose-Einstein distribution functions, respectively. By performing an analytic continuation $i\Omega_m\rightarrow	\Omega+i\eta$ and taking the imaginary part [ignoring the Drude term $\delta(\Omega)$], we obtain:
\begin{equation}
	\begin{aligned}
		\mathrm{Im}[\Xi^{\rm R}_{\rm{SE+V+AL}, J}(\Omega,T)]=&\frac{N|J|^2}{\Omega^2}\int \frac{d^2\bm{k_2}}{(2\pi)^2} \int \frac{d^2\bm{k_1}}{(2\pi)^2} 4k_{1x}k_{2x}\Big\{ \mathrm{Im}D(\epsilon_{\bm{k}_1+\bm{k}_2}-\epsilon_{\bm{k}_1}-\Omega+i\eta,\bm{k_2}) \\
		&\times [n_{\rm B}(\epsilon_{\bm{k}_1+\bm{k}_2}-\epsilon_{\bm{k}_1}-\Omega)+n_{\rm F}(\epsilon_{\bm{k}_1+\bm{k}_2})][n_{\rm F}(\epsilon_{\bm{k}_1}+\Omega)-n_{\rm F}(\epsilon_{\bm{k}_1})]	
	-(\Omega\rightarrow -\Omega)\Big\}.
	\end{aligned}
	\label{xiTJ}
\end{equation}
This is an odd function of $\Omega$. In the finite-temperature regime ($|\Omega|\ll T$), we have $n_{\rm F}(\epsilon\pm \Omega)=n_{\rm F}(\epsilon)\mp e^{\epsilon/T}n_{\rm F}^2(\epsilon)\frac{\Omega}{T}+o(\Omega^2/T^2)$. The dominant contribution comes from the region $|\epsilon|\ll T$, thus, we can approximate the distribution functions as $[n_{\rm B}(\epsilon)+n_{\rm F}(\epsilon)]\simeq T/\epsilon$ and $e^{\epsilon/T}n^2_{\rm F}(\epsilon)\simeq 1/4$.\\

The boson propagators crucial in dictating the transport
given their different dynamical exponent and the boson mass derived in Sec.~\ref{BTMJ},
specifically expression in Eq.(\ref{mTJ}) is used.

The main contribution of the imaginary part of the retarded polarization diagram comes from the following piece of integral,
 \begin{equation}
	\begin{aligned}
		\mathrm{Im}[\Xi^{\rm R}_{\rm{SE+V+AL}, J}(\Omega,T)]=-\frac{N|J|^2\tilde{J}^2}{2\Omega}\iint_\mathbb{C}  \frac{d^2\bm{k_2}}{(2\pi)^2} \frac{d^2\bm{k_1}}{(2\pi)^2} 4k_{1x}k_{2x}\frac{1}{[\bm{k}^2_2+m^2(T)]^2+\tilde{J}^4(\frac{\epsilon_{\bm{k}_1+\bm{k}_2}-\epsilon_{\bm{k}_1}}{\epsilon_{\bm{k}_2}})^2}\frac{1}{|\epsilon_{\bm{k}_2}|},
	\end{aligned}
\end{equation}
where the momentum integration region ${\mathbb C}$ is specified by following constraints,
\begin{equation}
\begin{aligned}
	&|\epsilon_{\bm{k}_1}|\ll T\\
	&|\epsilon_{\bm{k}_1+\bm{k}_2}|\ll T\\
	&|\Delta \epsilon(\bm{k}_1,\bm{k}_2)|\ll |\epsilon_{\bm{k}_2}|,
	\end{aligned}
	\label{tem4}
\end{equation} 
where we define $\Delta \epsilon(\bm{k}_1,\bm{k}_2)\equiv\epsilon_{\bm{k}_1+\bm{k}_2}-\epsilon_{\bm{k}_1}$. 
To carry out the momentum integral in the complex region,
we make another change of variables 
\begin{equation}
\begin{aligned}
& 2d^2\bm{k}_2dk_{1y}dk_{1x}k_{2x}=d\Delta \epsilon(\bm{k}_1,\bm{k}_2)dk_{1y}d^2\bm{k}_2\mathrm{sgn}(k_{2x}),\\ 
& |\epsilon_{\bm{k}_1}|=\Big|\big(\frac{\Delta \epsilon(\bm{k}_1,\bm{k}_2)-\epsilon_{\bm{k}_2}+2ak_{1y}k_{2y}}{2k_{2x}}\big)^2-ak_{1y}^2\Big|, \\
\end{aligned}
\end{equation} 
which leads to
\begin{equation}
\begin{aligned}
\mathrm{Im}[\Xi^{\rm R}_{\rm{SE+V+AL}, J}(\Omega,T)]\simeq &-\frac{N|J|^2\tilde{J}^2}{8\Omega\pi^2 } \int_{|\epsilon_{\bm{k}_2}|<4T} \frac{d^2\bm{k}_2}{(2\pi)^2} \int^{-\frac{k_{2y}}{2}+\frac{\sqrt{T}|k_{2x}|}{\sqrt{a|\epsilon_{\bm{k}_2}|}}}_{-\frac{k_{2y}}{2}-\frac{\sqrt{T}|k_{2x}|}{\sqrt{a|\epsilon_{\bm{k}_2}|}}} d k_{1y}\int^{|\epsilon_{\bm{k}_2}|}_{-|\epsilon_{\bm{k}_2}|}d\Delta \epsilon(\bm{k}_1,\bm{k}_2)\\
& \frac{\Delta \epsilon(\bm{k}_1,\bm{k}_2)-ak_{1y}^2+a(k_{1y}+k_{2y})^2-k_{2x}^2}{|k_{2x}||\epsilon_{\bm{k}_2}|}\frac{1}{[\bm{k}_2^2+m^2(T)]^2+\tilde{J}^4(\frac{\Delta \epsilon(\bm{k}_1,\bm{k}_2)}{\epsilon_{\bm{k}_2}})^2}\\
=&\frac{N|J|^2\sqrt{T}}{2\Omega\pi^2 \sqrt{a}}\int_{|\epsilon_{\bm{k}_2}|<4T} \frac{d^2\bm{k}_2}{4\pi^2}\frac{k_{2x}^2}{\sqrt{|\epsilon_{\bm{k}_2}|}}
  \frac{1}{\bm{k}_2^2+m^2(T)}\mathrm{ArcTan}[\frac{\tilde{J}^2}{\bm{k}_2^2+m^2(T)}]\\
 =&	\frac{N|J|^2\sqrt{T}\sqrt{m^2(T)}}{2\Omega\pi^4 \sqrt{a}}\int_0^{2\sqrt{T}/m(T)}dx\int_0^\infty d\theta \frac{x^2 \cosh(2\theta)}{x^2\cosh(2\theta)+1}\mathrm{ArcTan}(\frac{\tilde{J}^2/m^2}{x^2\cosh{2\theta}+1})\\
	\approx &\frac{N|J|^2\sqrt{T}\sqrt{m^2(T)}}{4\Omega\pi^4 \sqrt{a}}\int_0^{2\sqrt{T}/m(T)}dx \frac{i}{2}[\mathrm{Polylog}(2,-i\frac{\tilde{J}^2}{m^2(T)})-\mathrm{Polylog}(2,i\frac{\tilde{J}^2}{m^2(T)})]\\
	\approx &\left\{\begin{matrix}
  \frac{N|J|^2\tilde{J}^2T}{2\Omega\pi^4 \sqrt{a}m^2(T)}& \tilde{J}^2\ll m^2(T)\\
 \frac{N|J|^2T}{4\Omega\pi^3 \sqrt{a}}\ln \frac{\tilde{J}^2}{m^2(T)} &\tilde{J}^2\gg m^2(T)
\end{matrix}\right.
\end{aligned}
\label{sigmaT1}
\end{equation}
where $x^2\equiv k^2/m^2(T)$.

Equation~\eqref{sigmaT1} retains only $E\equiv|\epsilon_{\bm{k}_2}|<4T$;
the thermal constraints in Eq.~\eqref{tem4} also allow $E>4T$.
At $\Delta\epsilon=0$, setting $u=k_{1y}+k_{2y}/2$ gives
\begin{equation}
 \epsilon_{\bm{k}_1}
 =\epsilon_{\bm{k}_2}\left(\frac14-\frac{au^2}{k_{2x}^2}\right),
 \qquad
 \frac{|k_{2x}|}{2\sqrt a}\sqrt{1-\frac{4T}{E}}
 <|u|<
 \frac{|k_{2x}|}{2\sqrt a}\sqrt{1+\frac{4T}{E}}
 \quad(E>4T).
\end{equation}
The two allowed intervals have total width
$\frac{|k_{2x}|}{\sqrt a}[\sqrt{1+4T/E}-\sqrt{1-4T/E}]
\simeq4|k_{2x}|T/(\sqrt a E)$ for $E\gg T$.
Including the thermally allowed transfers $|\Delta\epsilon|\lesssim T$
therefore gives, up to numerical and fixed anisotropy factors,
\begin{equation}
 \begin{aligned}
 \delta\operatorname{Im}\Xi^{\rm R}_{\rm SE+V+AL,J}
 &\sim\frac{N|J|^2\tilde{J}^2T}{\Omega}
 \int_{E>4T}\frac{d^2\bm{k}_2}{(2\pi)^2}
 \frac{k_{2x}^2}{E^2}
 \int_{-T}^{T}\frac{d\Delta\epsilon}
 {[\bm{k}_2^2+m^2(T)]^2
   +\tilde{J}^4(\Delta\epsilon/E)^2}\\
 &\sim\frac{N|J|^2T}{\Omega}
 \int_{E>4T}\frac{d^2\bm{k}_2}{(2\pi)^2}
 \frac{k_{2x}^2}{E[\bm{k}_2^2+m^2(T)]}
 \arctan\frac{\tilde{J}^2T}{E[\bm{k}_2^2+m^2(T)]}.
 \end{aligned}
\end{equation}
The sharp thermal limits here are sufficient for estimating the logarithm.
Already in a fixed finite hyperbolic-angle sector, where
$\bm{k}_2^2\sim E$, the overdamped window
$\max\{4T,m^2(T)\}\ll E\ll\sqrt{\tilde{J}^2T}$ contributes
\begin{equation}
 \left.\delta\operatorname{Im}\Xi^{\rm R}_{\rm SE+V+AL,J}
 \right|_{\text{finite angle}}
 \sim\frac{N|J|^2T}{\Omega}
 \int_{\max\{4T,m^2(T)\}}^{\sqrt{\tilde{J}^2T}}\frac{dE}{E}
 =\frac{N|J|^2T}{\Omega}
 \ln\frac{\sqrt{\tilde{J}^2T}}{\max\{4T,m^2(T)\}}.
\end{equation}
For the thermal-mass scaling $m^2(T)\propto T\ln(T_0/T)$, with a fixed
reference scale $T_0$, this has the same $T\ln(T_0/T)/\Omega$ order as
Eq.~\eqref{sigmaT1}.

The conductivity contribution from the region retained in Eq.~\eqref{sigmaT1} is then expressed as,
\begin{equation}
	\mathrm{Re}[\sigma^{\rm J}(\Omega\ll T)]=-\frac{\mathrm{Im}[\Xi^{\rm R}_{\rm{SE+V+AL}, J}(\Omega,T)]}{\Omega}\sim 
	\left\{\begin{matrix}
-\frac{N^2|J|^4T}{N'\Omega^2m^2(T)} & m^2(T)\gg N|J|^2/N' \\
  -\frac{N|J|^2T}{\Omega^2}\ln(\frac{N|J|^2}{N' m^2(T)})& m^2(T)\ll N|J|^2/N'
\end{matrix}\right.
\label{restj}
\end{equation}
We obtain the second row in Fig.~4c in the main text. Substituting the boson thermal mass in Eq.~(\ref{mTJ}), we can further obtain:
\begin{equation}
	\mathrm{Re}[\sigma^{\rm J}(\Omega\ll T)]\sim 
	\left\{\begin{matrix}
-\frac{k_{\rm F}^4T}{\Omega^2}\ln T & T\gg k_{\rm F}^4/N|J|^2
 \\
  -\frac{N|J|^2T}{\Omega^2}\ln(\frac{k_{\rm F}^4}{N|J|^2T})& T\ll k_{\rm F}^4/N|J|^2
\end{matrix}\right.
\label{restj}
\end{equation}
\\

\subsection{$\sigma^{\rm J'}(\Omega,T)$ from spatially random interaction}
Now we consider the contribution to the conductivity arising from the interaction  $J'$. where the bosons are damped as described by Eq.(\ref{sdc-bos}). In this section, all the fermion-boson interaction vertices in Fig.~\ref{figs2} involve only spatially random interaction $J'$, and all calculations are performed in the static limit.\\

\subsubsection{Derivation on the $T=0$ optical conductivity} 
\label{sigmaJ'0}

The current-current correction functions perturbatively in $o(N|J'|^2)$ 
are shown in Fig.~\ref{figs2}. They are from the self-energy diagram,
\begin{equation}
	\begin{aligned}
		\Xi_{\rm SE,J'}(i\Omega_m,0)=&-NT\sum_n\int\frac{d^2 \bm{k}}{(2\pi)^2}\frac{4k_x^2}{i\omega_n+i\Omega_m-k_x^2+ak_y^2}\frac{1}{(i\omega_n-k_x^2+ak_y^2)^2}\Sigma_{\rm J'}(i\omega_n)+(\Omega_m\rightarrow -\Omega_m)\\
		=&\frac{2N}{\pi^2\sqrt{a} \Omega_m^2}T\sum_n\int d k_y	\Bigg\{ \sqrt{i\omega_n+i\Omega_m+k_y^2}\mathrm{ArcTanh}(\frac{\Lambda_{\rm UV}}{\sqrt{i\omega_n+i\Omega_m+k_y^2}}) \\
		&\quad \quad \quad \quad -\frac{i\omega_n+i\Omega_m+k_y^2}{\sqrt{i\omega_n+k_y^2}}\mathrm{ArcTanh}(\frac{\Lambda_{\rm UV}}{\sqrt{i\omega_n+k_y^2}})\Bigg\} \Sigma_{\rm J'}(i\omega_n) +(\Omega_m\rightarrow -\Omega_m)\\
		\simeq &-\frac{|J'|^2N\Lambda_{\rm U}^2\Lambda_\theta}{16\pi^5a}\int d\omega \frac{\omega}{\Omega_m^2}\ln(\frac{e\Lambda_{\rm U}^2}{\tilde{J}'^2|\omega|})[\mathrm{sgn}(\omega+\Omega_m)-\mathrm{sgn}(\omega)]+(\Omega_m\rightarrow -\Omega_m)\\
		=&\frac{|J'|^2N\Lambda_{\rm U}^2\Lambda_\theta}{16\pi^5a}\ln \frac{e^3\Lambda_{\rm U}^4}{\tilde{J}'^4\Omega_m^2},
	\end{aligned}	
	\label{SEJ'}
\end{equation}
the vertex diagram
\begin{equation}
	\begin{aligned}
		\Xi_{\rm V,J'}(i\Omega_m,0)=& -|J'|^2 N\int\frac{d\omega_1}{2\pi}\frac{d\omega_2}{2\pi}\int\frac{d^2\bm{k_1}}{(2\pi)^2}\frac{d^2\bm{k_2}}{(2\pi)^2}\frac{d^2\bm{k_3}}{(2\pi)^2}\frac{4k_{1x}k_{2x}}{i\omega_{1}+i\Omega_m-k_{1x}^2+ak_{1y}^2}\frac{1}{i\omega_1-k_{1x}^2+ak_{1y}^2}\\
		&\frac{1}{i\omega_2+i\Omega_m-k_{2x}^2+ak_{2y}^2}\frac{1}{i\omega_2-k_{2x}^2+ak_{2y}^2}\frac{1}{k_3^2-\Pi_{\rm J'}(i\omega_2-i\omega_1)}=0.
	\end{aligned}
\end{equation}
and the AL diagrams
\begin{equation}
\Xi_{\rm AL1,J'}(i\Omega_m,0)=0=\Xi_{\rm AL2,J'}(i\Omega_m,0).
\end{equation}
The vanishing contribution from AL diagrams can be readily
verified by making a change of variable: $k_x\rightarrow -k_x$.
As a result, the optical conductivity comes solely from the self-energy diagram in Fig.~\ref{figs2}b, which is same as the case of a conventional convex Fermi surface (without any VHS). The expression is given by,
\begin{equation}
	\mathrm{Re}[\sigma_{xx}^{\rm J'}(|\Omega|\gg T)]=-\frac{\mathrm{Im}[\Xi(i\Omega_m,0)_{i\Omega_m\rightarrow \Omega+i0^+}]}{\Omega}=-\frac{|J'|^2N\Lambda_{\rm U}^2\Lambda_\theta}{8\pi^4a}\frac{1}{|\Omega|}.
	\label{sigmaJJ}
\end{equation}
Again, the Hall conductivity $\sigma_{xy}^{\rm J'}(\Omega)=0$ due to spatial reflection symmetry.\\

\subsubsection{Boson thermal mass from interaction $J'$}
\label{BTMJ'}
Then, let's consider the boson thermal mass from spatially random interaction $J'$. 
Since spatial disorder involves scattering at impurity sites, it eliminates the momentum dependence of the Green’s function. Therefore, Irrespective of the electronic dispersion, whether characterized by the VHS saddle points or convex Fermi surface, the fermion Green's function is given by
\begin{equation}
G (i\omega)\equiv \int d^2\bm{k}G (i\omega,\bm{k})\sim -iD_{\rm F}\mathrm{sgn}(\omega),
\end{equation}
where $D_{\rm F}$ is the density of states, and for VHS, $D_{\rm F}\sim \Lambda_\theta\sim \ln\Lambda_{\rm U}^2$. Then, the one-loop polarization diagram $\Pi_0$ (Fig. \ref{figs1}a) already yields a $T$-dependent result 
\begin{equation}
 \Pi_0(0,0)=-\frac{N|J'|^2}{N'}T\sum_\omega \int \frac{d^2\bm{k}}{(2\pi)^2}G (i\omega,\bm{k})\frac{d^2\bm{p}}{(2\pi)^2}G (i\omega,\bm{p})\sim \frac{D_{\rm F}^2N|J'|^2}{N'}T\sum_\omega 1 \sim \frac{N|J'|^2D_{\rm F}^2 T}{N'},
\end{equation}
Thus, we arrive at the conclusion that the boson thermal mass square is linear-in-$T$ (up to log-correction) in both at VHS and on the convex Fermi surface regions, which reads
\begin{equation}
m^2(T)\sim \frac{N|J'|^2\Lambda_\theta^2 T}{N'}.
\label{mTJJ}
\end{equation}

\subsubsection{The finite temperature ($|\Omega| \ll T$) optical conductivity}

 At finite temperature, we start from the definition for the real part of the optical conductivity in Eq.~(\ref{Re_OC}) and conduct the calculation in the spectral representation.

The boson propagators crucial in dictating the transport
given their different dynamical exponent and the boson mass derived in Sec.~\ref{BTMJ'},
specifically expression in Eq.~(\ref{mTJJ}) is used.

Owing to the reflection symmetry, only the self-energy diagram contributes to $\sigma^{\rm J'}$ conductivity, which reads
\begin{equation}
	\begin{aligned}
		\Xi_{\rm{SE},J'}(i\Omega_m,T)=&-N|J'|^2T\sum_nT\sum_l\int\frac{d^2 \bm{k}}{(2\pi)^2}\frac{4k_x^2}{i\omega_n+i\Omega_m-k_x^2+ak_y^2}\frac{1}{(i\omega_n-k_x^2+ak_y^2)^2}\int \frac{d^2 \bm{p}}{(2\pi)^2}D(i\Omega_l,\bm{p})\\
		&\times \int \frac{d^2\bm{q}}{(2\pi)^2}\frac{1}{i\omega_n+i\Omega_l-q_x^2+a q_y^2}+(\Omega_m\rightarrow -\Omega_m)\\
	\end{aligned}	
	\label{SEJ'}
\end{equation}
Similarly, we carry out the summation over Matsubara frequency in the spectral representation. After performing the analytic continuation, we obtain the imaginary part of current correlation function as
\begin{equation}
	\begin{aligned}
	\mathrm{Im}[\Xi^{\rm R}_{\rm{SE},J'}(\Omega,T)]=&	-\frac{N|J'|^2}{\Omega^2}\int\frac{d^2 \bm{k}}{(2\pi)^2}\int\frac{d^2 \bm{p}}{(2\pi)^2}\int\frac{d^2 \bm{q}}{(2\pi)^2}4k_{x}^2\{\mathrm{Im}[D(\epsilon_{\bm{q}}+\Omega-\epsilon_{\bm{k}}+i\eta,\bm{p})][n_{\rm B}(\epsilon_{\bm{q}}+\Omega-\epsilon_{\bm{k}})+n_{\rm F}(\epsilon_{\bm{q}})]\\
	&\times [n_{\rm F}(\epsilon_{\bm{k}}-\Omega)-n_{\rm F}(\epsilon_{\bm{k}})]-(\Omega\rightarrow -\Omega)\}.
	\end{aligned}
\end{equation}
The dominant contribution comes from the momentum region,
\begin{equation}
\begin{aligned}
	&|\epsilon_{\bm{k}}|\ll T\\
	&|\epsilon_{\bm{q}}|\ll T.
	\end{aligned}
\end{equation}
Integrating over the momentums in these regions, we arrive at
\begin{equation}
\begin{aligned}
	\mathrm{Im}[\Xi^{\rm R}_{\rm{SE},J'}(\Omega,T)]&\simeq \frac{2N|J'|^2\tilde{J}'^2}{\Omega}\int\frac{d^2 \bm{p}}{(2\pi)^2}\int_{|\epsilon_{\bm{k}}|<T}\frac{d^2 \bm{k}}{(2\pi)^2}\int_{|\epsilon_{\bm{q}}|<T}\frac{d^2 \bm{q}}{(2\pi)^2}\frac{k_{x}^2}{[\bm{p}^2+m^2(T)]^2+\tilde{J}'^4(\epsilon_{\bm{q}}-\epsilon_{\bm{k}})^2}\\
	&=\frac{N|J'|^2\Lambda^2_{\rm U}\Lambda_\theta T}{\pi^4\Omega}\int\frac{d^2 \bm{p}}{(2\pi)^2}\frac{1}{\bm{p}^2+m^2(T)}\mathrm{ArcTan}(\frac{\tilde{J}'^2 T}{\bm{p}^2+m^2(T)})\\
	&=\frac{N|J'|^2\Lambda_{\rm U}^2\Lambda_\theta T}{4\pi^5\Omega}\times \left\{\begin{matrix}
 \frac{\tilde{J}'^2T}{m^2(T)} & m^2(T)\gg \tilde{J}'^2 T\\
\frac{\pi}{2}\ln \frac{m^2(T)}{\tilde{J}'^2 T} & m^2(T)\ll \tilde{J}'^2T
\end{matrix}\right.
\end{aligned}
\end{equation}

Using the Kubo formula, we obtain the conductivity:
\begin{equation}
	\mathrm{Re}[\sigma^{\rm J'}(\Omega\ll T)]=-\frac{\mathrm{Im}[\Xi^{\rm R}_{\rm{SE}, J'}(\Omega,T)]}{\Omega}\sim 
	\left\{\begin{matrix}
-\frac{N^2\Lambda_{\rm U}^2\Lambda_\theta |J'|^4T^2}{ N'\Omega^2 m^2(T)} & m^2(T)\gg N|J'|^2T/N' \\
  -\frac{N\Lambda_{\rm U}^2\Lambda_\theta |J'|^2T}{\Omega^2}\ln \frac{N'm^2(T)}{N|J'|^2T}& m^2(T)\ll N|J'|^2T/N'
\end{matrix}\right.
\end{equation}
Substituting the boson thermal mass in Eq.~(\ref{mTJJ}), we can further obtain:
\begin{equation}
	\mathrm{Re}[\sigma^{\rm J'}(\Omega\ll T)]=-\frac{\mathrm{Im}[\Xi^{\rm R}_{\rm{SE}, J'}(\Omega,T)]}{\Omega}\sim 
	\left\{\begin{matrix}
-\frac{N\Lambda_{\rm U}^2|J'|^2T}{ \Omega^2\Lambda_\theta} & \Lambda_\theta^2 \gg 1 \\
  -\frac{N\Lambda_{\rm U}^2\Lambda_\theta|J'|^2T}{\Omega^2}\ln \Lambda_\theta & \Lambda_\theta^2\ll 1
\end{matrix}\right.
\label{restjj}
\end{equation}
\\

\newpage
\section{Transport property with potential disorder $|w|\ne 0$}
In Sec.~\ref{v0con}, we present the optical conductivity and resistivity
induced at VHS saddle point subjected to spatially uniform $J$ and random $J^\prime$ Yukawa interactions, respectively, 
in the absence of potential disorder $w=0$.
In low-temperature limit $T\ll |\Omega|$, for both interactions ${\cal J}=J,J^\prime$, 
the optical conductivity (in finite frequency regime $\Omega\ne 0$) 
is proportional to $|\Omega|^{-1}$ without any residual resistivity.
At finite temperatures in the quantum-critical regime, it crosses over to the $T/\Omega^{2}$ form.
We summarize the results in Eqs.~(\ref{sigmaJ}), (\ref{sigmaJJ}), (\ref{restj}), and (\ref{restjj}) as
\begin{equation}
{\rm Re}\big[\sigma^{{\cal J}}(\Gamma=0)\big]
\sim
-{\cal J}^{2}N\frac{|\Omega|+T}{\Omega^{2}}.
\label{sigma_v0_T0}
\end{equation}
Here, the \(\Omega^{-2}\) denominator originates from the generic finite-frequency structure of the current response in the collisionless regime, whereas the numerator can be interpreted as the transport scattering rate,
\begin{equation}
\tau_{\rm tr}^{-1}\sim |\Omega|+T.
\end{equation}
These results arise from the two-loop self-energy and vertex (SE+V) diagrams in Figs.~\ref{figs2}b-c, together with the three-loop Aslamazov--Larkin (AL) diagrams in Figs.~\ref{figs2}d-e. For a generic point on a two-dimensional convex Fermi surface, the two contributions cancel each other, whereas this cancellation is avoided at the Lifshitz transition. In addition, the one-loop bubble diagram in Fig.~\ref{figs2}a does not contribute to the optical conductivity in the absence of potential disorder (\(w=0\)).

In the dc limit, Eq.~(\ref{sigma_v0_T0}) diverges and therefore fails to
predict a finite dc resistivity \(\rho(\Omega=0,T;\Gamma=0)\) at finite
temperatures.\\

In this section, our primary conclusion is that, in the presence of potential disorder $w\ne 0$,
the optical conductivity near a VHS at Lifshitz transition exhibits linear-in-$\omega$ and linear-in-$T$ forms at zero and finite temperatures, respectively.
Recall that the impurity scattering rate is defined as
\begin{equation}
\Gamma \equiv \frac{\Lambda_\theta}{2\pi}|w|^2 \sim  \ln \Lambda_{\rm U}^2 |w|^2.
\end{equation}
Therefore, it's reasonable to regard the impurity scattering rate as the largest energy scale 
and consider the low-frequency limit $\Gamma \gg |\Omega|$ henceforth.
At leading order ${\cal O}(\Gamma^{-1})$, the optical conductivity receive a finite contribution from the one-loop diagram in Fig.~\ref{figs2}a, which reads
\begin{equation}
{\rm Re} \sigma^{\cal J}_0( |\Omega| \ll \Gamma) \sim -\Lambda_{\rm U}^2 N \frac{1}{\Gamma}.
\end{equation}

In the low-temperature regime $T\ll |\Omega|$, the optical conductivity from the higher order diagrams, {\sl i.e.} SE+V+AL, is proportional to the $w=0$ result in Eq.~(\ref{sigma_v0_T0}) as
\begin{equation}
	{\rm Re} \sigma^{\cal J}_{\rm SE+V+AL}(|\Omega| \ll \Gamma)
	\sim {\cal J}^2 N \frac{|\Omega|+T}{\Gamma^2}.
	\label{vn0con}
\end{equation}
Intuitively, this amounts to replacing \(\Omega\) by \(\Omega+i\Gamma\) directly in the collisionless denominator of Eq.~\eqref{sigma_v0_T0}, while leaving the transport scattering rate $\tau_{\rm tr}^{-1}$ in the numerator unchanged.
The $\Gamma^{-2}$ scaling is consistent with the optical conductivity from generic 2D Fermi surface\cite{Chubukov2011},
despite that the impurity scattering rate takes a different form $\Gamma_{\rm FS} =2\pi D_{\rm F} w^2 $ 
with $D_{\rm F}$ being the density of state at the Fermi surface.
Moreover, we note that in addition to the charge transport dynamics at finite frequency, 
the potential disorder also significantly affects the Drude peak at $\omega=0$ leading to its complete depletion.
Recall that in the absence of potential disorder, the Drude peak originates from the $1/\Omega_m^2$ term,
which gives rise to the delta function $\delta(\Omega)$ after analytic continuation. 
Once the potential disorder is introduced, this term is given by $1/(\Omega_m^2+\Gamma^2)$.
Importantly, in the low-frequency limit $|\Omega|\ll \Gamma$, this term fails to generate the Drude peak;
instead, it leads to the scaling form $1/\Gamma^2$ in Eq.~(\ref{vn0con}).

Collecting terms from all the Feynman diagrams in Fig.~\ref{figs2}a-e, the conductivity is approximately written as,
\begin{equation}
\begin{aligned}
{\rm Re} \sigma^{\cal J}(|\Omega|\ll T; |\Omega| \ll \Gamma) \sim &  -\Lambda_U^2 N\frac{1}{\Gamma} + C N \frac{T}{\Gamma^2}  \ \simeq  -\Lambda_U^2 N\frac{1}{\Gamma+C\Lambda_U^{-2} T}\ ,\\
\end{aligned}
\label{sig_VHS}
\end{equation}
where the coefficient $C\sim {\cal J}^2$.
As a result, the resistivity exhibits the well-known linear-in-$T$ strange metal behavior with a residual constant term,
\begin{equation}
	\rho(T)\sim \Gamma/N\Lambda_U^2+C T/N\Lambda_U^4 + ...
\end{equation}
where $...$ represents higher order terms.

\subsection{Short-circuit issue}
So far, we have focused on the contribution from the Fermi-surface patch near
the VHS and shown that it displays strange-metal behavior in the presence of
either spatially uniform/random Yukawa interactions and potential disorder.
However, the full two-dimensional Fermi surface also contains large convex
regions with approximately linear dispersion. Therefore, in order to determine
the total resistivity, one has to take the entire Fermi surface into account.
A central question is then whether the relatively large resistivity generated
near the VHS can be short-circuited by the remaining, colder parts of the
Fermi surface. In this section, we show that, once potential disorder is
included, such a short-circuiting mechanism is avoided within a finite
temperature window.

We first recall the short-circuit problem introduced by Hlubina and
Rice~\cite{Hlubina1995} in the context of an antiferromagnetic
(AFM) quantum critical point, where the fluctuating order parameter carries a
finite momentum ${\bm Q}\neq 0$. In that case, only the Fermi-surface points
connected by ${\bm Q}$ are strongly scattered by the critical fluctuations.
These points become non-Fermi-liquid hot spots and give a contribution
$\rho(T)\sim T^\alpha$ with $\alpha<2$. By contrast, the remaining parts of
the Fermi surface remain conventional Fermi liquids, with
$\rho(T)\sim T^2$, and are therefore referred to as cold regions. Since the
cold regions have a much smaller scattering rate at low temperatures, they can
short-circuit the contribution from the hot spots, rendering the total
transport Fermi-liquid-like. Rosch subsequently pointed out that introducing a
small amount of potential disorder can strongly affect this conclusion and
help reveal the non-Fermi-liquid transport behavior~\cite{Rosch1999}.

Our situation is different from the AFM case in an important respect. Here the
critical boson carries zero momentum, ${\bm Q}=0$, and hence all points on the
Fermi surface are coupled to the critical fluctuations through the Yukawa
interaction. Nevertheless, the scattering remains highly anisotropic along the
Fermi surface. In the clean limit, the generic convex parts of the Fermi
surface acquire the usual non-Fermi-liquid self-energy
\begin{equation}
{\rm Im}\Sigma^J(i\omega_n)
\sim {\rm sgn}(\omega_n)|\omega_n|^{2/3},
\end{equation}
whereas the VHS saddle point is more singular,
\begin{equation}
{\rm Im}\Sigma^J(i\omega_n,{\bm k}_v)
\sim {\rm sgn}(\omega_n)|\omega_n|^{1/2}.
\end{equation}
Thus, the VHS region plays a role analogous to a hot region, while the generic
convex parts of the Fermi surface are relatively colder.

This anisotropy by itself does not guarantee that the VHS contribution
dominates the total resistivity. To see this, consider the linearized
Boltzmann equation, where the nonequilibrium distribution is written as
\begin{equation}
f_{\bm k}
=
n_F(\epsilon_{\bm k})
-
\Phi_{\bm k}
\frac{\partial n_F(\epsilon_{\bm k})}{\partial \epsilon_{\bm k}} .
\end{equation}
The standard variational ansatz is
\begin{equation}
\Phi_{\bm k} \sim {\bm v}_{\bm k}\cdot{\bm n},
\label{sa}
\end{equation}
where ${\bm n}$ is the direction of the applied electric field. However, this
ansatz does not necessarily minimize the resistivity in a clean system.
Hlubina and Rice~\cite{Hlubina1995} proposed a variational function
of the form
\begin{equation}
\Phi_{\bm k}
\sim
\frac{{\bm v}_{\bm k}\cdot{\bm n}}
{e^{\beta(\Delta-\phi)}+1},
\end{equation}
where $\phi$ measures the distance along the Fermi surface from a hot spot,
and $\Delta$ and $\beta$ are variational parameters. For $\beta=0$, this
reduces to the conventional ansatz up to an irrelevant overall factor. In
contrast, for $\beta\gg 1$ and $\Delta>0$, the denominator suppresses the
contribution from the hot region near $\phi=0$. The resistivity can therefore
be reduced by choosing a trial function that avoids the strongly scattered
region. In this sense, in the clean limit, the VHS contribution can in
principle be short-circuited by the colder convex parts of the Fermi surface.

Potential disorder changes this conclusion. Once impurity scattering is
included, the variational problem for the resistivity becomes
\begin{equation}
\rho(T)
=
\min_{\Phi_{\bm k}}
\left\{
\rho_{\rm ie}[\Phi_{\bm k}]
+
\rho_{\rm ee}[\Phi_{\bm k}]
\right\}.
\end{equation}
In the disorder-dominated regime, the leading contribution comes from
$\rho_{\rm ie}$, and hence the variational minimum is primarily fixed by the
impurity-scattering functional~\cite{Rosch1999}. The minimizing distribution
is then close to the standard ansatz in Eq.~\eqref{sa}. As a result, the
system can no longer substantially lower its resistivity by choosing a trial
function that suppresses the VHS region. The electron-electron contribution
$\rho_{\rm ee}$ is then evaluated with a distribution function essentially
fixed by impurity scattering, and the singular scattering near the VHS gives a
finite contribution to the total resistivity. Therefore, in the dirty limit,
the current is not short-circuited by the colder parts of the Fermi surface,
and the Hlubina-Rice short-circuiting problem is avoided within the relevant
finite temperature window.

Numerical support for this picture is provided by Lee's
study~\cite{PALee2021}. Lee considered a Fermi surface close to the
Brillouin-zone boundary, namely the regime in which the VHS is about to
emerge. His results support the conclusion that a temperature-linear
resistivity can persist down to zero temperature only when the Fermi surface
is sufficiently close to the Brillouin-zone boundary. Moreover, in the absence
of impurity scattering, this temperature-linear regime disappears. The
associated short-circuiting mechanism is consistent with that proposed by
Hlubina and Rice: the contribution from the region near the VHS, which
corresponds to the vicinity of the Umklapp scattering points in Lee's work, is
strongly suppressed.
\\

\subsection{$\sigma^{\rm J}(\Omega,T)$ from spatially uniform interaction}
In parallel with calculation in \ref{v0con}, we first consider the spatially uniform interaction $J$.

\subsubsection{Derivation on the $T=0$ optical conductivity}
To calculate the conductivity, we first derive a useful relation, meaning that Eq.~(\ref{con-for}) needs to be modified to 
\begin{equation}
	G_{\rm w}(i\omega_n+i\Omega_m,\bm{k})G_{\rm w}(i\omega_n,\bm{k})=\frac{G_{\rm w}(i\omega_n,\bm{k})-G_{\rm w}(i\omega_n+i\Omega_m,\bm{k})}{i\Omega_m+i\Gamma[\mathrm{sgn}(\omega_n+\Omega_m)-\mathrm{sgn}(\omega_n)]}
	\approx \frac{G_{\rm w}(i\omega_n,\bm{k})-G_{\rm w}(i\omega_n+i\Omega_m,\bm{k})}{i\Omega_m+2i\Gamma \mathrm{sgn}(\Omega_m)},
	\label{ggT0}
\end{equation}
in the presence of impurity scattering (potential disorder). Where the second equality arises in low-frequency limit, then $G_{\rm w}(i\omega_n,\bm{k})\approx 1/(i\Gamma\mathrm{sgn}(\omega_n)-\epsilon_{\bm k})=[-\epsilon_{\bm k}-i\Gamma\mathrm{sgn}(\omega_n)]/(\Gamma^2+\epsilon_{\bm k}^2)$ and $G_{\rm w}(i\omega_n,\bm{k})-G_{\rm w}(i\omega_n+i\Omega_m,\bm{k})\sim \mathrm{sgn}(\omega_n+\Omega_m)-\mathrm{sgn}(\omega_n)$, to obtain a non-zero result, the $\omega$  integral is restricted between  $0$  and  $|\Omega_m|$, where $\mathrm{sgn}(\omega_n+\Omega_m)-\mathrm{sgn}(\omega_n)$ yields $2\mathrm{sgn}(\Omega_m)$. By repeatedly using formula (\ref{ggT0}), we will find that the sum of all higher-loop diagrams in Fig.~\ref{figs2} contributes as follows:
\begin{equation}
	\begin{aligned}
		 \Xi_{{\rm AL+SE+V},J}^{\rm w}(i\Omega_m)=&-\frac{|J|^2N}{(|\Omega_m|+2\Gamma)^2} T\sum_{\omega_1}T\sum_{\Omega}\int \frac{d^2\bm{k}_1}{(2\pi)^2}\int \frac{d^2\bm{p}}{(2\pi)^2}v_{\bm{k}_1}\overline{\Delta v}[G_{\rm w}(i\omega_1-i\Omega_m/2,\bm{k}_1)-G_{\rm w}(i\omega_1+i\Omega_m/2,\bm{k}_1)]\\
 &\times G_{\rm w}(i\omega_1+i\Omega,\bm{k}_1+\bm{p})f^2_{\bm{k}_1+\bm{p}/2}[D(i\Omega-i\Omega_m/2,\bm{p})-D(i\Omega+i\Omega_m/2,\bm{p})],\\
	\end{aligned}
	\label{SE+V+ALv}
\end{equation}
with the group velocity $\overline{\Delta v}$ is the same as Eq.~(\ref{Delv}). 
In fact, this is precisely the substitution of the leading factor  $1/\Omega_m^2$ with $1/(|\Omega_m|+2\Gamma)^2$ in formula (\ref{SE+V+AL}).
Using Eqs.~(\ref{DD=D-D}) and (\ref{Delv}), Eq.~(\ref{SE+V+ALv}) can be simplified as Eq.~(b9) in the main text.
In the low-frequency limit $|\Omega_m|\ll \Gamma $, this corresponds to replacing  $1/\Omega_m^2$ with  $1/(2\Gamma)^2$ in formula (\ref{SE+V+AL}). 

For the swap channel,  $\Delta v =0$ remains, while in the extra channel,  $\Delta v_x = 2p_x$. Eq.~(\ref{SE+V+ALv}) can be rewritten as:
\begin{equation}
	\begin{aligned}
		 \Xi_{{\rm AL+SE+V},J}^{\rm w}(i\Omega_m)=&-\frac{4|J|^2N}{(|\Omega_m|+2\Gamma)^2} T\sum_{\omega_1}T\sum_{\Omega}\int \frac{d^2\bm{k}_1}{(2\pi)^2}\int \frac{d^2\bm{p}}{(2\pi)^2}p_xk_{1x}[G_{\rm w}(i\omega_1-i\Omega_m/2,\bm{k}_1)-G_{\rm w}(i\omega_1+i\Omega_m/2,\bm{k}_1)]\\
 &\times G_{\rm w}(i\omega_1+i\Omega,\bm{k}_1+\bm{p})[D(i\Omega-i\Omega_m/2,\bm{p})-D(i\Omega+i\Omega_m/2,\bm{p})]\\
 = &-\frac{2|J|^2N\mathrm{sgn}(\Omega_m)}{(|\Omega_m|+2\Gamma)^2}\int d\Omega \int_{-|\Omega_m|/2}^{|\Omega_m|/2}d\omega \int \frac{d^2\bm{k}}{(2\pi)^2}\int \frac{d^2\bm{p}}{(2\pi)^2}p_xk_{x}\frac{\Gamma/\pi}{\epsilon_{\bm k}^2+\Gamma^2}\frac{\Gamma/\pi}{\epsilon_{\bm{k}+\bm{p}}^2+\Gamma^2}\\
 &\times [\mathrm{sgn}(\omega+\Omega+\Omega_m/2)-\mathrm{sgn}(\omega+\Omega-\Omega_m/2)]D(i\Omega,\bm{p}).
  	\end{aligned}
  	\label{tem8}
\end{equation}

We will prove that when $\tilde{J}^2|\Omega|\gg \Gamma^2$, the leading-order optical conductivity is proportional to $|J|^2N|\Omega|/\Gamma^2$. This verifies the Eq.~(\ref{vn0con})  given at the beginning of this section. 

\begin{figure}[htbp]
		\centering
		\includegraphics[width=0.7\textwidth]{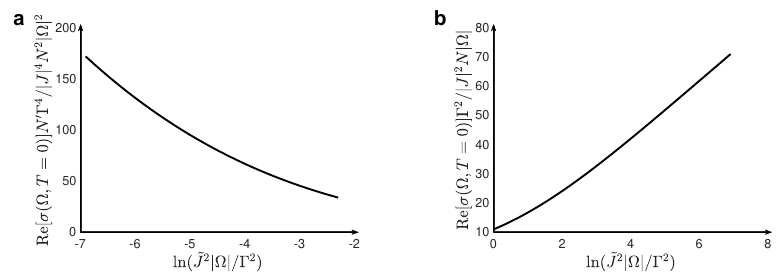}
		\caption{\textbf{The optical conductivity ${\rm Re}[\sigma^{\rm J}(\Omega,T=0)]$ in different frequency regimes at zero temperature.} \textbf{a} $\tilde{J}^2|\Omega|/\Gamma^2\in (10^{-3},10^{-1})\ll 1$. \textbf{b} $\tilde{J}^2|\Omega|/\Gamma^2\in (1,10^{3})> 1$.} 
		\label{Figs2}
\end{figure}

First, we compute the  $\bm{k}$-integral in Eq.~(\ref{tem8}) in a similar-fashion as Eq.~(\ref{gam=del}):
\begin{equation}
\begin{aligned}	
&\int \frac{d^2\bm{k}}{4\pi^2}k_x\frac{\Gamma/\pi}{\epsilon_{\bm k}^2+\Gamma^2}\frac{\Gamma/\pi}{\epsilon_{\bm{k}+\bm{p}}^2+\Gamma^2}\\
=&\int \frac{dk_+dk_-}{16\pi^2\sqrt{a}}(k_++k_-)\frac{\Gamma/\pi}{k_+^2k_-^2+\Gamma^2}\frac{\Gamma/\pi}{(k_++p_+)^2(k_-+p_-)^2+\Gamma^2}\\
=&\frac{\Gamma^2}{16\pi^4\sqrt{a}|\epsilon_{\bm p}|^3}\int dxdy\frac{xp_++yp_-}{x^2y^2+\Gamma^2/\epsilon_{\bm p}^2} \frac{1}{(x+1)^2(y+1)^2+\Gamma^2/\epsilon_{\bm p}^2}\\
=&\frac{\Gamma^2p_x}{8\pi^4\sqrt{a}|\epsilon_{\bm p}|^3}\int dxdy\frac{x}{x^2y^2+\Gamma^2/\epsilon_{\bm p}^2} \frac{1}{(x+1)^2(y+1)^2+\Gamma^2/\epsilon_{\bm p}^2}\\
=&-\frac{\Gamma p_x}{8\pi^3\sqrt{a}|\epsilon_{\bm p}|^2}\int dy\frac{\mathrm{Abs}(1+y)}{y^2(1+y)^2+\frac{\Gamma^2}{\epsilon_{\bm q}^2}\{1+2y(1+y)+2\mathrm{Abs}[y(1+y)]\}}\\
= &-\frac{p_x}{8\pi^2\sqrt{a}|\epsilon_{\bm p}|} f(\Gamma/|\epsilon_{\bm p}|)\\
\approx &\left\{\begin{matrix}
 -\frac{p_x}{8\pi^2\sqrt{a}|\epsilon_{\bm p}|} & \Gamma\ll |\epsilon_{\bm p}|\\
 -\frac{p_x}{16\pi^3\sqrt{a}\Gamma}\ln \frac{4e\Gamma}{|\epsilon_{\bm p}|} & \Gamma\gg |\epsilon_{\bm p}|
\end{matrix}\right.,
\end{aligned}
\label{tem3}
\end{equation}
Where $f(\Gamma/|\epsilon_{\bm p}|)$ has been defined in Eq.~(\ref{gam=del}): when $\Gamma\ll |\epsilon_{\bm p}|$,  $f(\Gamma/|\epsilon_{\bm p}|)=1$; otherwise, when  $\Gamma\gg |\epsilon_{\bm p}|$, $f(\Gamma/|\epsilon_{\bm p}|)=\frac{|\epsilon_{\bm p}|}{2\pi \Gamma}\ln\frac{4e\Gamma}{|\epsilon_{\bm p}|}$. Note that the result in the limit $\Gamma\ll |\epsilon_{\bm p}|$ is exactly the same as the one obtained by directly applying the approximation (\ref{vapp}) in the calculations. Then, we compute the remaining frequency and momentum integrals:
\begin{equation}
	\begin{aligned}
\Xi_{{\rm AL+SE+V},J}^{\rm w}(i\Omega_m)
 = &\frac{|J|^2N}{4\pi^2\sqrt{a}(|\Omega_m|+2\Gamma)^2}\int_0^{|\Omega_m|} d\Omega (|\Omega_m|-\Omega) \int \frac{d^2\bm{p}}{(2\pi)^2}\frac{p_x^2}{|\epsilon_{\bm p}|}f(\frac{\Gamma}{|\epsilon_{\bm p}|})\frac{1}{\bm{p}^2+\tilde{J}^2\frac{|\Omega|}{|\epsilon_{\bm p}|}f(\frac{\Gamma}{|\epsilon_{\bm p}|})}\\
  = &\frac{|J|^2N}{4\pi^2\sqrt{a}(|\Omega_m|+2\Gamma)^2}\int \frac{d^2\bm{p}}{(2\pi)^2}p_x^2\{\frac{\bm{p}^2|\epsilon_{\bm p}|}{\tilde{J}^4f(\frac{\Gamma}{|\epsilon_{\bm p}|})}\ln[1+\frac{\tilde{J}^2|\Omega_m|}{\bm{p}^2|\epsilon_{\bm p}|}f(\frac{\Gamma}{|\epsilon_{\bm p}|})]+\frac{|\Omega_m|}{\tilde{J}^2}\ln[\frac{1}{e}+\frac{\tilde{J}^2|\Omega_m|}{e\bm{p}^2|\epsilon_{\bm p}|}f(\frac{\Gamma}{|\epsilon_{\bm p}|})]\}\\
  \end{aligned}
  \label{wsig}
\end{equation}
After performing the analytic continuation, the optical conductivity is obtained as:
\begin{equation}
	\begin{aligned}
		\sigma_{\rm w}^{\rm J}(|\Omega|\ll \Gamma)=&-\frac{\mathrm{Im}[\Xi^{\rm w}_{\rm SE+V+AL}(i\Omega_m,0)]|_{i\Omega_m\rightarrow \Omega+i0^+}}{\Omega}\\
		\approx &\frac{|J|^2N\Gamma^2}{2^4\pi^2\sqrt{a}\Gamma^4\Omega}\int \frac{d^2\bm{p}}{(2\pi)^2}p_x^2\{\frac{\bm{p}^2|\epsilon_{\bm p}|}{\tilde{J}^4f(\frac{\Gamma}{|\epsilon_{\bm p}|})}\mathrm{ArcTan}[\frac{\tilde{J}^2\Omega}{\bm{p}^2|\epsilon_{\bm p}|}f(\frac{\Gamma}{|\epsilon_{\bm p}|})]+\frac{\Omega}{2\tilde{J}^2}\ln[\frac{1}{e^2}+\frac{\tilde{J}^4\Omega^2}{e^2\bm{p}^4|\epsilon_{\bm p}|^2}f^2(\frac{\Gamma}{|\epsilon_{\bm p}|})]\}\\
		= &\frac{|J|^2N\Gamma^4}{2^5\pi^4\sqrt{a}\tilde{J}^2\Gamma^4}\int_0^\infty dxd\theta \cosh(2\theta)x\{\frac{x^2\cosh(2\theta)\Gamma^2}{\tilde{J}^2\Omega f(1/x)}\mathrm{ArcTan}[\frac{\tilde{J}^2\Omega}{x^2\Gamma^2\cosh(2\theta)}f(1/x)]\\
		&+\frac{1}{2}\ln[\frac{1}{e^2}+\frac{\tilde{J}^4\Omega^2}{e^2\cosh(2\theta)^2\Gamma^4x^4}f^2(1/x)]\}
	\end{aligned}
\end{equation}
 where $x\equiv p^2/\Gamma$. Similarly, we perform analytical calculations in the two limits in Eq.~(\ref{tem3}) and subsequently provide the precise numerical results by using the exact boson self-energy $\frac{1}{4\pi^2\sqrt{a}|\epsilon_{\bm p}|} f(\frac{\Gamma}{|\epsilon_{\bm p}|})$ (\ref{gam=del}). When $\Gamma \ll |\epsilon_{\bm p}|$, $f(1/x)=1$, thus:
\begin{equation}
	\begin{aligned}
		\sigma_{\rm w}^{\rm J}(|\Omega|\ll \Gamma)=&\frac{|J|^2N}{2^5\pi^4\sqrt{a}\tilde{J}^2}\int_1^\infty dxd\theta \cosh(2\theta)x\{\frac{x^2\cosh(2\theta)\Gamma^2}{\tilde{J}^2\Omega }\mathrm{ArcTan}[\frac{\tilde{J}^2\Omega}{x^2\Gamma^2\cosh(2\theta)}]	+\frac{1}{2}\ln[\frac{1}{e^2}+\frac{\tilde{J}^4\Omega^2}{e^2\cosh(2\theta)^2\Gamma^4x^4}]\}\\
		=&\frac{|J|^2N\Gamma^2}{2^7\pi^4\sqrt{a}\tilde{J}^4\Omega}\int d\theta \cosh^2(2\theta)[\frac{\tilde{J}^2\Omega}{\cosh(2\theta)\Gamma^2}+(\frac{\tilde{J}^4\Omega^2}{\cosh^2(2\theta)\Gamma^4}-1)\mathrm{ArcTan}(\frac{\tilde{J}^2\Omega}{\cosh(2\theta)\Gamma^2})\\
		&-\frac{\tilde{J}^2\Omega}{\cosh(2\theta)\Gamma^2}\ln(1+\frac{\tilde{J}^4\Omega^2}{\cosh^2(2\theta)\Gamma^4})]\\
		\approx &\left\{\begin{matrix}
\frac{|J|^2N\tilde{J}^2\Omega^2}{1536\pi^3\sqrt{a}\Gamma^4} & \tilde{J}^2|\Omega|\ll \Gamma^2\\
 \frac{|J|^2N|\Omega|}{512\pi^3\sqrt{a}\Gamma^2}\ln\frac{2\tilde{J}^2|\Omega|}{e^{3/2}\Gamma^2} & \tilde{J}^2|\Omega|\gg \Gamma^2
\end{matrix}\right.
	\end{aligned}
	\label{condis}
\end{equation}
In the opposite limit $\Gamma \gg |\epsilon_{\bm p}|$,  $f(1/x)=x\ln(4e/x)/(2\pi)$, thus:
\begin{equation}
	\begin{aligned}
		\sigma_{\rm w}^{\rm J}(|\Omega|\ll \Gamma)\approx&\frac{|J|^2N\Gamma^4}{2^{10}\pi^5\sqrt{a}\tilde{J}^4\Gamma^2\Omega}\int_0^1 dx \frac{x^2}{\ln(4e/x)}[\frac{2\tilde{J}^2\Omega\ln(4e/x) }{\Gamma^2x}+(\frac{4\tilde{J}^4\Omega^2 \ln(4e/x)^2}{\Gamma^4x^2}-1)\mathrm{ArcTan}(\frac{2\tilde{J}^2\Omega \ln(4e/x)}{\Gamma^2x})\\
		&-\frac{2\tilde{J}^2\Omega\ln(4e/x) }{\Gamma^2x}\ln(1+\frac{4\tilde{J}^4\Omega^2 \ln(4e/x)^2}{\Gamma^4x^2})]\\
		\approx &\left\{\begin{matrix}
\frac{|J|^2N\tilde{J}^2\Omega^2}{384\pi^5\sqrt{a}\Gamma^4}\ln \frac{\Gamma^2e^{11/6}}{2\tilde{J}^2|\Omega|} & \tilde{J}^2|\Omega|\ll \Gamma^2\\
\frac{|J|^2N\ln(4e^2)|\Omega|}{512\pi^4\sqrt{a}\Gamma^2} & \tilde{J}^2|\Omega|\gg \Gamma^2
\end{matrix}\right.
	\end{aligned}
	\label{tem9}
\end{equation}
As shown in Fig.~\ref{Figs2}a and b, we numerically calculated the optical conductivity by substituting the complete boson self-energy $\frac{1}{4\pi^2\sqrt{a}|\epsilon_{\bm p}|} f(\frac{\Gamma}{|\epsilon_{\bm p}|})$ (\ref{gam=del})  and formula  (\ref{tem3}). It was found that in the $\tilde{J}^2|\Omega|\gg \Gamma^2$ limit, the optical conductivity ${\rm Re}[\sigma^{\rm J}]$ is proportional to the frequency, up to a logarithmic correction. 
In the $\tilde{J}^2|\Omega|\ll \Gamma^2$ limit,  the results shown in Fig.~\ref{Figs2}a,  Eqs.~(\ref{condis}) and (\ref{tem9}) are very small, and within the framework of perturbation theory, it is at the order of $\mathcal{O}(N^2|J|^4)$. A larger contribution comes from the real part provided by the integral in Eq.~(\ref{wsig}), while the prefactor of the integral contributes to the imaginary part. Thus, we obtain ${\rm Re}[\sigma_{\rm w}^{\rm J}(\tilde{J}^2|\Omega|\ll \Gamma^2)]\sim \frac{N|J|^2\Omega^2}{\Gamma^3}\ln \frac{\Lambda_{\rm U}^2}{\Gamma}$.

So, based on Eq.~(\ref{sigmaJ}, \ref{condis}, \ref{tem9}),  the optical conductivity contribution from the interaction can be expressed as:
\begin{equation}
		{\rm Re}[\sigma^{\rm J}_{\rm w}(\frac{\Gamma^2}{\tilde{J}^2}<|\Omega|\ll \Gamma)]\sim {\rm sgn}(\Gamma-|\Omega|) \frac{|J|^2N|\Omega|}{\max\{\Gamma^2,\Omega^2\}}, \ \ \ \ {\rm Re}[\sigma^{\rm J}(\Gamma=0, |\Omega|)]\sim {\rm sgn}(\Gamma-|\Omega|)\frac{|J|^2N|\Omega|}{\max\{\Gamma^2,\Omega^2\}}		
\end{equation}
where we have omitted the logarithmic dependence. So, to summarize, in the limit $\frac{\Gamma^2}{\tilde{J}^2}<|\Omega|$, there is alway:
\begin{equation}
		{\rm Re}[\sigma^{\rm J}_{\rm w}(\frac{\Gamma^2}{\tilde{J}^2}<|\Omega|)]\sim {\rm sgn}(\Gamma-|\Omega|)\frac{|J|^2N|\Omega|}{\max\{\Gamma^2,\Omega^2\}}.	
		\label{tem10}
\end{equation}

Specifically, the one-loop diagram Fig.~\ref{figs2}a gives a conductivity that is proportional to $1/\Gamma$:
\begin{equation}
\begin{aligned}
	\Xi_0^{\rm w}(i\Omega_m,0)=&-NT\sum_n\int \frac{d^2\bm{k}}{(2\pi)^2}4k_x^2G_{\rm w}(i\omega_n+i\Omega_m,\bm{k})G_{\rm w}(i\omega_n,\bm{k})\\
	\approx &-NT\sum_n\int \frac{d^2\bm{k}}{(2\pi)^2}4k_x^2\frac{\frac{1}{i\omega_n+i\Gamma{\rm sgn}(\omega_n)-\epsilon_{\bm k}}-\frac{1}{i\omega_n+i\Omega_m+i\Gamma{\rm sgn}(\omega_n+\Omega_m)-\epsilon_{\bm k}}}{i\Omega_m+i\Gamma [\mathrm{sgn}(\omega_n+\Omega_m)-\mathrm{sgn}(\omega_n)]}\\
	=&-\frac{2N}{\pi\sqrt{a}} T\sum_n\int_0^{\Lambda_{\rm U}} dk_x k_x^2\frac{\frac{1}{\sqrt{i\omega_n+i\Gamma{\rm sgn}(\omega_n)-k_x^2}}-\frac{1}{\sqrt{i\omega_n+i\Omega_m+i\Gamma{\rm sgn}(\omega_n+\Omega_m)-k_x^2}}}{i\Omega_m+i\Gamma [\mathrm{sgn}(\omega_n+\Omega_m)-\mathrm{sgn}(\omega_n)]}\\
	\approx &-\frac{N\Lambda_{\rm U}^2}{\pi\sqrt{a}} T\sum_n\frac{\mathrm{sgn}(\omega_n+\Omega_m)-\mathrm{sgn}(\omega_n)}{\Omega_m+\Gamma [\mathrm{sgn}(\omega_n+\Omega_m)-\mathrm{sgn}(\omega_n)]}\\
	=& -\frac{N\Lambda_{\rm U}^2}{\pi^2\sqrt{a}(|\Omega_m|+2\Gamma)}|\Omega_m|
\end{aligned}	
\end{equation}
Thus, the contribution of the single-loop diagram to the conductivity is:
\begin{equation}
		\sigma^0(|\Omega|\ll\Gamma)=-\frac{\mathrm{Im}[\Xi^{\rm w}_0(i\Omega_m,0)]|_{i\Omega_m\rightarrow \Omega+i0^+}}{\Omega}=-\frac{N\Lambda_{\rm U}^2}{2\pi^2\sqrt{a}\Gamma}
		\label{tem10}
\end{equation}


\subsubsection{Boson thermal mass from interaction $J$}
\label{BTMJGam}
Similar to the discussion in Sec.~3.1.2, even in the presence of impurity scattering, the contribution from the VHS up to the leading order at $o(|J|^4)$ has no effect on the thermal mass.
For the convex Fermi surface, Ref.~\cite{Patel2023} demonstrates that $m^2(T)\simeq \tilde{J}^2 T\ln (T)/\Gamma$. As a result, the total effective thermal mass takes a form, 
\begin{equation}
m^2(T)\simeq \frac{\tilde{J}^2}{\Gamma} T\ln T.
\label{mTJGam}
\end{equation}

\subsubsection{The finite temperature ($|\Omega| \ll T$) optical conductivity}
The main difference from the zero-temperature case is that, in the presence of potential disorder, the Matsubara frequency components of the electronic Green’s function cannot be neglected as it is temperature-dependent. The electronic Green’s function can be expressed as:
\begin{equation}
	G_{\rm w}(i\omega_n,\bm{k})=\frac{1}{i\omega_n+i\Gamma\mathrm{sgn}(\omega_n)-\epsilon_{\bm k}}.
	\label{egT}
\end{equation}
At this point, Eq.~(\ref{con-for}) needs to be modified to:
\begin{equation}
\begin{aligned}
	G_{\rm w}(i\omega_n+i\Omega_m,\bm{k})G_{\rm w}(i\omega_n,\bm{k})=\frac{G_{\rm w}(i\omega_n,\bm{k})-G_{\rm w}(i\omega_n+i\Omega_m,\bm{k})}{i\Omega_m+i\Gamma [\mathrm{sgn}(\omega_n+\Omega_m)-\mathrm{sgn}(\omega_n)]}\\
\end{aligned}
	\label{ggT}	
\end{equation}

We first calculate in the limit $\Gamma\ll T$, where the Green’s function on the right-hand side of the above equation can be approximated as in the case without potential disorder. For the self-energy diagrams Fig.~\ref{figs2}b, repeatedly using Eqs.~(\ref{ggT}), we obtain:
\begin{equation}
\begin{aligned}
	\Xi_{\rm SE,J}^{\rm w}(i\Omega_m,\Gamma\ll T)=&-NT\sum_n\int \frac{d^2\bm{k}}{(2\pi)^2} v_{\bm k}^2 G_{\rm w}(i\omega_n,\bm{k})[G_{\rm w}(i\Omega_m+i\omega_n,\bm{k})]^2\Sigma_{\rm J}(i\omega_n+i\Omega_m,\bm{k})+(\Omega_m\rightarrow -\Omega_m)\\
	=& NT\sum_n\int \frac{d^2\bm{k}}{(2\pi)^2} v_{\bm k}^2 \frac{G_{\rm w}(i\omega_n,\bm{k})-G_{\rm w}(i\Omega_m+i\omega_n,\bm{k})}{\{i\Omega_m+i\Gamma[\mathrm{sgn}(\omega_n+\Omega_m)-\mathrm{sgn}(\omega_n)]\}^2}[\Sigma_{\rm J}(i\omega_n,\bm{k})-\Sigma_{\rm J}(i\omega_n+i\Omega_m,\bm{k})]\\
	\approx &N\int \frac{d^2\bm{k}}{(2\pi)^2}\int \frac{d\mu}{\pi} v_{\bm k}^2 \frac{\mathrm{Im}\Sigma^{\rm R}(\mu,\bm{k})}{(|\Omega_m|+2\Gamma)^2}[n_{\rm F}(\epsilon_{\bm k})-n_{\rm F}(\mu)][\frac{2}{\epsilon_{\bm k}-\mu}-\frac{1}{\epsilon_{\bm k}-\mu+i\Omega_m}-\frac{1}{\epsilon_{\bm k}-\mu-i\Omega_m}]
	\end{aligned}
\end{equation}
Its imaginary part is given by:
\begin{equation}
	\mathrm{Im}\Xi_{\rm SE,J}^{\rm w}(|\Omega|\ll \Gamma\ll T)\approx \frac{N}{4\Gamma^2}\int \frac{d^2\bm{k}}{(2\pi)^2} v_{\bm k}^2 \mathrm{Im}\Sigma^{\rm R}(\Omega+\epsilon_{\bm k},\bm{k})[n_{\rm F}(\epsilon_{\bm k})-n_{\rm F}(\epsilon_{\bm k}+\Omega)]-(\Omega\rightarrow -\Omega)
\end{equation}
The primary difference compared to the case with potential disorder $|w| = 0$  (no impurities) is the absence of the Drude peak.

The vertex diagram in Fig.~\ref{figs2}c gives us
\begin{equation}
\begin{aligned}
	\Xi_{\rm V,J}^{\rm w}(i\Omega_m,\Gamma\ll T)=&-N|J|^2T\sum_l T\sum_n\int \frac{d^2\bm{k_2}}{(2\pi)^2} \int \frac{d^2\bm{k_1}}{(2\pi)^2} v_{{\bm k}_1}v_{{\bm k}_1+{\bm k}_2} G_{\rm w}(i\omega_n,\bm{k_1})G_{\rm w}(i\Omega_m+i\omega_n,\bm{k_1})D(i\Omega_l,\bm{k_2})\\
	&\times G_{\rm w}(i\omega_n+i\Omega_l+i\Omega_m,\bm{k_1}+\bm{k_2})G_{\rm w}(i\omega_n+i\Omega_l,\bm{k_1}+\bm{k_2})\\
	=&N|J|^2T\sum_l T\sum_n\int \frac{d^2\bm{k_2}}{(2\pi)^2}\int \frac{d\mu}{\pi} \int \frac{d^2\bm{k_1}}{(2\pi)^2} v_{{\bm k}_1}v_{{\bm k}_1+{\bm k}_2} \frac{G_{\rm w}(i\omega_n,\bm{k_1})-G_{\rm w}(i\Omega_m+i\omega_n,\bm{k_1})}{i\Omega_m+i\Gamma[\mathrm{sgn}(\omega_n+\Omega_m)-\mathrm{sgn}(\omega_n)]}\frac{\mathrm{Im}D^{\rm R}(\mu,\bm{k_2})}{i\Omega_l-\mu}\\
	&\times \frac{G_{\rm w}(i\omega_n+i\Omega_l,\bm{k_1}+\bm{k_2})-G_{\rm w}(i\omega_n+i\Omega_l+i\Omega_m,\bm{k_1}+\bm{k_2})}{i\Omega_m+i\Gamma[\mathrm{sgn}(\omega_n+\Omega_l+\Omega_m)-\mathrm{sgn}(\omega_n+\Omega_l)]}\\
	\approx &N|J|^2T\sum_n\int \frac{d^2\bm{k_2}}{(2\pi)^2}\int \frac{d\mu}{\pi} \int \frac{d^2\bm{k_1}}{(2\pi)^2} v_{{\bm k}_1}v_{{\bm k}_1+{\bm k}_2} \frac{G^0(i\omega_n,\bm{k_1})-G^0(i\Omega_m+i\omega_n,\bm{k_1})}{i\Omega_m+i\Gamma[\mathrm{sgn}(\omega_n+\Omega_m)-\mathrm{sgn}(\omega_n)]}\mathrm{Im}D^{\rm R}(\mu,\bm{k_2})\\
	&\times \{\frac{n_{\rm B}(\mu)}{i\Omega_m+i
	\Gamma[\mathrm{sgn}(\omega_n+\Omega_m)-\mathrm{sgn}(\omega_n)]}+\frac{n_{\rm F}(\epsilon_{\bm{k}_1+\bm{k}_2})}{i\Omega_m+i\Gamma\mathrm{sgn}(\Omega_m)}\}(\frac{1}{i\omega_n+\mu-\epsilon_{\bm{k}_1+\bm{k}_2}}\\
	&-\frac{1}{i\omega_n+i\Omega_m+\mu-\epsilon_{\bm{k}_1+\bm{k}_2}})\\
	\approx &-N|J|^2\int \frac{d^2\bm{k_2}}{(2\pi)^2}\int \frac{d\mu}{\pi} \int \frac{d^2\bm{k_1}}{(2\pi)^2} v_{{\bm k}_1}v_{{\bm k}_1+{\bm k}_2} \frac{n_{\rm F}(\epsilon_{\bm{k}_1})-n_{\rm F}(\epsilon_{\bm{k}_1+\bm{k}_2}-\mu)}{(|\Omega_m|+2\Gamma)^2}\mathrm{Im}D^{\rm R}(\mu,\bm{k_2})\\
	&\times [n_{\rm B}(\mu)+n_{\rm F}(\epsilon_{\bm{k}_1+\bm{k}_2})][\frac{2}{\epsilon_{\bm{k}_1}+\mu-\epsilon_{\bm{k}_1+\bm{k}_2}}-\frac{1}{\epsilon_{\bm{k}_1}+i\Omega_m+\mu-\epsilon_{\bm{k}_1+\bm{k}_2}}-\frac{1}{\epsilon_{\bm{k}_1}-i\Omega_m+\mu-\epsilon_{\bm{k}_1+\bm{k}_2}}]\\
		\end{aligned}
\end{equation}
Its imaginary part is given by:
\begin{equation}
\begin{aligned}
	\mathrm{Im}\Xi_{\rm V,J}^{\rm w}(|\Omega|\ll \Gamma\ll T)\approx &-\frac{N|J|^2}{4\Gamma^2}\int \frac{d^2\bm{k_2}}{(2\pi)^2} \int \frac{d^2\bm{k_1}}{(2\pi)^2} v_{{\bm k}_1}v_{{\bm k}_1+{\bm k}_2} [n_{\rm F}(\epsilon_{\bm{k}_1})-n_{\rm F}(\epsilon_{\bm{k}_1}+\Omega)]\mathrm{Im}D^{\rm R}(\epsilon_{\bm{k}_1+\bm{k}_2}-\epsilon_{\bm{k}_1}-\Omega,\bm{k_2})\\
	&\times [n_{\rm B}(\epsilon_{\bm{k}_1+\bm{k}_2}-\epsilon_{\bm{k}_1}-\Omega)+n_{\rm F}(\epsilon_{\bm{k}_1+\bm{k}_2})]-(\Omega \rightarrow-\Omega)\\
	\end{aligned}
\end{equation}
Noting  $\mathrm{Im}\Sigma^{\rm R}(\epsilon_{\bm{k}_1}+\Omega,\bm{k}_1)\approx |J|^2\int \frac{d^2\bm{k_2}}{(2\pi)^2}[n_{\rm B}(\epsilon_{\bm{k}_1+\bm{k}_2}-\epsilon_{\bm{k}_1}-\Omega)+n_{\rm F}(\epsilon_{\bm{k}_1+\bm{k}_2})]\mathrm{Im}D^{\rm R}(\epsilon_{\bm{k}_1+\bm{k}_2}-\epsilon_{\bm{k}_1}-\Omega,\bm{k_2})$. The sum of the self-energy diagram and the vertex diagram contributes to the optical conductivity of the extra channel. The main contribution comes from the following regions:
\begin{equation}
\begin{aligned}
	&|\epsilon_{\bm{k}_1}|\ll T\\
	&|\epsilon_{\bm{k}_1+\bm{k}_2}|\ll T\\
	&|\Delta \epsilon(\bm{k}_1,\bm{k}_2)|\ll |\epsilon_{\bm{k}_2}|/f(\Gamma/|\epsilon_{\bm{k}_2}|).
	\end{aligned},
\end{equation}
The last equality actually comes from $|\Omega|\ll |\epsilon_{\bm{k}_2}|/f(\Gamma/|\epsilon_{\bm{k}_2}|)$, which can be divided into two regions: (1) $\Gamma\ll |\epsilon_{\bm k}|, \ \ |\Omega| \ll |\epsilon_{\bm k}|$ ;(2) $\Gamma\gg |\epsilon_{\bm k}|, \ \ |\Omega| \ll \Gamma$. Therefore, we need to analyze why the damping takes the form of $|\Omega|f(\Gamma/|\epsilon_{\bm{k}_2}|)/|\epsilon_{\bm{k}_2}|$ only in these two regions. For (1), the Fermi propagator can be approximated using Eq.~(\ref{vapp}), leading to a damping term of $|\Omega|f(\Gamma/|\epsilon_{\bm{k}_2}|)/|\epsilon_{\bm{k}_2}|$. For (2), the damping term is given by the result of Eq.~(\ref{tem5}), which is also $|\Omega|f(\Gamma/|\epsilon_{\bm{k}_2}|)/|\epsilon_{\bm{k}_2}|$. Outside regions (1) and (2), there is only $\Gamma \ll |\epsilon_{\bm k}|\ll |\Omega|$ or $|\epsilon_{\bm k}|\ll \Gamma \ll |\Omega|$, which is equivalent to requiring the use of the Fermi Green’s function $1/(i\omega_n-\epsilon_{\bm k})$ to compute the bosonic energy $\Pi_{\rm J}(|\Omega|,{\bm k}=0)$. The result, given by Eq.~(\ref{PiJ}), is zero.

The total conductivity contributed by the extra channel is given by:
\begin{equation}
\begin{aligned}
\mathrm{Re}[\sigma^{\rm J}_{\rm w}(\Omega=0,\Gamma\ll T)]\simeq &-\frac{N|J|^2\tilde{J}^2}{32\Gamma^2\pi^2 } \int_{|\epsilon_{\bm{k}_2}|<4T} \frac{d^2\bm{k}_2}{(2\pi)^2} \int^{-\frac{k_{2y}}{2}+\frac{\sqrt{T}|k_{2x}|}{\sqrt{a|\epsilon_{\bm{k}_2}|}}}_{-\frac{k_{2y}}{2}-\frac{\sqrt{T}|k_{2x}|}{\sqrt{a|\epsilon_{\bm{k}_2}|}}} d k_{1y}\int^{|\epsilon_{\bm{k}_2}/|f(\Gamma/|\epsilon_{\bm{k}_2}|)}_{-|\epsilon_{\bm{k}_2}|/f(\Gamma/|\epsilon_{\bm{k}_2}|)}d\Delta \epsilon(\bm{k}_1,\bm{k}_2)\\
& \frac{\Delta \epsilon(\bm{k}_1,\bm{k}_2)-ak_{1y}^2+a(k_{1y}+k_{2y})^2-k_{2x}^2}{|k_{2x}||\epsilon_{\bm{k}_2}/f(\Gamma/|\epsilon_{\bm{k}_2}|)|}\frac{1}{[\bm{k}_2^2+m^2(T)]^2+\tilde{J}^4[\frac{\Delta \epsilon(\bm{k}_1,\bm{k}_2)}{\epsilon_{\bm{k}_2}/f(\Gamma/|\epsilon_{\bm{k}_2}|)}]^2}\\
& =\frac{N|J|^2\sqrt{T}}{8\Gamma^2\pi^2 \sqrt{a}}\int_{|\epsilon_{\bm{k}_2}|<4T} \frac{d^2\bm{k}_2}{4\pi^2}\frac{k_{2x}^2}{\sqrt{|\epsilon_{\bm{k}_2}|}}
  \frac{1}{\bm{k}_2^2+m^2(T)}\mathrm{ArcTan}[\frac{\tilde{J}^2}{\bm{k}_2^2+m^2(T)}]
\end{aligned}
\end{equation}
Thus, the DC conductivity contributed by the interactions is:
\begin{equation}
	\mathrm{Re}[\sigma^{\rm J}_{\rm w}(\Omega=0,\Gamma\ll T)]\sim 
	\left\{\begin{matrix}
\frac{N^2|J|^4T}{N'\Gamma^2m^2(T)} & m^2(T)\gg N|J|^2/N' \\
  \frac{N|J|^2T}{\Gamma^2}\ln(\frac{N|J|^2}{N'm^2(T)})& m^2(T)\ll N|J|^2/N'
\end{matrix}\right.
\end{equation}
This yields Eq.~(11) in the main text.
By substituting the bosonic thermal mass in Eq.~(\ref{mTJGam}), the above expression can be further simplified to:
\begin{equation}
	\mathrm{Re}[\sigma^{\rm J}_{\rm w}(\Omega=0,\Gamma\ll T)]\sim 
\frac{N|J|^2}{\Gamma}\ln T.
\end{equation}

In the opposite limit $\Gamma\gg T$, the Green’s function on the right-hand side of Eq.~(\ref{ggT}) can be approximated as $G_{\rm w}(i\omega_n,{\bm k})\approx 1/[i\Gamma\mathrm{sgn}(\omega_n)-\epsilon_{\bm k}]$. In this case, the entire calculation process is no different from the zero-temperature case, except that the frequency integral needs to be replaced with a summation. Ultimately, we obtain:
\begin{equation}
	\begin{aligned}
\Xi_{{\rm AL+SE+V},J}^{\rm w}(i\Omega_m,\Gamma\gg T)
 \approx &\frac{|J|^2NT}{2\pi\sqrt{a}(|\Omega_m|+2\Gamma)^2}\sum_{n=0}^{|m|} (|\Omega_m|-\Omega_n) \int \frac{d^2\bm{p}}{(2\pi)^2}\frac{p_x^2}{|\epsilon_{\bm p}|}f(\frac{\Gamma}{|\epsilon_{\bm p}|})\frac{1}{\bm{p}^2+m^2(T)+\tilde{J}^2\frac{|\Omega_n|}{|\epsilon_{\bm q}|}f(\frac{\Gamma}{|\epsilon_{\bm p}|})}\\
 =&- \frac{|J|^2N}{4\pi^2\sqrt{a}(|\Omega_m|+2\Gamma)^2\tilde{J}^2}\int \frac{d^2\bm{p}}{(2\pi)^2}p_x^2\{|\Omega_m|[1+{\rm Polygamma}(0,b)\\
 &-{\rm Polygamma}(0,1+|m|+b)]+2\pi Tb[{\rm Polygamma}(0,1+b)-{\rm Polygamma}(0,1+|m|+b)]\}
  \end{aligned}
\end{equation}
where ${\rm Polygamma}$ is the polygamma function, and $b=[{\bm p}^2+m^2(T)]|\epsilon_{\bm p}|/2\pi T\tilde{J}^2f(\Gamma/|\epsilon_{\bm p}|)$. In the following calculations, we only provide the result in quantum critical regime $m^2(T)\ll \tilde{J}^2$ [$m^2(T)\rightarrow 0$]. After performing the analytic continuation, the conductivity is obtained as:
\begin{equation}
	\begin{aligned}
		&\mathrm{Re}[\sigma^{\rm J}_{\rm w}(\Omega=0,\Gamma\gg T)]\\
		=&-\frac{\mathrm{Im}[\Xi^{\rm w}_{\rm SE+V+AL}(i\Omega_m,0)]|_{i\Omega_m\rightarrow \Omega+i0^+}}{\Omega}\\
		= &\frac{|J|^2N}{16\pi^2\sqrt{a}\Gamma^2\tilde{J}^2}\int \frac{d^2\bm{p}}{(2\pi)^2}p_x^2[b{\rm Polygamma}(1,1+b)+{\rm Polygamma}(0,1+b)-{\rm Polygamma}(0,b)-1]\\
		= &\frac{|J|^2N}{32\pi^4\sqrt{a}\tilde{J}^2}\int_0^\infty dx d\theta  x[\tilde{b}{\rm Polygamma}(1,1+\tilde{b})+{\rm Polygamma}(0,1+\tilde{b})-{\rm Polygamma}(0,\tilde{b})-1]\\
		\approx & \frac{|J|^2N\Lambda_\theta T}{32\pi^3\sqrt{a}\Gamma^2}\ln\frac{\Lambda_{\rm U}^2}{\Gamma}.
	\end{aligned}
	\label{tem11}
\end{equation}
where $\tilde{b}=x^2/f(\sqrt{\cosh(\theta)}/x)/(2\pi\tilde{J}^2T/\Gamma^2)$. This corresponds to Eq.~(10) in the main text

Specifically, the one-loop diagram Fig.~\ref{figs2}a gives a conductivity (\ref{tem10}) that is proportional to $1/\Gamma$.

Combining Eqs.~(\ref{tem11}), and (\ref{tem10}), we obtain the total resistivity in the quantum critical region $m^2(T) \ll \tilde{J}^2$ (i.e. $T\ll \Gamma$) as:
\begin{equation}
	\rho(\Omega\ll \Gamma, T\ll \Gamma)=\frac{1}{\mathrm{Re}\sigma^0(\Omega\ll \Gamma,T\ll \Gamma)+\mathrm{Re}\sigma^{\rm J}_{\rm w}(\Omega\ll \Gamma,T\ll \Gamma)}\sim
 -\frac{\Gamma}{N\Lambda_{\rm U}^2}-\frac{|J|^2 T}{N\Lambda_{\rm U}^4}\ln\frac{\Lambda_{\rm U}^2}{\Gamma}
\end{equation}
This yields Eq.~(12) in the main text.

\subsection{$\sigma^{\rm J'}(\Omega,T)$ from spatially disordered interaction}

Now we consider the contribution to the conductivity arising from the interaction  $J'$. where the bosons are damped as described by Eq.(\ref{sdc-bos}). In this section, all the fermion-boson interaction vertices in Fig.~\ref{figs2} involve only spatially disordered interaction $J'$, and all calculations are performed in the low-frequency limit.\\

\subsubsection{Derivation on the $T=0$ optical conductivity} 
Similar to the case without impurities, only the self-energy diagram contributes to the conductivity.
\begin{equation}
	\begin{aligned}
		\Xi_{\rm SE,J'}^{\rm w}(i\Omega_m)=&-NT\sum_n\int\frac{d^2 \bm{k}}{(2\pi)^2}4k_x^2G_{\rm w}(i\omega_n+i\Omega_m,\bm{k})G^2_{\rm w}(i\omega_n,\bm{k})\Sigma_{\rm J'}(i\omega_n)+(\Omega_m\rightarrow -\Omega_m)\\
		=&NT\sum_n\int\frac{d^2 \bm{k}}{(2\pi)^2}4k_x^2\frac{G_{\rm w}(i\omega_n+i\Omega_m,\bm{k})+G_{\rm w}(i\omega_n-i\Omega_m,\bm{k})-2G_{\rm w}(i\omega_n,\bm{k})}{(|\Omega_m|+2\Gamma)^2}\Sigma_{\rm J'}(i\omega_n)\\
		=&\frac{|J'|^2N\Lambda_\theta\Omega_m^2}{16\pi^4\sqrt{a}(|\Omega_m|+2\Gamma)^2}\ln(\frac{e^3\Lambda_{\rm U}^4}{\tilde{J'}^4\Omega_m^2})\int \frac{d^2 \bm{k}}{(2\pi)^2}4k_x^2\frac{\Gamma}{\Gamma^2+\epsilon_{\bm k}^2}\\
		=&\frac{|J'|^2N\Lambda_{\rm U}^2\Lambda_\theta\Omega_m^2}{16\pi^5a(|\Omega_m|+2\Gamma)^2}\ln(\frac{e^3\Lambda_{\rm U}^4}{\tilde{J'}^4\Omega_m^2})
	\end{aligned}	
\end{equation}
We find that this result is the optical conductivity without impurities multiplied by $\Omega^2/\Gamma^2$. So the total resistivity can thus be expressed as:
\begin{equation}
	\rho(|\Omega|\ll \Gamma,T=0)=\frac{1}{\mathrm{Re}\sigma(\Omega,T=0)}\sim-\frac{1}{\frac{N\Lambda_{\rm U}^2}{\Gamma}-\frac{|J'|^2N\Lambda_{\rm U}^2\Lambda_\theta|\Omega|}{\Gamma^2}}\approx -\frac{\Gamma}{N\Lambda_{\rm U}^2}-\frac{|J'|^2\Lambda_\theta}{N\Lambda_{\rm U}^2}|\Omega|.
\end{equation}

\subsubsection{The finite temperature ($|\Omega| \ll T$) optical conductivity} 
Similar to the process without disorder, here we only need to make the substitution  (\ref{vn0con}), and the other processes remain entirely the same. Thus, the final resistivity is obtained as follows:
\begin{equation}
	\rho(\Omega=0,T)=\frac{1}{\mathrm{Re}\sigma(\Omega=0,T)}\sim\left\{\begin{matrix}
-\frac{\Gamma}{N\Lambda_{\rm U}^2}-\frac{|J'|^4T^2}{ N'\Lambda_{\rm U}^2 m^2(T)} & m^2(T)\gg N|J'|^2T/N' \\
  -\frac{\Gamma}{N\Lambda_{\rm U}^2}-\frac{|J'|^2T}{N\Lambda_{\rm U}^2}\ln \frac{N'm^2(T)}{N|J'|^2T}& m^2(T)\ll N|J'|^2T/N'
\end{matrix}\right..
\end{equation}

\newpage
\section{NFLs with both $J$ and $J^\prime$ types Yukawa interaction while the potential disorder $|w|=0$}

In this section we show the NFL and MFL are stable solutions when both $J$ and $J^\prime$ interactions are switched on.
The NFL solution is found in the low energy limit, while it crossovers to the MFL at higher energies.\\

We start with the bosonic damping terms derived in Eq.(\ref{sic-bos-1})
and Eq.(\ref{sdc-bos}). The total boson self-energy correction in the presence of both $J$ and $J^\prime$ interactions is given by,
\begin{equation}
	\Pi(i\Omega_m,\bm{q})-\Pi(0,0)= -\tilde{J}^2\frac{|\Omega_m|}{|q_x^2-aq_y^2|}-\tilde{J}'^2|\Omega_m|.
	\label{bo-bo}
\end{equation}

Then, we evaluate the fermion self-energy using this damped boson in the large-$N$ limit. The total fermion self-energy is divided into two pieces $\Sigma(i\omega_n,0)=\Sigma_1(i\omega_n,0)+\Sigma_2(i\omega_n)$, where
\begin{equation}
	\begin{aligned}
 		\Sigma_1(i\omega_n,0)=|J|^2T\sum_m \Big[&\int_\mathbb{A} \frac{d^2 \bm{q}}{(2\pi)^2}\frac{1}{\bm{q}^2-\tilde{J}^2\frac{|\Omega_m|}{q_x^2-a q_y^2}+\tilde{J}'^2|\Omega_m|}\frac{1}{i\omega_n+i\Omega_m-q_x^2+a q_y^2}\\
 		&+\int_\mathbb{B}\frac{d^2 \bm{q}}{(2\pi)^2}\frac{1}{\bm{q}^2+\tilde{J}^2\frac{|\Omega_m|}{q_x^2-a q_y^2}+\tilde{J}'^2|\Omega_m|}\frac{1}{i\omega_n+i\Omega_m-q_x^2+a q_y^2}\Big],
 	\end{aligned}
 	\label{bo-fer1}
\end{equation}
and
\begin{equation}
	\begin{aligned}
		\Sigma_2(i\omega_n)=&|J'|^2T\sum_m \Big[\int_\mathbb{A} \frac{d^2 \bm{k}}{(2\pi)^2}\frac{1}{\bm{k}^2-\tilde{J}^2\frac{|\Omega_m|}{k_x^2-a k_y^2}+\tilde{J}'^2|\Omega_m|}+\int_\mathbb{B}\frac{d^2 \bm{k}}{(2\pi)^2}\frac{1}{\bm{k}^2+\tilde{J}^2\frac{|\Omega_m|}{k_x^2-a k_y^2}+\tilde{J}'^2|\Omega_m|}\Big]\\
 		&\quad \quad\quad \times \int \frac{d^2\bm{q}}{(2\pi)^2}\frac{1}{i\omega_n+i\Omega_m-q_x^2+a q_y^2}.	
	\end{aligned}
	\label{bo-fer2}
\end{equation}

Similar to the calculation routines in Sec.~\ref{J}, we obtain the integrals for the $\mathbb{A}$ and $\mathbb{B}$ momentum regions separately,
\begin{equation}
	\begin{aligned}
		&\int_\mathbb{A} \frac{d^2 \bm{q}}{(2\pi)^2}\frac{1}{\bm{q}^2-\tilde{J}^2\frac{|\Omega_m|}{q_x^2-a q_y^2}+\tilde{J}'^2|\Omega_m|}\frac{1}{i\omega_n+i\Omega_m-q_x^2+a q_y^2}\\
		&\simeq  \left\{\begin{matrix}
 					(\frac{1}{16|\tilde{J}|\sqrt{|\Omega_m}|}-\frac{\tilde{J}'^2\sqrt{a}}{16\pi\tilde{J}^2})[\ln \gamma-i\mathrm{sgn}(\omega_n+\Omega_m)]) &  \ \ \ 4\tilde{J}^2\gg a\tilde{J}'^4|\Omega_m|\\
					\frac{1}{8\pi\sqrt{a}\tilde{J}'^2|\Omega_m|}\ln\frac{a\tilde{J}'^4|\Omega_m|}{\tilde{J}^2}[\ln \gamma-i\mathrm{sgn}(\omega_n+\Omega_m)] & \ \ \  4\tilde{J}^2\ll a\tilde{J}'^4|\Omega_m|.
				\end{matrix}\right.				
	\end{aligned}
	\label{bo-fer1-1}
\end{equation}
The distinction between different regions essentially comes down to comparing the bosonic damping terms. Since the coefficient in front of the boson dispersion $\bm{q}^2$ is 1, $\tilde{J}^{\prime 4}|\Omega_m|^4$ and $4\tilde{J}^2|\Omega_m|$ are on equal footing. When $\tilde{J}^{\prime 4}|\Omega_m|^4\ll 4\tilde{J}^2|\Omega_m|$, Landau damping dominates, and vice versa. Applying the same low-frequency condition, we carry out the integral over the domain $\mathbb{B}$ is 
\begin{equation}
	\begin{aligned}
		&\int_\mathbb{B}\frac{d^2 \bm{q}}{(2\pi)^2}\frac{1}{\bm{q}^2+\tilde{J}^2\frac{|\Omega_m|}{q_x^2-a q_y^2}+\tilde{J}'^2|\Omega_m|}\frac{1}{i\omega_n+i\Omega_m-q_x^2+a q_y^2}\\
				 & \quad \simeq\left\{\begin{matrix}
 					-(\frac{1}{16|\tilde{J}|\sqrt{a}\sqrt{|\Omega_m}|}-\frac{\tilde{J}'^2}{16\pi\sqrt{a}\tilde{J}^2})[\ln \gamma+i\mathrm{sgn}(\omega_n+\Omega_m)]) &  \ \ \ 4\tilde{J}^2\gg \tilde{J}'^4|\Omega_m|,\\
					-\frac{1}{8\pi\sqrt{a}\tilde{J}'^2|\Omega_m|}\ln\frac{\tilde{J}'^4|\Omega_m|}{\tilde{J}^2}[\ln \gamma+i\mathrm{sgn}(\omega_n+\Omega_m)]&\ \ \ 4\tilde{J}^2\ll \tilde{J}'^4|\Omega_m|.
				\end{matrix}\right.
	\end{aligned}
	\label{bo-fer1-2}
\end{equation}
Eq.~(\ref{bo-fer1-1},\ref{bo-fer1-2}) have almost the same form, with a minor difference in the sgn, which can result into exact cancellation when $a=1$. The integrals over the domain $\mathbb{A}$ and $\mathbb{B}$ in Eq. (\ref{bo-fer2}) are

\begin{equation}
	\begin{aligned}
		& \int_\mathbb{A}  \frac{d^2 \bm{k}}{(2\pi)^2}\frac{1}{\bm{k}^2-\tilde{J}^2\frac{|\Omega_m|}{k_x^2-a k_y^2}+\tilde{J}'^2|\Omega_m|}+\int_\mathbb{B}\frac{d^2 \bm{k}}{(2\pi)^2}\frac{1}{\bm{k}^2+\tilde{J}^2\frac{|\Omega_m|}{k_x^2-a k_y^2}+\tilde{J}'^2|\Omega_m|}\\
		 &\quad \simeq \left\{\begin{matrix}
 				 \frac{\mathrm{ArcTan}(\sqrt{a})}{2\pi^2} \ln \frac{4a\Lambda_{\rm U}^4}{\tilde{J}^2|\Omega_m|}+ \frac{\mathrm{ArcTan}(1/\sqrt{a})}{2\pi^2} \ln \frac{4\Lambda_{\rm U}^4}{\tilde{J}^2|\Omega_m|}& \ \ \  4\tilde{J}^2\gg \mathrm{max}\{a\tilde{J}'^4|\Omega_m|,\tilde{J}'^4|\Omega_m|\}\\
 				\frac{\mathrm{ArcTan}(\sqrt{a})}{2\pi^2} \ln \frac{16\Lambda_{\rm U}^4}{\tilde{J}'^4|\Omega_m|^2}+ \frac{\mathrm{ArcTan}(1/\sqrt{a})}{2\pi^2} \ln \frac{16\Lambda_{\rm U}^4}{\tilde{J}'^4|\Omega_m|^2}& \ \ \  4\tilde{J}^2\ll \mathrm{min}\{a\tilde{J}'^4|\Omega_m|,\tilde{J}'^4|\Omega_m|\}.
\end{matrix}\right.
	\end{aligned}
	\label{bo-fer2-1}
\end{equation}

After summing over Matsubara frequency, the fermion self-energy at zero temperature is derived for low and high energy regimes.
In the low energy regime set by,
\begin{equation}
\frac{N'|J|^2}{N\pi\sqrt{a}}\gg \max\{\frac{|J'|^4\Lambda_\theta^4|\Omega_m|}{4a^2\pi^6},\frac{|J'|^4\Lambda_\theta^4|\Omega_m|}{4a\pi^6}\}
\end{equation}
the complete form of the fermion self energy is given by
\begin{equation}
	\begin{aligned}
\Sigma(i\omega_n,0)-\Sigma(0,0)=&-i\frac{|J|}{4\sqrt{\pi}}\sqrt{\frac{N'}{N}}[a^{1/4}+a^{-1/4}]\mathrm{sgn}(\omega_n)|\omega_n|^{1/2}-i\frac{|J'|^2\Lambda_\theta \mathrm{ArcTan}(\sqrt{a})}{8\pi^4\sqrt{a}}[\omega_n+\omega_n\ln\frac{16\pi a^{3/2}N'\Lambda_{\rm U}^4}{N|J|^2|\omega_n|}]\\
&-i\frac{|J'|^2\Lambda_\theta \mathrm{ArcTan}(1/\sqrt{a})}{8\pi^4\sqrt{a}}[\omega_n+\omega_n\ln\frac{16\pi a^{1/2}N'\Lambda_{\rm U}^4}{N|J|^2|\omega_n|}]
	\end{aligned}
	\label{fi}
\end{equation}
In the high energy regime set by,
\begin{equation}
\frac{N'|J|^2}{N\pi\sqrt{a}}\ll \max\{\frac{|J'|^4\Lambda_\theta^4|\Omega_m|}{4a^2\pi^6},\frac{|J'|^4\Lambda_\theta^4|\Omega_m|}{4a\pi^6}\}
\end{equation}
the complete form of the fermion self energy is given by
\begin{equation}
	\begin{aligned}
	\Sigma(i\omega_n,0)-\Sigma(0,0)=&-i\frac{|J|^2\pi N'\sqrt{a}}{8N|J'|^2\Lambda^2_\theta}(\ln\frac{N\Lambda^4_\theta |J'|^4|\omega_n|}{\pi^5\sqrt{a}N'})^2-i\frac{|J|^2\pi N'\sqrt{a}}{8N|J'|^2\Lambda^2_\theta}(\ln\frac{N\Lambda^4_\theta |J'|^4|\omega_n|}{\pi^5a^{3/2}N'})^2
-i\frac{|J'|^2\Lambda_\theta \mathrm{ArcTan}(\sqrt{a})}{4\pi^4\sqrt{a}}[\omega_n\\
&+\omega_n\ln\frac{8\pi^3aN'\Lambda_{\rm U}^2}{N|J'|^2\Lambda^2_\theta|\omega_n|}]-i\frac{|J'|^2\Lambda_\theta \mathrm{ArcTan}(1/\sqrt{a})}{4\pi^4\sqrt{a}}[\omega_n+\omega_n\ln\frac{8\pi^3aN'\Lambda_{\rm U}^2}{N|J'|^2\Lambda^2_\theta|\omega_n|}]
	\end{aligned}
	\label{se}
\end{equation}

These expression can simplified in respective energy regimes:

1) When the bosonic frequency $|\Omega_m|\ll N'|J|^2/N|J'|^4$, we take the first limit (\ref{fi}), and the dominant contribution to the boson self-energy comes from the Landau damping term $\tilde{J}^2|\Omega_m|/|\epsilon_{\bm k}|$. In this region, if  the fermionic frequency $|\omega_n|\ll N'|J|^2/N|J'|^4$, the system exhibits NFL behavior described by Eq.(\ref{sic-fer-5}); In the opposite limit $|\omega_n|\gg N'|J|^2/N|J'|^4$, it exhibits MFL behavior described by Eq.(\ref{sdc-fer-2}).

2) When the bosonic frequency $|\Omega_m|\gg N'|J|^2/N|J'|^4$, we take the second limit (\ref{se}), and the dominant contribution to the boson self-energy comes from the damping term $\tilde{J}^{'2}|\Omega_m|$.  In this case, the low-frequency limit requires $|\omega_n|\gg|\Omega_m|$,and only MFL behavior is exhibited.  Despite the presence of a new scaling with fermion self-energy $\sim \ln |\omega_n|$ in (\ref{se}), it does not manifest.

\end{document}